\documentclass[11pt,letterpaper]{article}

\usepackage[includeheadfoot,
            marginratio={1:1,2:3}, 
            width=412pt,
            height=688pt,]{geometry}
            \usepackage{tikz}
\usepackage[utf8]{inputenc}
\usepackage{amsmath}
\usepackage{amsfonts}
\usepackage{amssymb}
\usepackage{amsthm}
\usepackage{graphicx}
\usepackage{booktabs}
\usepackage{verbatim}
\usepackage{empheq}
\usepackage{paralist}
\usepackage{cite}
\usepackage[hidelinks]{hyperref}
\usepackage{xcolor}
\usepackage{tensor}
\usepackage{adjustbox}
\usepackage{subcaption}
\usepackage{braket}
\usepackage{tikz}
\usepackage{tikz-cd}
\usepackage{yfonts}
\usepackage{enumitem}
\usepackage{physics}
\usepackage{slashed}
\usepackage{mdframed}
\usepackage{simpler-wick}
\usepackage{hhline}

\usetikzlibrary{decorations.pathreplacing,calc}

\newcommand{\nc}{\newcommand}
\nc{\lb}{\llbracket}
\nc{\rb}{\rrbracket}
\nc{\gl}{\llbracket}
\nc{\gr}{\rrbracket}
\nc{\bbR}{\mathbb{R}}
\nc{\bbC}{\mathbb{C}}
\nc{\bbZ}{\mathbb{Z}}
\nc{\cO}{\mathcal{O}}
\nc{\cS}{\mathcal{S}}
\nc{\cM}{\mathcal{M}}
\nc{\cT}{\mathcal{T}}
\nc{\cX}{\mathcal{X}}
\nc{\cQ}{\mathcal{Q}}
\nc{\cA}{\mathcal{A}}
\nc{\cD}{\mathcal{D}}
\nc{\cL}{\mathcal{L}}
\nc{\cC}{\mathcal{C}}
\nc{\cG}{\mathcal{G}}
\nc{\cF}{\mathcal{F}}
\nc{\cI}{\mathcal{I}}
\nc{\cN}{\mathcal{N}}
\nc{\pd}{\partial}
\nc{\la}{\lambda}

\newcommand\beq{\begin{equation}}
\newcommand\eeq{\end{equation}}
\nc{\del}{\partial}
\nc{\tri}{\hspace{-3.5pt}\vartriangle\hspace{-3.5pt}}
\nc{\blacktri}{\blacktriangle}
\nc{\eq}[1]{\begin{equation}
                     \begin{split} #1 \end{split}
                     \end{equation}}
\nc{\ul}{\underline}
\nc{\ov}{\overline}
\nc{\fa}{\hat}
\nc{\fb}{\MakeUppercase}
\nc{\fc}{\tilde }
\nc{\Lie}{{\cal L}} 
\nc{\lambdabar}{{\mkern0.75mu\mathchar '26\mkern -9.75mu\lambda}}

\newmuskip\pFqmuskip

\newcommand*\pFq[7][8]{
  \begingroup 
  \pFqmuskip=#1mu\relax
  \mathchardef\normalcomma=\mathcode`,
  \mathcode`\,=\string"8000
  \begingroup\lccode`\~=`\,
  \lowercase{\endgroup\let~}\pFqcomma
  {}_{#2}{#3}_{#4}{\left[\left.\genfrac..{0pt}{}{#5}{#6}\right|#7\right]}
  \endgroup
}
\newcommand{\pFqcomma}{{\normalcomma}\mskip\pFqmuskip}

\allowdisplaybreaks[2]

\theoremstyle{definition}
\newtheorem{definition}{Definition}[section]

\begin{document}

\vspace*{1.5cm}
\begin{center}
{\LARGE
\textbf{On the Complexity of Effective Theories}}\\[.4cm]
{\Large
\textbf{-- Seiberg-Witten theory --}}

\end{center}

\vspace{0.35cm}
\begin{center}
 {\hspace{-0.05cm}\bf Martin Carrascal$^{1,3}$, Ferdy Ellen$^1$, Thomas W.~Grimm$^{1,2}$} and
 {\bf David Prieto$^{1,2}$}
\end{center}

\vspace{.25cm}
\begin{center} 
\emph{
$^1$\, Institute for Theoretical Physics, Utrecht University,
\\
Princetonplein 5, 3584 CC Utrecht, 
The Netherlands\\
$^2$\, Center of Mathematical Sciences and Applications,\\
Harvard University, Cambridge, MA 02138, USA \\
$^3$\, Department of Mathematical Sciences, University of Liverpool,\\ 
Liverpool, L69 7ZL, United Kingdom}\\
\end{center}

\vspace{2.5cm}


\begin{abstract}
\noindent
Motivated by the idea that consistent quantum field theories should admit a finite description, we investigate the complexity of effective field theories using the framework of effective o-minimality. Our focus is on quantifying the geometric and logical information required to describe moduli spaces and quantum-corrected couplings. As a concrete setting, we study pure $\mathcal{N}=2$ super-Yang-Mills theory along its quantum moduli space using Seiberg-Witten elliptic curves. We argue that the complexity computation should be organized in terms of local cells that cover the near-boundary regions where additional states become light, each associated with an appropriate duality frame. These duality frames are crucial for keeping the global complexity finite: insisting on a single frame extending across all such limits would yield a divergent complexity measure. This case study illustrates how tame geometry uses dualities to yield finite-complexity descriptions of effective theories and points towards a general framework for quantifying the complexity of the space of effective field theories.
\bigskip

\end{abstract}

\clearpage

\tableofcontents


\newpage

\parskip=.2cm

\section{Introduction}

A guiding theme of the recent works \cite{Grimm:2021vpn,Douglas:2022ynw,Douglas:2023fcg,Grimm:2023xqy,Grimm:2024elq} is to ask which physical effective field theories admit tame descriptions in the framework of o-minimal structures. Originating in mathematical logic, o-minimality provides a rigorously controlled notion of tame geometry \cite{van1998tame}: allowed sets and functions have finite geometric complexity and exclude pathologies such as wild oscillations or infinitely many connected components. This makes o-minimality a natural language for expressing finiteness and regularity properties expected of physically admissible theories.
In this work we bring this framework to bear on a class of effective theories that has been studied in great depth in the past: four-dimensional $\cN=2$ pure super-Yang–Mills. The geometric solution of Seiberg and Witten \cite{Seiberg:1994rs,Seiberg:1994aj} provides exactly the leverage we need, since it encodes the low-energy theory in period integrals varying on a family of auxiliary complex curves, making the effective field theory data amenable to an o-minimal analysis. We use this structure to quantify, patch by patch on the Coulomb branch, the complexity of the individual effective descriptions and of their defining data.

While o-minimality has been developed over decades into a robust qualitative language of tameness, only recently has a genuinely quantitative approach emerged that assigns explicit complexities to tame sets and tracks how they propagate under the basic operations of tame geometry. These quantitative notions are known as effective and sharp o-minimality \cite{binyamini2023tameness,binyamini2022sharplyominimalstructuressharp} and equip tame objects with a numerical complexity consisting of one or multiple positive integers. 
Yet, beyond the defining axioms, concrete structures with a workable, explicitly controlled complexity theory have remained scarce (in contrast to the sizable list of known o-minimal structures). 
Binyamini’s recent advance \cite{binyamini2024log} aims at filling this gap by introducing Log–Noetherian (LN) functions and proving that they generate an effectively o-minimal structure $\mathbb{R}_{\mathrm{LN}}$
 endowed with an explicit notion of complexity (a format $\cF$). 
 Passing to the so-called Pfaffian closure $\mathbb{R}_{\mathrm{LN},\mathrm{PF}}$, this structure is sufficiently large to assign a complexity to period integrals and sets a solid foundation to assign a
 complexity to the effective field theories arising from $\cN=2$ super Yang-Mills theory.

In $\mathcal{N}=2$ $SU(2)$ super Yang-Mills, the Coulomb branch is a one-complex-dimensional moduli space (the $u$--plane) parameterized by the vacuum expectation value of the complex scalar fields in the super-multiplet. Seiberg and Witten recast the low--energy theory geometrically: to each point one associates an elliptic curve with a distinguished differential; its two independent periods encode the effective coupling, central charges, and BPS spectrum~\cite{Seiberg:1994rs,Seiberg:1994aj}. The $u$-plane contains singular loci where specific BPS states become massless, so there is no single global Lagrangian; instead one has multiple effective theories, each valid on a natural patch (electric at weak coupling, magnetic near the monopole point, and dyonic near the dyon point). These theories are connected by duality transformations in $SL(2,\mathbb{Z})$ which rotate the lattice of electric and magnetic changes. With this information one can glue the local descriptions into a coherent global solution which dictate how couplings transform between patches. This patch-wise structure is exactly what we will exploit when assigning and controlling complexity.

This work shows that the physical concepts of having multiple effective theories in patches of the quantum Coulomb branch, naturally maps to the complex cell decomposition of \cite{binyamini2024log} (see \cite{binyamini2019complex} for the underlying construction). Three punctured discs adapt to the the three singular points of the Coulomb branch, and we find that three additional discs are needed to cover the remaining parts of the moduli space. Within each disc one can then compute the complexity $\cF$ of the fully quantum corrected gauge-coupling function.\footnote{This formalism allows us to extend the complexity computation performed in \cite{Grimm:2023xqy} on a real slice of the moduli space to the full complex moduli space. While \cite{Grimm:2023xqy} uses Pfaffian complexity, we now need the more involved complexity notion for $\mathbb{R}_{\mathrm{LN},\mathrm{PF}}$.} 
A key observation is that these local bounds remain finite only as long as one does not extend the discs to reach another singular point. This basic observation shows that the complexity only remains finite if one allows for a change of duality frame and a new local description. By gluing these finitely many patches together, each with controlled complexity, we then obtain an overall finite complexity for the effective couplings on the full moduli space. This remarkable behavior matches with the recent observations on the need of duality symmetries in hyperbolic field spaces \cite{Delgado:2024skw} and their connection with tame geometry \cite{Grimm:2025lip}. 

This paper is organized as follows. In section \ref{ch:tameness-eff-o-minimality} we first recall some basics on tameness formulated through o-minimal structures. We then introduce effective o-minimal structures with a complexity measure $\cF$ and give its explicit form for the Log-Noetherian structure $\bbR_{\rm LN}$ and its Pfaffian extension $\bbR_{\rm LN,PF}$. 
This section is accompanied by a technical appendix~\ref{Effective_appendix}, which lays out more of the mathematical structure and describes the connection of the complexity into logic.
The explicit computation of the format for elliptic curves is then presented in section \ref{effectiveformatofellipticcurves}, where we compute $\cF$ for the period map in the individual patches of the Coulomb branch. The applications of these results to 
$SU$(2) $\cN=2$ super Yang-Mills theory can then be found 
in section~\ref{complex_SW}. We will conclude this work by giving an outlook towards further application of these ideas, and comment on the complexity bounds that we expect to arise when coupling super Yang-Mills theory to gravity.

\section{Tameness and effective o-minimality} \label{ch:tameness-eff-o-minimality}

In this section we introduce many of the mathematical concepts relevant throughout this work. We begin with a 
brief recap of tameness defined via o-minimal structures in section~\ref{sec-tame}. This part will be short and we refer to the excellent textbook \cite{van1998tame} for further basic results on tame geometry. Sections~\ref{sec-eff-omin}-\ref{ch:example-computations} review more recent developments and introduce the concepts of effective o-minimality (section \ref{sec-eff-omin}) and the precise complexity measure $\cF$ introduced by Binyamini in \cite{binyamini2024log} in section \ref{sec-logNoeth}. Since this material is likely new to most of the readers, we have chosen to give a rather detailed exposition, include several examples in section~\ref{ch:example-computations}, and summarize a more in-depth mathematical approach in appendix~\ref{Effective_appendix}. To get an intuitive 
understanding, we note that one can think of $\cF$ as being the amount of information needed to define a set or a function.

\subsection{Tameness from o-minimal structures}\label{sec-tame}

Our discussion of complexity starts by a quick review of tameness in the context of o-minimality. O-minimality, a mathematical subject originally from model theory (mathematical logic), is the study of subsets of $\mathbb{R}^n$ that satisfy some generalized finiteness properties. These properties will be necessary (but not sufficient) in order to assign a complexity to them later. To be more precise, we introduce the notion of an o-minimal structure as a collection $\mathcal{S} = (\mathcal{S}_m)_{m \in \mathbb{N}}$, where each $\mathcal{S}_m$ is itself a collection of subsets of $\mathbb{R}^m$. Every $\mathcal{S}_m$ must be closed under finite unions, intersections and complements, as well as Cartesian products and linear projections. Moreover, $\mathcal{S}$ should contain all algebraic sets (zero sets of polynomials). The generalized finiteness comes from the o-minimality axiom, which states that $\mathcal{S}_1$ (i.e.~the collection of subsets of $\mathbb{R})$ must consist of \textit{finite} unions of points and intervals. Due to the closure under set-theoretic operations, this finiteness then extends to the other $\mathcal{S}_m$'s as well. 

Given an o-minimal structure $\mathcal{S}$, sets included in the structure are said to be tame or definable in $\mathcal{S}$. Moreover, one also defines a function $f:A\to B$ to be tame if its graph $\text{Graph}(f)$ is a tame set in $\mathcal{S}$.

There are many examples of different o-minimal structures. The simplest one is known as $\mathbb{R}_{\text{alg}}$, generated by semi-algebraic sets (i.e.~sets defined by polynomial (in)equalities). A more elaborate one, that is of relevance in physics (see e.g. \cite{Douglas:2022ynw, Grimm_2022, Grimm:2021vpn}), is known as $\mathbb{R}_{\text{an,exp}}$. It is the smallest structure containing the graph of the exponential function, as well as the graphs of all analytic functions, restricted to compact domains. Note that this restriction to compact domains is crucial because analytic
functions on infinite domains can contain infinitely many zeroes, which means that $\mathcal{S}_1$ should contain an infinite union of points, in contradiction with the o-minimality axiom.

While the structure $\mathbb{R}_{\text{an,exp}}$ is sufficiently large for certain physical applications, it is \textit{too} large to admit a consistent notion of complexity for each set within it. Therefore, we will be focusing on a smaller structure known as $\mathbb{R}_{\text{LN,PF}}$, introduced in \cite{binyamini2024log}, which is still large enough to contain all period mappings of algebraic varieties, meaning it still contains many functions found in physics. This structure will suffice to discuss Seiberg-Witten theory, but also contains, for example, finite-loop Feynman amplitudes, seen as functions of the masses and external momenta  \cite{Douglas:2022ynw}.

\subsection{Effective o-minimality}\label{sec-eff-omin}

Having set the basic playing field for our complexity formalism, we will now refine further to \textit{effective} o-minimal structures: these are the structure that admit a notion of complexity, quantified by a \textit{format} $\mathcal{F}$ \cite{binyamini2024log}. This format can be seen as a measure of the following things (which are all related):
\begin{itemize}
    \item It characterizes the number of logical components necessary to write down a formula that defines a set. For example, a polynomial is made out of basic building blocks involving the multiplication function $\cdot$ and the addition function $+$. The format gives a rough measure of how many times these functions appear and therefore how many real coefficients must be specified to fix the polynomial. More generally, complicated functions can also be assigned such a format quantifying their complexity, often depending on systems of differential equations. The larger and more complicated the system, the higher the complexity of the function, again measuring the number of logical components necessary. A more in-depth discussion on the connection between logic and effective o-minimality can be found in appendix \ref{Effective_appendix}.
    \item It characterizes how often the derivative of a function can change sign. This is done by the so-called cellular decomposition: any set in an (effective) o-minimal structure can be split up into a finite number of (possibly unbounded) cells, whose walls are themselves built from functions in the structure. Crucially, for a function $f:\mathbb{R}^n \to \mathbb{R}$, there is a cell decomposition such that the function is differentiable on them and any discontinuities in $f$ or its derivatives occur on the boundaries between adjacent cells. Then on any cell, the set $\{(x_1, \dots, x_n, y)\, |\, y=\partial_i f , y>0\}$ is in the structure, and will have a number of connected components given by some primitive recursive function\footnote{A primitive recursive function $f: \mathbb{N}^n \to \mathbb{N}$ is a function defined recursively only using finitely many times the basic operations of addition by one, projection of a specific coordinate, composition of other primitive recursive functions, and recursion using other primitive recursive functions.} of the format $\mathcal{F}$ of the function, and the number of cells on which we perform this procedure is also bounded by a primitive recursive function of the format $\mathcal{F}$ \cite{binyamini2022sharplyominimalstructuressharp, binyamini2024log}.
    \item It gives an estimate on the volume of functions or sets within some ambient space. To be more precise, if $B^n(r)$ is an $n$-dimensional ball, and $A \subset \mathbb{R}^n$ is some definable set with format $\mathcal{F}$ and of dimension $l$, then this volume satisfies the Gabrielov bound \cite{Gabrielov-1968,yomdin-2004}:
    \begin{equation}
        \text{Vol}_l(A \cap B^n(r)) \leq C(n,l,\mathcal{F}) \cdot r^l,
    \end{equation}
    where $C(n,l,\mathcal{F})$ is a primitive recursive function of $n,l,\mathcal{F}$. In other words, the format measures how much the function or set `wiggles around'.
\end{itemize}

\paragraph{Defining effective o-minimal structures.} Let us now discuss the defining features of these structures. An effective o-minimal structure $\cS$ is an o-minimal structure endowed with a filtration $(\Omega_\mathcal{F})_{\mathcal{F} \in \mathbb{N}}$ such that each set $A$ in the structure belongs to some $\Omega_{\mathcal{F}}$. We then say that $A$ has format $\mathcal{F}$. Having a filtration means that $A \in \Omega_{\cF}$ automatically also belongs to $\Omega_{\mathcal{F}+1}$ as well. We also require that the format transfers consistently through the set-theoretic operations. A final key condition is to ensure that the format can be used to give universal bounds on the number of connected components of a set. Together, we impose the  
conditions:
\begin{itemize}
    \item[(a)] The filtration is increasing  $\Omega_\cF \subset \Omega_{\cF+1}$, and exhaustive $\cup_\cF \Omega_\cF = \cS$.   
    \item[(b)] If $A,B \subset \bbR^n$ and $A \in \Omega_{\cF(A)}$, $B \in \Omega_{\cF(B)}$ then
    \begin{equation} \label{ax-b1}
         A \cup B,\ A \cap B,\ A \times B \in \Omega_{\max\{\mathcal{F}(A), \mathcal{F}(B)\} +1}\,,
    \end{equation}
    while 
    \beq \label{ax-b2}
      \mathbb{R}^n \setminus A, \pi_k(A) \in \Omega_{\mathcal{F}(A)+1}\,,
    \eeq
    where $\pi_k$ is the projection to the first $k$ coordinates.
\item[(c)] There exists of a universal (primitive recursive) function $\mathcal{E}(\mathcal{F})$ such that for each $A \subset \bbR$ with format $\cF$ the number of connected components of $A$ is bounded by $\mathcal{E}(\cF)$. 
\end{itemize}

Let us make a few comments to unpack these abstract axioms. First, condition (c) upgrades the defining o-minimality statement, that each $A\subset \bbR$ has only finitely many connected components, into a quantitative (effective) bound. 
Second, (a) implies that the format assigned to a set is not unique, yet every set $A$ of an effective o-minimal structure admits a minimal format $\cF_{\rm min}(A) \in \mathbb{N}$. In practice, conditions \eqref{ax-b1}, \eqref{ax-b2} provide only consistency requirements on $\cF_{\rm min}$ and explicitly finding $\cF_{\rm min}(A)$ can be notoriously difficult. Nevertheless,  \eqref{ax-b1}, \eqref{ax-b2} can be used to infer consistent format assignments for $A \cup B$, $A \cap B$, $A \times B$ via 
\beq    
\label{eq:formatrules-1}
         \mathcal{F} (A \cup B) = \mathcal{F} (A \cap B) = \mathcal{F} (A \times B) = \max\{\mathcal{F} (A), \mathcal{F} (B)\} +1\,,
\eeq
 and for $\mathbb{R}^n \setminus A$, $\pi_k(A)$ via
\beq
        \label{eq:formatrules-2}
        \mathcal{F}(\mathbb{R}^n \setminus A) =\mathcal{F}(\pi_k(A)) = \mathcal{F}(A)+1\,.
\eeq
Note that although the formats on the right‑hand side may be minimal, the corresponding set formats on the left need not be, as illustrated by choosing $A=B$.

\paragraph{Format of logical formulas.} Since o-minimality originated in the field of mathematical logic, the above concepts also have equivalent formulations in terms of logical formulas instead of sets. To make the translation, one notes that any logical formula $\phi (x)$ defines a set as $\{ x \in \mathbb{R}^n \, | \, \phi(x) \}$. So, similarly to how we assign formats to sets, we can also assign formats to formulas, and it will turn out that this is often more convenient. 
The definition of effective o-minimal structures can then alternatively proceed by first defining the notion of an  \textit{atomic formula} and its associated format. Atomic formulas may look different depending on the o-minimal structure, and specifying all of them constitutes specifying the entire o-minimal structure. They form the basic building blocks of all other formulas, and are then combined using logical symbols, namely conjunction $\wedge$ (`and'), disjunction $\vee$ (`or'), negation $\neg$ (`not'), as well as the existential quantifier $\exists$ (`there exists').\footnote{The universal quantifier $\forall$ (`for all') can be written as $\neg \exists \neg$ so we do not need to consider it separately.}  Similarly to equations \eqref{eq:formatrules-1} and \eqref{eq:formatrules-2}, we demand that the format of formulas $\phi_1, \phi_2$ behaves as
    \begin{equation}
    \label{eq:format-formulas1}
        \mathcal{F}(\phi_1 \wedge \phi_2) = \mathcal{F}(\phi_1 \vee \phi_2) = \max\{\mathcal{F}(\phi_1), \mathcal{F}(\phi_2)\} +1\,,
    \end{equation}
    and
    \begin{equation}
    \label{eq:format-formulas2}
        \mathcal{F}(\neg \phi_1) = \mathcal{F}(\exists x_1,\dots,x_n: \phi_1(x_1,\dots,x_n,y)) = \mathcal{F}(\phi_1)+1\,.
    \end{equation}
This shows that we can think of the $+1$ in \eqref{eq:format-formulas1} and \eqref{eq:format-formulas2} as counting the number of logical operations $\wedge,\vee,\neg,\exists$ applied to the formulas $\phi_1,\phi_2$. A more in-depth discussion on logical formulas and their formats can be found in appendix \ref{Logic-semialg}.

\paragraph{The effective o-minimal structure $\bbR_{\rm alg}$.} The simplest example of an effective o-minimal structure is $\mathbb{R}_{\text{alg}}$. There are various 
ways to endow this structure with a format filtration $(\Omega_\mathcal{F})_{\mathcal{F} \in \mathbb{N}}$.\footnote{Note that in contrast to the axioms 
of sharp o-minimality \cite{binyamini2023tameness,binyamini2022sharplyominimalstructuressharp}, there is no axiom on the format for polynomials for effective o-minimal structures.} We will discuss a number of possibilities 
in the following:
\begin{itemize}
  \item A simple format assignment can be inferred from the sharp o-minimality axioms for algebraic sets \cite{binyamini2023tameness,binyamini2022sharplyominimalstructuressharp}. This amounts to setting 
\begin{equation} \label{F-poly1}
    \mathcal{F}(\{P(x_1,\dots,x_n) =0\}) = \max\{n,\text{deg}\, P\}\,,
\end{equation}
where $P$ is a polynomial. Given this assignment, one obtains the format for polynomial inequalities from the projection rule presented in \eqref{eq:formatrules-2}. The corresponding primitive recursive function $\mathcal{E}$ can be deduced from B\'ezout's theorem ensuring that $\mathcal{E}(\mathcal{F}) =C \mathcal{F}^{\mathcal{F}}$ is indeed a universal bound on the connected components for $C$ a sufficiently large universal constant. Note that one can upper-bound \eqref{F-poly1} by formats that account for a broader information set, i.e. including the size of the coefficients of the polynomial, by setting 
\beq
  \mathcal{F}(\{P(x_1,\dots,x_n) =0\}) = n + \text{deg}\,P + \Vert P\Vert\,,
\eeq
where $\Vert P \Vert$ is the sum of the absolute values of the coefficients of $P$ rounded up to the next integer. This shows that effective o-minimal structures might have complexity notions that are far from optimal when it comes to bounding the number of connected components. Ideally, one aims to find the measures that provide the tightest universal bounds. 

\item A second format assignment arises from first specifying atomic formulas in $\bbR_{\rm alg}$ and then assigning them a format. In a structure with symbols $(=,>,\cdot,+, a)$, where $a\in \mathbb{R}$ is a constant, the atomic formulas consist of combinations of these symbols.\footnote{Of course, $=$ and $>$ can only appear once in such an atomic formula. An expression like $x=y=z$ should be written as $x=y \wedge y=z$.}
We assign a format to all of them:
\begin{equation} \label{eq:formats-atomic-symbols}
    \mathcal{F}(=)=\mathcal{F}(>)=\mathcal{F}(\cdot) = \mathcal{F}(+)=\mathcal{F}(a)=1.
\end{equation}
We then declare the format of an atomic formula to be the sum of the formats of all the symbols appearing in it (variables do not add to the format). For example, the formula $y=3 \cdot x_1 + 5 \cdot x_1 \cdot x_2$ has format 7, since it contains once the symbol $=$, once the symbol $+$, three times the symbol $\cdot$, and two constants $3$ and $5$. Note that even for polynomials this number might not be the minimal format one can assign to a formula, since there might exist a simpler representations using conjunction and \eqref{eq:format-formulas1}.
While this format assignment is more unwieldy than \eqref{F-poly1}, it allows for generalizations to larger structures. We include a discussion on this logic-based approach in appendix~\ref{Logic-semialg}. 

\end{itemize}

\paragraph{Beyond $\bbR_{\rm alg}$ -- including non-polynomial functions.}
In order to generalize beyond $\mathbb{R}_{\text{alg}}$ and include more interesting functions in our effective o-minimal structure, we will be looking at functions that satisfy certain differential equations. To see why, consider the o-minimal structure $\mathbb{R}_{\text{an,exp}}$. A generic restricted analytic function of one variable is written as
\begin{equation}
    f(x) = \sum_n a_n x^n\,.
\end{equation}
In this expression, the coefficients $a_n$ are a priori undetermined and all independent of each other. Hence, fully specifying such a function in general requires specifying an infinite number of coefficients, which will naturally have infinite complexity. One can make this argument more rigorous, as done in \cite{binyamini2023tameness}, but the upshot is that the structure $\mathbb{R}_{\text{an,exp}}$ admits no format filtration and is therefore not effective o-minimal. On the other hand, many functions that are used in physics \textit{are} analytic, and the fact that they satisfy differential equations is key in assigning them a finite complexity. For example, the exponential function satisfies the differential equation $f'=f$, meaning that its coefficients are related as
\begin{equation}
    a_{n+1} = \frac{a_n}{n+1}\,,
\end{equation}
and therefore we only need to specify $a_0$ in addition to this relation. If we want to consider such functions on unrestricted domains, however, satisfying a generic differential equation is not enough to guarantee even plain o-minimality, let alone finite complexity. For example, the sine function satisfies $f''=f$, but taking the intersection of its graph with the $x$-axis, we obtain a set with an infinite number of connected components, in contradiction to the o-minimality axiom. Thus, there are two options available to extend the effective o-minimal structure $\mathbb{R}_{\text{alg}}$:
\begin{itemize}
    \item Keep the domain restricted and allow for general differential equations. This approach is taken by Binyamini in \cite{binyamini2024log} to define $\mathbb{R}_{\text{LN}}$, the effective o-minimal structure of Log-Noetherian functions.
    \item Allow for an unbounded domain, but restrict the types of differential equations that can be satisfied. An example of this is given by Pfaffian functions, which satisfy so-called Pfaffian chains of differential equations and fit into an effective o-minimal structure known as $\mathbb{R}_{\text{PF}}$. This structure, first investigated by Khovanskii in \cite{khovanskiĭfewnomials} before the modern notion of o-minimality, was proven to be o-minimal by Wilkie in \cite{Wilkie1999ATO}. Its complexity properties as an o-minimal structure are explored in detail in \cite{binyamini2020effectivecylindricalcelldecompositions,binyamini2022sharplyominimalstructuressharp,binyamini2023tameness} and physical applications are investigated in \cite{Grimm:2023xqy,Grimm:2024elq,Grimm:2024mbw}.
\end{itemize}
Finally, these approaches can be combined into a structure known as $\mathbb{R}_{\text{LN,PF}}$, the Pfaffian extension of Log-Noetherian functions. This effective o-minimal structure will be the focus of this paper, as it contains the period maps of algebraic varieties, making it a suitable replacement for $\mathbb{R}_{\text{an,exp}}$ in many situations, with the added advantage that we are able to assign complexities to sets and functions appearing within it.

\subsection{Log-Noetherian Complexity}\label{sec-logNoeth}

\begin{figure}[h]
  \vspace{-5pt}
  \centering
 \includegraphics[width=10cm]{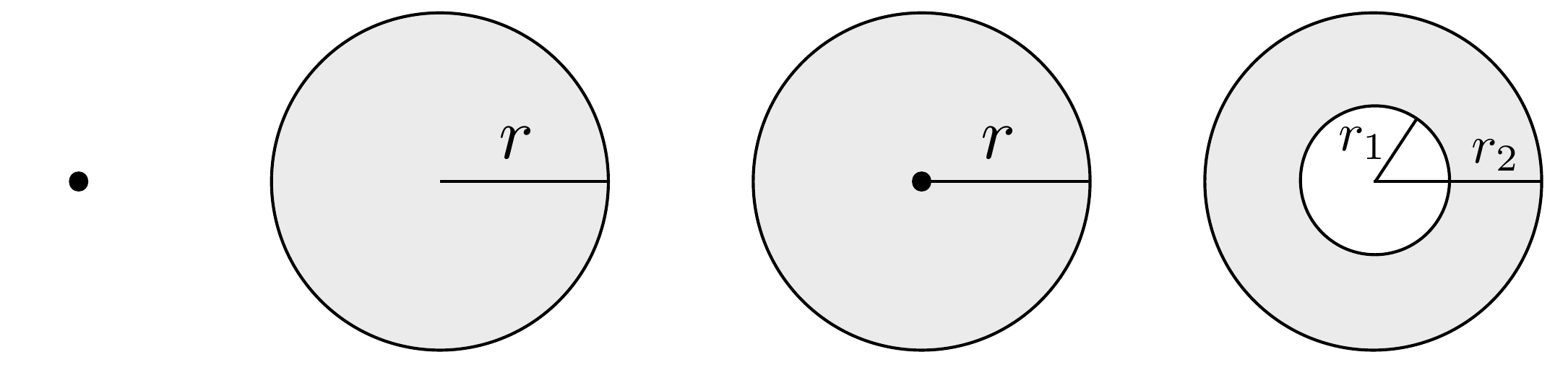}
    \caption{Basic building blocks (cells) for LN-functions: point, disc $D(r)$, punctured disc $D_\circ(r)$, annulus $A(r_1,r_2)$.}
    \label{figblocks}
\end{figure}

In order to construct the effective o-minimal structure $\mathbb{R}_{\text{LN,PF}}$, which will be of interest for our physical applications, we first start with a discussion of Log-Noetherian functions, which generate $\mathbb{R}_{\text{LN}}$. As mentioned, these are functions on a restricted domain $\mathcal{C}$ that obey some differential equations. In full generality, the domain $\mathcal{C}$ is defined using cellular constructions as discussed in appendix \ref{ch:app-lncells}. This construction allows for very general domains by making use of a fibration structure. In summary, one starts from the basic building blocks of a point, a disc $D(r)$, a punctured disc $D_\circ(r)$, and an annulus $A(r_1,r_2)$, and fibers these over each other (e.g.~by allowing the radii $r,r_1,r_2$ to change along the coordinates of the base as LN functions).\footnote{We will explain in appendix~\ref{ch:app-lncells} that in such fibrations the radius functions $r,r_1,r_2$ are generally complex LN-functions and the discs and annuli are evaluated at $|r|,|r_1|,|r_2|$. For our discussion in the main text, real and constant radii will be sufficient.} We display these individual building blocks in 
Figure~\ref{figblocks}.

In the following, we will simplify the setting to product domains only involving complex discs and punctured discs, since this suffices for the examples considered in this work.\footnote{The fact that annuli need to be included as building blocks is essential for many examples \cite{binyamini2019complex}. Otherwise, the number of punctured discs would need to grow to infinity whenever two singularities or boundaries approach each other.} 
Hence, we take $\mathcal{C}$ to be of the form
\begin{equation} \label{simple_cell}
    \mathcal{C} = D_\circ (r_1) \times \cdots \times D_\circ  (r_n) \times D(r_{n+1}) \times \cdots \times D(r_{n+m})\,,
\end{equation}
where $D_\circ(r) = \{z  \in \mathbb{C} \, | \, 0< |z| < r\}$ denote the punctured discs and $D(r) = \{ z \in \mathbb{C} \, | \, |z|< r\}$. Note that these, in contrast to the formalism discussed up until now, are complex sets instead of real ones. In order to construct a Log-Noetherian function $F: \mathcal{C} \to \mathbb{C}$, we find a system of differential equations known as a Log-Noetherian (LN) chain $(F_1, \dots, F_N) : \mathcal{C} \to \mathbb{C}$ as follows:
\begin{equation}
\label{eq:ln-chain}
\begin{gathered}
    z_i \frac{\partial F_1}{\partial z_i} = G_{1i} (F_1, \dots, F_N)\,, \quad r_j\frac{\partial F_1}{\partial z_{j}} = G_{1j} (F_1,\dots,F_N)\,,\\
    \vdots\\
    z_i \frac{\partial F_N}{\partial z_i} = G_{Ni} (F_1, \dots, F_N)\,, \quad r_j\frac{\partial F_N}{\partial z_{j}} = G_{Nj} (F_1,\dots,F_N)\,,
\end{gathered}
\end{equation}
where $z_i$ are the coordinates corresponding to the punctured discs, $z_j$ are those corresponding to the regular discs which have radii $r_j$, and $G_{kl}$ are polynomials in all of the functions in the chain. Note that in particular, the coordinates should also be seen as functions. A Log-Noetherian (LN) function is then a function $F: \mathcal{C} \to \mathbb{C}$ defined as a polynomial of the functions in the chain:
\begin{equation} \label{eq:LN-function}
    F = G(F_1,\dots,F_N)\,.
\end{equation}

In order to assign a format to such an LN function, we first assign one to the LN chain defining it as follows:\footnote{In \cite{binyamini2024log}, a more subtle construction of cells using so-called $\delta$-extensions is used, which is also reflected in the format. In this work, we omit the dependence on this potential extension.} 
\begin{equation}\label{formatLNchain}
    \mathcal{F}(F_1,\dots,F_N)  = \cF(\cC) + N + \sum_{k,l} \text{deg } G_{kl} + \Vert G_{kl} \Vert + \sup_{\substack{i=1, \dots, N\\ \mathbf{z}\in \mathcal{C}}} |F_i (\mathbf{z})|\,,
\end{equation}
where the norm $\Vert \cdot \Vert$ of a polynomial is the sum of the absolute values of its coefficients. To be precise, the format is given by the least integer upper bound to the RHS of \eqref{formatLNchain}. This expression depends on the format of the cell $\cC$, which in the simple case \eqref{simple_cell} is evaluated to be 
\beq
   \cF(\cC)=n+m+\sum_{k=1}^{n+m} \lceil r_k \rceil\ ,
\eeq
where $\lceil \cdot \rceil$ is the ceiling function (see appendix \ref{Effective_appendix} for the general formula). When this format is not an integer, we round it up as well. The format of the LN function is then given in terms of the format of the chain as
\begin{equation} \label{eq:format-ln-func}
    \mathcal{F}(F) = \mathcal{F}(F_1,\dots,F_N) + \text{deg } G + \Vert G \Vert\,.
\end{equation}
Note that this expression has explicit dependence on both the coefficients appearing in the LN chain, the size of the domain, and the supremum of the functions used in the construction of the chain. The supremum term in particular also ensures that the functions $F_i$ cannot have poles on the LN cell, as this would yield an infinite format. Thus, LN functions are bounded functions on bounded domains. 

\paragraph{Completing the format filtration -- atomic formulas of $\mathbb{R}_{\text{LN}}$.}
The format assignment \eqref{eq:format-ln-func} does not yet generate a proper format filtration for the entire structure $\mathbb{R}_{\text{LN}}$, because this structure contains some sets that are not graph of functions of the form \eqref{eq:LN-function}. Namely, while every (semi-)algebraic set should be part of any o-minimal structure by the axioms of o-minimality, such sets may be unbounded. In contrast, LN-functions like \eqref{eq:LN-function} are bounded functions on bounded domains. Thus, to complete the structure and provide a full format filtration, we also need to assign a format to polynomials. Recall that specifying the atomic formulas in a structure together with their formats is enough to generate a format filtration for the entire structure. Therefore, we take as atomic formulas
\begin{equation}
    y=P(x_1,\dots,x_n), \qquad y>P(x_1,\dots,x_n), \qquad (x,y) \in \text{Graph}(F),
\end{equation}
where $P$ is a polynomial and $F$ an LN-function of the form \eqref{eq:LN-function} on some cell $\mathcal{C}$, seen as a real function $(\text{Re } F, \text{Im } F)$ with real variables $(x,y)$ such that $z=x+iy$.\footnote{This represents a slight departure from \cite{binyamini2024log}, where only real LN-function on the real part of their graphs are considered instead. However, in the same work it was shown that the graphs of complex LN functions are also included in the structure, and their format is given by some primitive recursive function of the format of the corresponding real LN-function on the real part of its domain. Therefore, our modification maintains effective o-minimality.} The format for the atomic formula $(x,y) \in \text{Graph}(F)$ is then given by equation \eqref{eq:format-ln-func}, and the format of the formulas $y=P(x_1,\dots,x_n)$ and $y>P(x_1,\dots,x_n)$ are obtained by specifying the formats of all the symbols $=,>,+,\cdot,a$ to be equal to 1, and counting how many of these symbols appear in the atomic formula, as in the discussion around equation \eqref{eq:formats-atomic-symbols}. So, in summary we obtain
\begin{equation} \label{eq:format-polyal}
    \mathcal{F}(y \geqq P(x)) = \text{\# of symbols } (=,>,+,\cdot,a).
\end{equation}

While assigning a format to polynomials by counting the symbols in their expressions gives \textit{a} format, it does not necessarily yield the lowest format, similarly to how different LN-chains may give different formats for the same LN-function.  This is because the polynomial may be expressed in a more efficient way. For example, consider the polynomial
\begin{equation}
    y=2x_1 \cdot x_3^2 + 6\cdot  x_2 + x_4,
\end{equation}
which naively would have a format of $8$ since it contains once the symbol $=$, 3 times the symbol $\cdot$, and twice the symbol $+$, as well as two constants. However, we can introduce some auxiliary variables $t_1, \dots, t_5$ to write this expression in terms of shorter polynomials as
\begin{equation}
    y=  t_1 + x_4 \wedge t_1 = t_2 + t_3 \wedge t_2 = 2 \cdot t_4 \wedge t_4 = x_1 \cdot t_5 \wedge t_5 = x_3 \cdot x_3 \wedge t_3 = 6 \cdot x_2.
\end{equation}
These simpler formulas have formats $2,2,3,2,2,3$ respectively, and repeated use of the rule \eqref{eq:format-formulas1} yields a total format $\mathcal{F}=5$ instead.\footnote{There are two caveats here. First, introducing these auxiliary variables should be accompanied by an existence quantifier $\exists t_1,\dots,t_5$. Since we will be using results such as this as intermediate steps, we have not added it here, instead we include a single existence quantifier at the very end. Secondly, the order in which we perform the rule \eqref{eq:format-formulas1} is relevant and we have performed it in an optimal way here. In appendix \ref{Conj_app} we give a structured way to deal with the resulting nested maxima.}

This subtlety makes assigning polynomials a minimal format hard. For the relatively simple polynomials that will appear in this work, we can manually decompose them into conjunctions of simple formulas of the form $y=x+z$, $y=x \cdot z$ et cetera, and assign them a format using equations \eqref{eq:format-formulas1} and \eqref{eq:format-formulas2}. We will discuss some examples of this in section \ref{ch:example-computations}, and discuss more general considerations in the appendix.

\paragraph{Pfaffian extension.} In order to extend our structure to include functions that may be unbounded (and be defined on unbounded domains) that are \textit{not} polynomials, we now discuss the \textit{Pfaffian extension} of the structure $\mathbb{R}_{\text{LN}}$. The resulting structure will be called $\mathbb{R}_{\text{LN,PF}}$, and the functions within it LN,PF-functions. Such LN,PF-functions are real functions, on real domains of the form:
\begin{equation}
\label{eq:LNPF-domains}
    G = I_1 \times \cdots \times I_n \subset \mathbb{R}^n\,,
\end{equation}
where $I_k= (a_k,b_k)$ is a possibly infinite interval (so $a_k,b_k$ can be $\pm \infty$). Once again, the domains can also be more complicated as elaborated upon in appendix \ref{Effective_appendix}. An LN,PF-function $f: G \to \mathbb{R}$ is then constructed using (again) a system of differential equations. However, this time it is an LN,PF-chain $(\zeta_1,\dots,\zeta_N)$ of differential equations, satisfying a triangular system of the form
\begin{equation}
\label{eq:pfaffian-chain}
    \begin{split}
        \frac{\partial \zeta_1}{\partial x_k} &= P_{1k}(\zeta_1)\,,\\
        \frac{\partial \zeta_2}{\partial x_k} &= P_{2k} (\zeta_1, \zeta_2)\,,\\[-.2cm]
        &\quad \vdots\\[-.2cm]
        \frac{\partial \zeta_N}{\partial x_k} &= P_{Nk} (\zeta_1,\dots,\zeta_N)\,,
    \end{split}
\end{equation}
where, instead of just polynomials, the $P_{jk}$ are now real analytic functions  definable in $\mathbb{R}_{\text{LN}}$ and hence they are either LN functions or polynomials. Their domains should also have a triangular structure such that they contain the images of $(\zeta_1,\dots,\zeta_N)$. In other words, $P_{jk}$ should map some domain $X_j \to \mathbb{R}$ such that $\zeta_1(G) \times \cdots \times \zeta_j(G) \subset X_j$. An LN,PF-function $f: G \to \mathbb{R}$ is then defined as
\begin{equation} \label{def-f}
    f = P(\zeta_1,\dots,\zeta_N)\,,
\end{equation}
where $P: X_N \equiv X \to \mathbb{R}$ is again a real analytic function in $\mathbb{R}_{\text{LN}}$. The structure $\mathbb{R}_{\text{LN,PF}}$ is then generated by the graphs of LN,PF-functions. Note that since $P_{jk}$ can be polynomials, this structure automatically includes the structure $\mathbb{R}_{\text{Pfaff}}$, which was already shown to be very useful for physical applications \cite{Grimm:2023xqy,Grimm:2024elq,Grimm:2024mbw}. 

Since this is a new structure, we should also specify an appropriate format filtration. Once again, we assign a format to an LN,PF-function by first specifying the format of an LN,PF-chain as 
\begin{equation}
    \label{eq:lnpfchainformat}
    \mathcal{F}^{\text{LN,PF}} (\zeta_1, \dots, \zeta_N) = N + \mathcal{F}^{\text{LN}} (G) + \sum_{j,k} \mathcal{F}^{\text{LN}} (P_{jk})\ ,
\end{equation}
where $\mathcal{F}^{\text{LN}}$ denote the LN-formats given by equations \eqref{eq:format-ln-func} and \eqref{eq:format-polyal}. The format of $f$ defined in \eqref{def-f} is then given by
\begin{equation}
    \label{eq:lnpffuncformat}
    \mathcal{F}^{\text{LN,PF}} (f) = \mathcal{F}^{\text{LN,PF}} (\zeta_1, \dots, \zeta_N) + \mathcal{F}^{\text{LN}} (P)\ .
\end{equation}
Finally, while polynomials are LN,PF-functions and in principle do not need to be added by hand, we add them in in exactly the same way as we did for the structure $\mathbb{R}_{\text{LN}}$ to keep computations tractable. This is a slight departure from \cite{binyamini2024log}, but it does maintain effective o-minimality.

From now on, we will permanently work in $\mathbb{R}_{\text{LN,PF}}$ to find the formats of period maps and therefore we will drop the superscript `LN,PF' on the formats, leaving the underlying structure implicit. If we want to refer to formats in the structure $\mathbb{R}_{\text{LN}}$, we use the term `LN-format' and the symbol $\mathcal{F}^{\text{LN}}$.

\subsection{Example computations of complexities in $\mathbb{R}_{\rm LN,PF}$} \label{ch:example-computations}

In order to illustrate how to do concrete computations of complexities of functions in $\mathbb{R}_\text{LN}$ and $\mathbb{R}_{\text{LN,PF}}$, we provide here a few examples, starting with polynomials, which will be necessary for the later calculations. We also provide a computation of the format of the real as well as complex logarithm, since the latter will be necessary later for our computation of the format of period mappings of elliptic curves.

\paragraph{LN-Format of semi-algebraic sets.}
We start with the LN-formats of polynomials. It is important to have a proper understanding of this case, since the functions $P_{jk}$ in a Pfaffian chain of the form \eqref{eq:pfaffian-chain} are often polynomials, and the domain $G$ is often a semi-algebraic set. As discussed previously, such sets have to be constructed using the atomic formulas discussed above \eqref{eq:formats-atomic-symbols}. We list some simple examples:
\begin{itemize}
    \item The set $\mathbb{R}$ is the set corresponding to the formula $x=x$, and since $x=x$ is an atomic formula already, it has LN-format $\mathcal{F}^{\text{LN}}(\mathbb{R})=1$.
    \item An open interval $(a,\infty)$ corresponds to a formula $x>a$ which contains the symbols $>$ and $a$ so it has an LN-format of 2. An open interval $(a,b)$ corresponds to $x<b \ \wedge \ x>a$ which is the conjunction of two formulas that have LN-format 2, for a total LN-format of $\mathcal{F}^{\text{LN}}((a,b))= \max\{2,2\}+1=3$.
    \item A line $y=a x$ contains a constant, an $=$ symbol and an $\cdot$ symbol so it has an LN-format $\mathcal{F}^{\text{LN}}(y=ax)=3$. For the graph $y=ax^2$ we use the decompositon $y=a \cdot t \ \wedge \ t= x \cdot x$, which thus has LN-format $ \mathcal{F}^{\text{LN}}(y=ax^2) = \max\{3,2\}+1=4$.
    \item A general quadratic equation $y=ax^2 + bx + c$ is written as a conjunction $y=t_1 + c \ \wedge \ t_1 = t_2 + t_3 \ \wedge \ t_2 = ax^2 \ \wedge \ t_3 = bx$. These formulas have LN-formats $3,2,4,3$ respectively, so we should take the maxima in an optimal way:
    \begin{equation}
        \mathcal{F}^{\text{LN}}(y=ax^2+bx+c) = \max\{4,\max\{3,\max\{3,2\}+1\}+1\}+1=6.
    \end{equation}
    In appendix \ref{Conj_app} we present a general formula for how to calculate such nested maxima optimally.
\end{itemize}
We will also need to consider complex polynomials. These are constructed analogously, but we should start from complex addition and multiplication instead:
\begin{itemize}
    \item $z_1 = z_2 + z_3$ with $z_1,z_2,z_3 \in \mathbb{C}$ can be written as the conjunction of its real and imaginary part:
    \begin{equation}
        (z_1 = z_2 + z_3) \iff (x_1 = x_2 + x_3) \ \wedge \ (y_1 = y_2 + y_3),
    \end{equation}
    where $z_i=x_i + i y_i$. This is a conjunction of two formulas of format 2, so complex multiplication has a format $\mathcal{F}^{\text{LN}} (z_1 = z_2 + z_3) = 3$.
    \item Complex multiplication $z_1 = z_2 \cdot z_3$ is somewhat more involved:
    \begin{equation}
        (z_1 = z_2 \cdot z_3) \iff (x_1 = x_2 x_3 - y_2 y_3) \ \wedge \ (y_1 = x_2 y_3+x_3 y_2),
    \end{equation}
    which can be checked to have LN-format $\mathcal{F}^{\text{LN}}(z_1 = z_2 \cdot z_3) = 6$. Likewise, replacing $z_3$ by some complex constant $\alpha$ results also in an LN-format $\mathcal{F}^{\text{LN}}(z_1 = \alpha z_2) = 6$.
\end{itemize}
With these basic building blocks, other complex polynomials can be constructed in the same way as real polynomials. From this point on, whenever we assign formats to polynomials, we will have split them up into these smaller formulas, with the appropriate conjunctions taken.

\paragraph{Format of the exponential function.}
Let us consider a simple function definable in $\bbR_{\rm LN,PF}$
that is not definable in $\bbR_{\rm LN}$; namely  $f(x)=e^x$ on the unbounded domain $G =(-\infty,\infty)$. 
Clearly, we can define $e^x$ using the Pfaffian chain
\begin{equation}
    \frac{\partial \zeta_1}{\partial x} = \zeta_1 \equiv P_1 (\zeta_1)\,,
\end{equation}
with solution $\zeta_1 = e^x$. The image of the domain $G$ is $X=  (0,\infty)\subset \mathbb{R}$. Since it will turn out that taking an unrestricted domain for $P_1$ leads to a lower format, we take $X_1=\mathbb{R}$.
The polynomial $P_1 (t_1)=t_1$ has LN-format 1.
The function $f(x)=e^x$ is then given as $f(x)=P(\zeta_1)=\zeta_1$, where $P$ is the same polynomial as $P_1$ and thus also has a format of 1. The format of $G=\mathbb{R}$ is 1, as discussed previously. All together, filling in equation \eqref{eq:lnpffuncformat} yields
\begin{equation}
    \mathcal{F}(e^x) = 1+1+1+1=4\,.
\end{equation}
The crucial part to note here is that in order to be able to assign formats to unbounded sets in the $\mathbb{R}_{\text{LN,PF}}$ setting, we need to relate them to intervals and polynomials, as these are the objects that we know are definable over their full unbounded domain.

\paragraph{Format of the real logarithm.}
For a slightly more involved example, we now focus on the logarithm $f(x) = \ln x$ on the domain $G=(0,\infty)$. This fits into the Pfaffian chain
\begin{equation}
\frac{\partial \zeta_1}{\partial x} = -\zeta_1^2 = P_1 (\zeta_1)\,, \qquad \frac{\partial \zeta_2}{\partial x}{ = \zeta_1} = P_2 (\zeta_1,\zeta_2)\,,
\end{equation}
with solutions
\begin{equation}
    (\zeta_1,\zeta_2) = (1/x,\ln x)\,,
\end{equation}
such that $X = X_1 \times X_2 \supset (0,\infty) \times \mathbb{R}$. Once again, we extend these domains to $X_d=\mathbb{R}^d$ where $d=1,2$. The function $f(x)$ is then given by the LN-function $P$ as $f(x) = P(\zeta_1,\zeta_2) = \zeta_2$. The format of the domain $G$ is equal to 2.  $P_1:t_1 \mapsto -t_1^2$ is a polynomial with format 4, and both $P_2$ as well as $P$ map $(t_1,t_2) \mapsto t_2$ and are given by the graph $\mathbb{R} \times \{y=t_2\}$ so they each have format 2.
Then for the format of the chain and of $f$ we have:
\begin{equation}
    \begin{split}
    \mathcal{F}(\zeta_1,\zeta_2) &= 2+2+4+2=10\,,\\
    \mathcal{F}(\ln x) &= 10+2=12\,.
    \end{split}
\end{equation}

\paragraph{Format of the complex logarithm.} 
Next, we will be computing the format of the complex logarithm $\log z$ on a domain $D_\circ (r)$, with a branch cut along the negative real axis.\footnote{This domain is not of the form \eqref{eq:LNPF-domains} but is still allowed as it is a fibration $G = (-r,r) \odot (-\sqrt{r^2-x^2}, \sqrt{r^2-x^2})$ following the notation discussed in appendix \ref{Effective_appendix}. We assign the same format to a domain with a branch cut as we assign to the full disc.}
As Pfaffian extensions are only defined over the reals, we will split this function up into its real and imaginary part, and also consider it as a function of two real variables $(x,y)$ instead of the complex variable $z$. Hence we get:
\begin{equation}
    \log z = \ln \sqrt{x^2+y^2} + i\, \text{Arg} (x,y)\,.
\end{equation}
We start with the real part, given by $f(x,y) = \ln \sqrt{x^2 + y^2}$, on the domain $D_\circ (r)$. This fits into the following Pfaffian system of differential equations:
\begin{align*}
    \frac{\partial \zeta_1}{\partial x} &= 1 \equiv P_{1x}\,; \qquad &\frac{\partial \zeta_2}{\partial y} &= 1 \equiv P_{2y}\,;\\
    \frac{\partial \zeta_3}{\partial x} &= -2\zeta_1 \zeta_3^2 \equiv P_{3x}\,; \qquad &\frac{\partial \zeta_3}{\partial y} &= -2\zeta_2 \zeta_3^2 \equiv P_{3y}\,;\\
    \frac{\partial \zeta_4}{\partial x} &= \zeta_1\zeta_3 \equiv P_{4x}\,; \qquad &\frac{\partial \zeta_4}{\partial y} &=\zeta_2 \zeta_3 \equiv P_{4y}\,;
\end{align*}
with all other derivatives being zero. This is solved by
\begin{equation}
    (\zeta_1,\zeta_2,\zeta_3,\zeta_4) = \left(x,y,\frac{1}{x^2+y^2}, \ln\sqrt{x^2+y^2} \right),
\end{equation}
and we choose the domains of $P_{jk}$ such that they cover the images of $\zeta_j (G)$ as $X_j = \mathbb{R}^j$ again.
The function $f$ is given by $f(x,y) = P(\zeta_1,\zeta_2,\zeta_3,\zeta_4) = \zeta_4$.
The functions $P_{jk}$ again are polynomials, and have formats $2,3,5,5,3,3$, 
respectively, and $P$ has format 4.\footnote{Note that the domain of all the $P_j$'s has to contain the image of $\zeta_1(G) \times \dots \times \zeta_j(G)$. For example, the graph of the polynomial $P_2:\mathbb{R}^2 \to \mathbb{R}$ is given by $\mathbb{R} \times \{P_{2y}=1\}$ which is why it has format 3.} The LN-format of the domain $G$ (defined by $x^2+y^2<1 \wedge x^2+y^2>0$) is 5.
Altogether we find for the format of the chain and of $f$ respectively:
\begin{equation}
    \mathcal{F}(\zeta_1,\zeta_2,\zeta_3,\zeta_4) =30, \quad
    \mathcal{F}(f) =34\,.
\end{equation}
For the complex part of the logarithm (still on the domain $G= D_\circ  (r)$ with a branch cut), which is given by $g(x,y) = \text{Arg}(x,y)$, we recycle $\zeta_1,\zeta_2, \zeta_3$ from the previous chain and add:
\begin{align*}
    \frac{\partial \tilde{\zeta_4}}{\partial x} &= -\zeta_2\zeta_3 \equiv \tilde{P}_{4x}\,; \qquad &\frac{\partial \tilde{\zeta_4}}{\partial y} &= -\zeta_1\zeta_3 \equiv \tilde{P}_{4y}\,;
\end{align*}
with solution $\tilde{\zeta}_4 = \text{Arg} (x,y)$.
The function $g$ is then given by $g(x,y) = P(\zeta_1,\zeta_2,\zeta_3,\tilde{\zeta}_4) = \tilde{\zeta}_4$. The polynomials $\tilde{P}_{4x}$ and $\tilde{P}_{4y}$ have format 5 each, and $P$ has format 4, so we obtain
\begin{equation}
    \mathcal{F} (\zeta_1,\zeta_2,\zeta_3,\zeta_4) =34 \, , \quad 
    \mathcal{F}(g) =38\,.
\end{equation}
It still remains to merge the real and complex part of the logarithm together. In order to do so, we note that this amounts to the formula $z_1 = f(x,y) \ \wedge \ z_2 = g(x,y)$ and find:
\begin{equation}
    \mathcal{F}(\log z)= 39\,.
\end{equation}
As periods generically contain logarithmic singularities, we will need this result in the next section to calculate their formats.

\paragraph{LN,PF-format of an LN function.} 
As a final example, we will show how to convert the LN-format of an LN-function into an LN,PF-format. This will be necessary later in order to combine the holomorphic part of the period mapping (which is LN) with the part containing logarithmic singularities (which, as shown, is LN,PF). 

Suppose we have an LN-function $F(z)$ of 1 complex variable $z$ on a domain $D_\circ (r)$. As in the complex logarithm, this domain has LN-format 5. For the LN chain we take only the coordinates
\begin{equation}
    \frac{\partial \zeta_1}{\partial x} = 1\, , \quad \frac{\partial \zeta_2}{\partial y} = 1\,,
\end{equation}
solved by $(\zeta_1, \zeta_2) = (x,y)$. The function $F(z)$ is then split up into its real and imaginary parts $F^{R,I}$. Both fit into the chain as $F^{R,I}(x+iy) = P^{R,I}(\zeta_1,\zeta_2)$ and since they are projections of an LN-function, their format is the same as the LN function itself.\footnote{We again postpone the projection to the end of our more involved calculations later.} The LN,PF-format of $F^{R,I}$ is then given by 
\begin{equation}
    \mathcal{F}^{\rm LN,PF}(F^{R,I}) = 12+\mathcal{F}^{\rm LN} (F)\,,
\end{equation}
and as before, combining the real and imaginary parts simply adds 1 to the format, hence we obtain
\begin{equation}\label{eq:embeddingintoRLNPF}
    \mathcal{F}^{\rm LN,PF} (F) = 13 + \mathcal{F}^{\text{LN}} (F)\, .
\end{equation}

\section{Effective format of elliptic curves}\label{effectiveformatofellipticcurves}
In this section we discuss the moduli spaces of elliptic curves, as well as how to assign a complexity to the period maps defining them. We will see that the Picard-Fuchs equations provide an LN-chain for the holomorphic part of the periods, and that the monodromy/duality group corresponding to the elliptic curve plays a key role. We keep the discussion here general and consider different families of elliptic curves that appear in physical theories. The results obtained here will then be applied to the particular case of $\mathcal{N}=2$ Seiberg-Witten theory in section \ref{complex_SW}.

\subsection{Elliptic Curves}
We consider complex elliptic curves, which as smooth manifolds are diffeomorphic to tori. Any such curve admits a description as a double cover of the Riemann sphere $\mathbb{P}^1$, given by the equation\footnote{The notation $\mathbb{P}^n$ will always refer to complex projective space.}
\begin{equation}
    y^2= P(x)\,,
\end{equation}
where $P$ is a polynomial of degree 3 or 4. With a change of variables, these curves can always be brought into Legendre form:
\begin{equation}
\label{eq:legendre-family}
    \mathcal{E}_\lambda :\quad  y^2= x(x-1)(x-\lambda)\,,
\end{equation}
with $\lambda \in \mathbb{P}^1$. At $\lambda=0,1,\infty$ the equation degenerates, and so the family of curves is parametrized by $\lambda \in \mathbb{P}^1\setminus\{0,1,\infty\}$, the thrice-punctured Riemann sphere.

\begin{figure}[h!]
    \centering
    \includegraphics[width=0.4\linewidth]{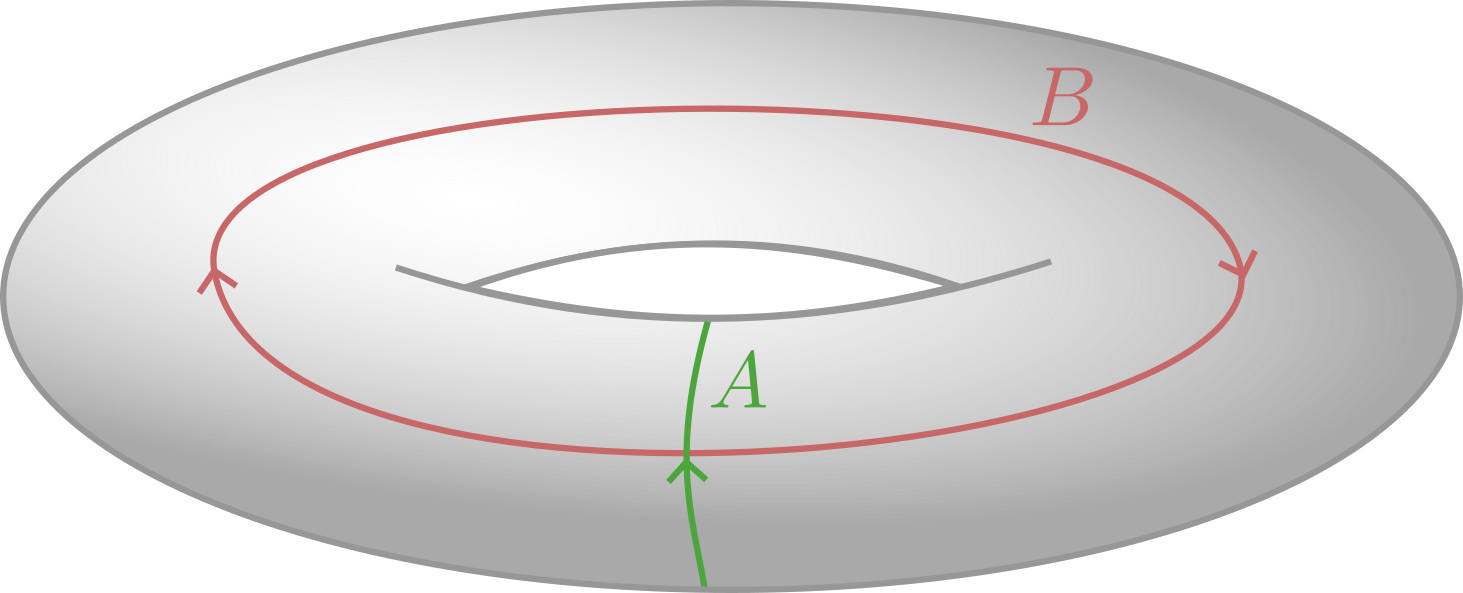}
    \caption{The homology of an elliptic curve.}
    \label{fig:homology-torus}
\end{figure}

\paragraph{(Co)homology and the period matrix.} In order to study the complexity of the moduli space of such elliptic curves, we characterize it by means of the period mapping, which specifies variation in their (co)homology. Thus, we start by discussing the homology and cohomology of the family of elliptic curves.
Since the elliptic curve is diffeomorphic to a torus, its homology is parametrized by two cycles, which we call the $A$ and $B$ cycles as in figure \ref{fig:homology-torus}, with intersection form
\begin{equation}
    \eta = \begin{pmatrix}
        0 & 1 \\ -1 & 0
    \end{pmatrix}\,.
\end{equation}
The cycles $A,B$ constitute a basis of the integer homology group $H_1 (\mathcal{E}_\lambda,\mathbb{Z})$.

The cohomology group $H^1 (\mathcal{E}_\lambda, \mathbb{C})$ can be endowed with a richer structure when linking it with the complex structure of the elliptic curve. In particular, it enjoys a Hodge decomposition:
\begin{equation}
    H^1 (\mathcal{E}_\lambda, \mathbb{C}) = H^{1,0} \oplus H^{0,1}\,,
\end{equation}
where $H^{1,0}$ is the part spanned by holomorphic forms and $H^{0,1}$ is spanned by anti-holomorphic forms. For our purposes, we instead consider the Hodge filtration $(F^1,F^0)$, defined as $H^{1,0} = F^1 \subset F^0 = H^1 (\mathcal{E}_\lambda, \mathbb{C})$.
A basis compatible with the Hodge filtration is given by
\begin{equation}
    \omega_1(\lambda) = \frac{dx}{\sqrt{x(x-1)(x-\lambda)}}\,,\quad  \omega_2(\lambda) = \lambda \frac{d \omega_1}{d \lambda}\,,
\end{equation}
such that $\omega_1 \in F^1$ and $\omega_2 \in F^0$.
Having found a basis for both the cohomology and homology groups, we can pair them together to form the period matrix of the system:
\begin{equation}
    \mathbf{\Pi}(\lambda) = \begin{pmatrix}
        \oint_A \omega_1(\lambda) & \oint_B \omega_1 (\lambda)\\ \oint_A \omega_2(\lambda) & \oint_B \omega_2(\lambda)
    \end{pmatrix}.
\end{equation}

\paragraph{Monodromy.}
When varying $\mathbf{\Pi}$ along the space $\mathbb{P}^1 \setminus \{0,1,\infty\}$, we can go around one of the singular points. When encircling any of these singularities $\lambda_0 \in \{ 0,1,\infty\}$, the entries of the period matrix are mixed according to the \textit{monodromy} of the system by
\begin{equation}
    \mathbf{\Pi}((\lambda-\lambda_0) \text{e}^{2\pi i})= \mathbf{\Pi}(\lambda-\lambda_0) \cdot M_{\lambda_0}\, ,
\end{equation}
where $M_{\lambda_0}$ is the monodromy matrix associated to $\lambda_0$. These matrices generate the monodromy group $\Gamma$, which is a finite-index arithmetic subgroup of the group $SL(2,\mathbb{Z})$. Note that since enclosing all three of the singularities is equivalent to enclosing none of them, the product of all three monodromy matrices must be the identity matrix, and thus only two monodromy matrices are linearly independent. In the case of the Legendre family, the associated monodromy group is known as $\Gamma(2)$ and it is defined as
\begin{equation}\label{definitionGamma2}
    \Gamma(2) = \left\{ A \in SL(2,\mathbb{Z}) \ \middle| \  A = I \ \text{mod } 2 \right\}.
\end{equation}
In the physical application discussed later, this group will be identified as the duality group, because it distinguishes different theories that are valid around different singularities (and thus have different associated monodromy).

\paragraph{The period map.} Given the period matrix $\mathbf{\Pi}$ and the monodromy group $\Gamma$, we can construct the \textit{period map}, which fully characterizes the family of elliptic curves. It is a map
\begin{equation}
    \tau: \mathbb{P}^1\setminus\{0,1,\infty\} \to D/\Gamma\,,
\end{equation}
where $D$ is known as a (Griffiths) period domain, which itself is a quotient $D=SL(2,\mathbb{R})/U(1) \cong \mathbb{H}$, the Poincar\'e upper half-plane. The space $D/\Gamma$ is what we will call the moduli space, and it inherits a nontrivial metric from the Fubini-Study metric associated to $\mathbb{P}^1$. This is the standard Poincar\'e metric
\begin{equation}
    ds^2 = \frac{d\tau d\overline{\tau}}{(\text{Im}\hspace{0.5mm}\tau)^2}.
\end{equation}
The period map $\tau$ is given by the ratio of the entries of the top row of the period matrix, i.e. 
\begin{equation}
\label{eq:periodmap}
    \tau (\lambda) = \frac{\oint_A \omega_1 (\lambda)}{\oint_B \omega_1 (\lambda)}\,,
\end{equation}
on which a monodromy $M =$ {\footnotesize $ \Big( \! \! \begin{array}{cc}
    a &\! b \\   c &\! d
\end{array} \! \!\Big)$} acts via the M\"obius transformation $\tau \to \frac{a \tau + b}{c \tau + d}$.

\paragraph{Picard-Fuchs equations.}
When we take another (logarithmic) derivative of the form $\omega_2$, the result will again be a form in $H^1(\mathcal{E}_\lambda, \mathbb{C})$. However, since $(\omega_1,\omega_2)$ form a basis for this space, this second derivative must be expressible as a combination of $\omega_1$ and $\omega_2 = \lambda \partial_\lambda \omega_1$. Thus, there is a differential equation governing the behavior of the form $\omega_1$ as we vary along the space $\mathbb{P}^1\setminus \{0,1,\infty\}$, known as the Picard-Fuchs equation. Taking the contour integral around either of the cycles $A$ or $B$ transforms this into a differential equation for the columns of the period matrix. For the Legendre family \eqref{eq:legendre-family}, this differential equation is given as \cite{Carlson:period-mappings}
\begin{equation}
\label{eq:picard-fuchs-legendre}
    \left(\lambda (1-\lambda) \partial_\lambda^2 + (1-2\lambda) \partial_\lambda - \frac 1 4 \right) \Pi_{A,B} = 0,
\end{equation}
where $\Pi_{A,B} = \oint_{A,B} \omega_1$ are the entries of the first row of the period matrix.

\paragraph{Generalization to different families.}
Beyond the Legendre family with monodromy group $\Gamma(2)$, one can also consider families of elliptic curves with different monodromy groups. We focus here on some simple cases, namely those where $\Gamma$ is given by a subgroup of $SL(2,\mathbb{Z})$, defined as
\begin{equation}\label{eq:def_Gamma1(4)}
    \Gamma_1 (n) = \left\{ \begin{pmatrix}
        a & b \\ c & d
    \end{pmatrix} \in SL(2,\mathbb{Z})\  \middle| \ a,d = 1 \ \text{mod } n, \  c = 0  \ \text{mod } n \right\}\,,
\end{equation}
for $n=1,2,3,4$, where $n=1$ is the maximal case $\Gamma=SL(2,\mathbb{Z})$.\footnote{Note $\Gamma(2)$ can be easily connected with $\Gamma_{1}(4)$ through a rescaling of the complex structure parameter. This can be understood explicitly by looking at the Möbius transformations associated to their respective generators acting over $\mathbb{H}$ (parametrized by $\tau$). The matrices
\begin{equation}
   S=\begin{pmatrix}
    1&&0\\
    2&&1\\
 \end{pmatrix},\quad T=\begin{pmatrix}
    1&&2\\
    0&&1\\
 \end{pmatrix},\quad\tilde{S}=\begin{pmatrix}
    1&&0\\
    4&&1\\
 \end{pmatrix},\quad\tilde{T}=\begin{pmatrix}
    1&&1\\
    0&&1\\
 \end{pmatrix}\,,
\end{equation}
generate the $\Gamma(2)=\langle S,T\rangle$ and $\Gamma_{1}(4)=\langle\tilde{S},\tilde{T}\rangle$. Under a rescaling of the form $\tau\rightarrow 2\tau$, one has that $S\rightarrow\tilde{S}$ and $T\rightarrow\tilde{T}$, mapping both pictures onto each other.} The families of curves associated to $\Gamma=\Gamma_1(1)$ have a well-known interpretation in the context of S-duality, for example in F-theory \cite{Vafa_1996} and specific types of $\mathcal{N}=2$ SCFT's \cite{Gaiotto_2012}. These groups are generated by the action of the monodromy around the points $\lambda=0,1,\infty \in \mathbb{P}^1$, given by \cite{vandeheisteeg:2024chartingcomplexstructurelandscape,Schimannek:2022ellipticcurves}
\begin{equation}
\label{eq:monodromies-pf}
     M_0 = \begin{pmatrix}
        1 & 1\\
        0 & 1
    \end{pmatrix}, \quad
    M_1 = \begin{pmatrix}
        1 & 0 \\ -n & 1
    \end{pmatrix}, \quad
    M_\infty=(M_0 M_1)^{-1} = \begin{pmatrix}
        1 - n & -1 \\ n & 1
    \end{pmatrix},
\end{equation}
and the corresponding Picard-Fuchs equations are of hypergeometric type. The local exponents are found from the eigenvalues $\lambda_i$ of the monodromy matrices as $\alpha_i=\text{Arg}(\lambda_i)/2\pi$. The monodromies around $z=0,1$ have eigenvalues $(1,1)$ so the local exponents vanish, while the monodromy around $z=\infty$ yields $(\alpha_1,\alpha_2) = (\frac 1 6, \frac 5 6)$, $(\frac 1 4 , \frac 3 4)$, $(\frac 1 3, \frac 2 3 )$, $(\frac 1 2, \frac 1 2)$ for $n=1,2,3,4$ respectively. Thus, the Picard-Fuchs equation is given by
\begin{equation}
\label{eq:picfuchs-logderivatives}
    \left(\theta^2 - z(\theta+\alpha_1)(\theta+\alpha_2)\right) \Pi_{A,B} =0, \quad {\rm with} \quad \theta = z \partial_z.
\end{equation}

\paragraph{Solutions of the Picard-Fuchs equation.}
Due to the existence of monodromies, equation \eqref{eq:picfuchs-logderivatives} does not have a unique solution by itself. Instead, we have to adapt a solution basis to the monodromies \eqref{eq:monodromies-pf}. A convenient form of the solution is given in terms of hypergeometric functions:
\begin{equation}
\label{eq:frob-sols}
    \Pi_0 (z) = \phantom{.}_2F_1 (\alpha_1, \alpha_2, 1; z); \quad \Pi_1 (z) = \frac{i}{\sqrt{n}}\phantom{.}_2 F_1 (\alpha_1, \alpha_2, 1; 1-z),
\end{equation}
where the hypergeometric $\phantom{\,}_2 F_1$ function is defined as a power series
\begin{equation}
    \phantom{\,}_2 F_1(a,b,c;z) = \sum_{k=0}^\infty \frac{(a)_k (b)_k}{(c)_k} \frac{z^k}{k!},
\end{equation}
and $(x)_k = \prod_{j =0}^{k-1} (x+j)$ is the rising Pochhammer symbol. This series converges on a disc $|z|<1$ centered around the origin, but the hypergeometric function $\phantom{\,}_2 F_1$ is taken to be its analytic continuation beyond this disc.\footnote{This is done by means of Kummer's connection formulas as found in e.g. \cite{NIST:DLMF} which relate hypergeometric functions to each other on patches where they all converge. One then extends the original hypergeometric function beyond its radius of convergence by defining it to be a combination of these other hypergeometric functions which are valid on different discs in the complex plane.} This analytic continuation is exactly such that the monodromies \eqref{eq:monodromies-pf} are respected, as we will now show.

Around the $z=0$ point, $\Pi_0$ is holomorphic, while $\Pi_1$ is of the form
\begin{equation}
    \Pi_1(z) = \frac{\Pi_0(z)}{2\pi i} \log z + (\text{holom. piece)}\, ,
\end{equation}
such that under a monodromy $z \to e^{2\pi i}z$ these periods transform as $\Pi_0 \to \Pi_0$ and $\Pi_1 \to \Pi_0 + \Pi_1$ as required.\footnote{Note that the monodromy matrices in \eqref{eq:monodromies-pf} are written to act on the period matrix as $\mathbf{\Pi} \cdot M_i$. If we want to consider their action on the period \textit{vector} $\begin{pmatrix} \Pi_0 \\ \Pi_1 \end{pmatrix}$, then we should take the transpose of $M_i$.} For the $z=1$ point, we can also expand
\begin{equation}
    \Pi_0 (1-z) = -\frac{n}{2\pi i} \Pi_1(1-z) \log(1-z)+ (\text{holom. piece)},
\end{equation}
and in this case $\Pi_1(1-z)$ is holomorphic. Thus, under monodromy $\Pi_0 \to \Pi_0 - n\, \Pi_1$ and $\Pi_1 \to \Pi_1$.

For the period around $z=\infty$ we see a difference between the families of curves, because for the family corresponding to monodromy group $\Gamma_1(4)$ this point, in addition to the other two singularities, is a point of maximally unipotent monodromy, which means that the monodromy is once again logarithmic. For the other families of curves, the monodromy is due to fractional powers of $z$ (which could in principle be removed by going to a finite cover). More specifically, for $\Gamma_1 (n)$ with $n=1,2,3$ we have
\begin{align}
    \Pi_0(1/z) &= f_0 (1/z) z^{-\alpha_1} + f_1 (1/z) z^{-\alpha_2},\\
    \Pi_1(1/z) &= \frac{i}{\sqrt{n}}\left( e^{-\pi i \alpha_1} f_0 (1/z) z^{-\alpha_1} - e^{\pi i \alpha_1} f_1(1/z) z^{-\alpha_2}\right),
 \end{align}
with $f_0, f_1$ holomorphic functions of $1/z$. These periods can be checked to have monodromy behavior $\Pi_0 \to (1-n)\Pi_0 + N \Pi_1, \Pi_1 \to \Pi_1 - \Pi_0$. For $\Gamma_1(4)$ the expansion instead looks like
\begin{align}
    \Pi_0 (1/z) &= -z^{-1/2}\left(\tilde{f}_0(z) \left(1 + 8 \frac{\log 2}{2\pi i} \right) - \frac{\tilde{f}_1 (z)}{2\pi i} - 2\tilde{f}_0 (z) \frac{\log z^{-1}}{2\pi i}\right),\\
    \Pi_1 (1/z) &= z^{-1/2} \left( -4\tilde{f}_0 (z) \frac{\log 2 }{2\pi i} + \frac 1 2 \frac{\tilde{f}_1(z)}{2\pi i} - \tilde{f}_0(z) \frac{\log z^{-1}}{2\pi i} \right),
\end{align}
where $\tilde{f}_{0,1}$ are again holomorphic functions such that $\Pi_0 \to -3\Pi_0 + 4\Pi_1$ and $\Pi_1 \to -\Pi_0 + \Pi_1$ for a loop around $1/z=0$.

\subsection{Format of period maps}\label{section:format_of_period_maps}
We are now ready to find the complexity of the period map \eqref{eq:periodmap} for the monodromy groups $\Gamma_1(n)$ with $n=1,2,3,4$. Our strategy is as follows. First, we need to find a covering of the space $\mathbb{P}^1\setminus\{0,1,\infty\}$ in terms of (punctured) discs, so we can consider local period mappings around each one of the singularities $0,1,\infty$. We find that we can split up the entries of the period matrix into a holomorphic part and a multivalued part, which is either logarithmic or has a fractional exponent. The holomorphic part will satisfy a Log-Noetherian system of differential equations extracted from the Picard-Fuchs equation, whereas the logarithm is definable in $\mathbb{R}_{\text{LN,PF}}$ as shown in section \ref{ch:example-computations}, and we combine them appropriately. Finally, if we want to consider a complexity for the global period map instead of only locally around a singularity, we need to make sure that the space is fully covered and that we match each of the solutions by mapping the images of each of the cells to an appropriate fundamental domain.

\paragraph{Constructing a cover.}
In order to use the Log-Noetherian framework described in section 2, we will need to first study the period mapping on local domains of the (simplified) cellular form described in \eqref{simple_cell}. To do so, we need to find coordinate patches in $\mathbb{P}^1\setminus\{0,1,\infty\}$, endowed with the Fubini-Study metric, such that discs in these patches cover the entire space. The geodesic distance between two points $(z_1, z_2)$ is given by \cite{Bengtsson:2006rfv}
\begin{equation}\label{eq:geodesic_distance}
    d(z_1,z_2) = \arccos\sqrt{\frac{(1+z_1 \overline{z_2})(1+\overline{z_1}z_2)}{(1+\overline{z_1}z_1)(1 + \overline{z_2}z_2)}}\,.
\end{equation}
Notably, when one constructs an open ball on the sphere using this distance, it also results in an open disc under stereographic projection (which is what the Fubini-Study metric does). Therefore, there is a convenient mapping between patches of the space $\mathbb{P}^{1} \setminus\{0,1,\infty\}$ given by open balls, and discs in the affine chart.\footnote{However, the radius and basepoint of the disc are not the same as those of the open ball, so we have to be careful that the puncture is located in the center of the disc, not in the center of the open ball.} 

We cover the space by 6 cells: 3 of them are necessarily punctured discs centered around $0,1,\infty$. These cells cannot be so large as to contain more than one singularity, since that would mean the monodromy is not well defined on the patch. Moreover, as we will see later, the complexity on that patch would become infinite as well. In order to cover the rest of the space, we take (non-punctured) discs centered at the points $-1,i,-i$. These represent parts of the moduli space from which any of the adjacent local solutions can be extended as they have trivial monodromy.

\begin{figure}[h]
     \centering
     \begin{subfigure}{0.4\textwidth}
     \includegraphics[width=\textwidth]{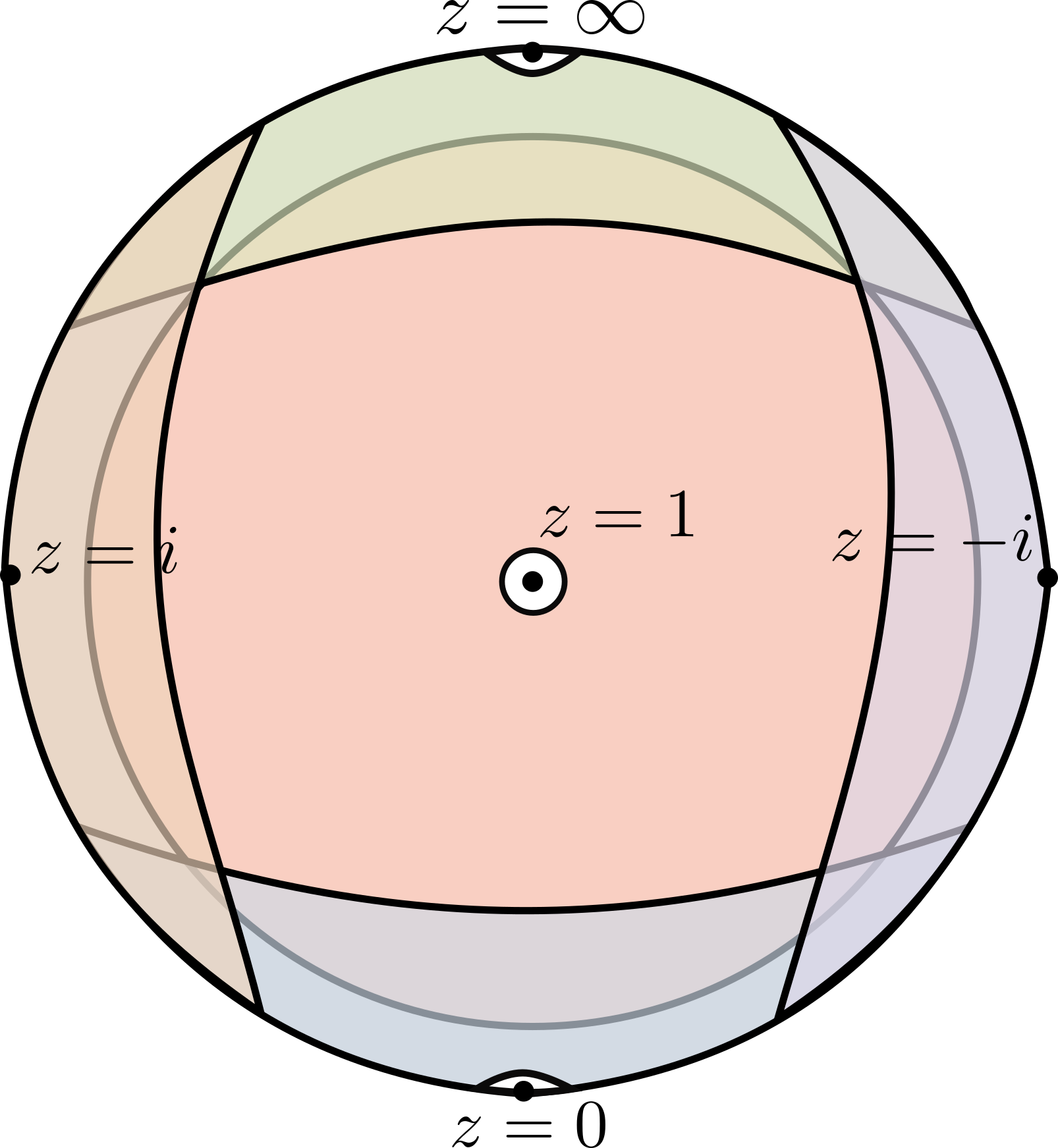}
     \end{subfigure}\hspace{1cm}
     \begin{subfigure}{0.45\textwidth}
         \includegraphics[width=\textwidth]{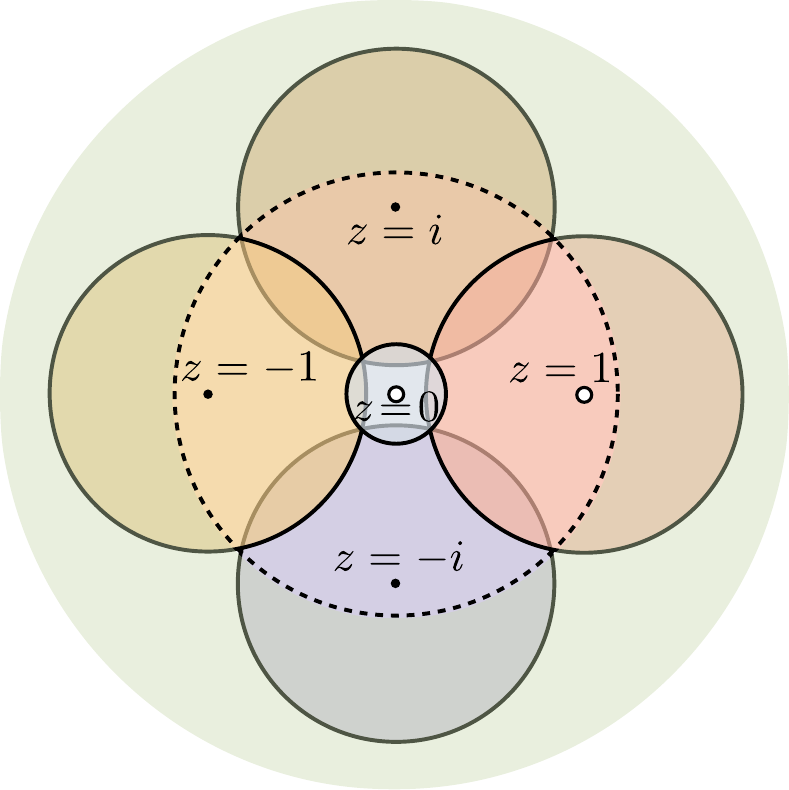}
     \end{subfigure}
    \caption{The covering of the thrice-punctured Riemann sphere $\mathbb{P}^1\setminus\{0,1,\infty\}$ (left), as well as its image in affine coordinates on the plane (right). The dashed line indicates the disc around $z=\infty$, which maps to the complement of a disc under stereographic projection.}
    \label{fig:riemannsphere-covering}
\end{figure}

As is well-known, a full cover of $\mathbb{P}^1$ requires at least two charts, since one has to include either one of the poles. For the `standard' coordinates used so far, the north pole had coordinate $z=\infty$ and so was excluded. To remedy this, we choose for the open ball around infinity the opposite chart, which is related to the previous one by $z \to 1/z$. In figure~\ref{fig:riemannsphere-covering} the covering is shown both on the space $\mathbb{P}^1\setminus\{0,1,\infty\}$ 
and on its standard affine chart, where the disc
around $z=\infty$ becomes the complement of a disc.

\paragraph{Extracting the multivalued part.}
This construction is based on the proofs of theorem 22 and section 10.3 of \cite{binyamini2024log}.
In order to find a format for the period map, we rewrite the Picard-Fuchs equation \eqref{eq:picfuchs-logderivatives} as a linear differential equation for the period matrix as
\begin{equation} \label{eq:pf-as-connection}
    z \partial_z \begin{pmatrix}
        \Pi_A & \Pi_B \\ z \partial_z \Pi_A & z \partial_z \Pi_B
    \end{pmatrix} = \begin{pmatrix}
        0 & 1 \\ \alpha_1 \alpha_2 \frac{z}{1-z} & \frac{z}{1-z}
    \end{pmatrix} \begin{pmatrix}
        \Pi_A & \Pi_B \\ z \partial_z \Pi_A & z \partial_z \Pi_B
    \end{pmatrix}.
\end{equation}
We want to use the monodromy to construct local solutions around the singularities $z=0,1,\infty$ which look like
\begin{equation}
    \mathbf{\Pi} (z) = Y(z)\cdot z^{N/2\pi i},
\end{equation}
where $z$ is a local coordinate such that the relevant singularity is located at $z=0$, and $N$ is a log-monodromy matrix, i.e.~a matrix such that
\begin{equation}
    M = e^{N}.
\end{equation}
Note that such a log-monodromy matrix is not unique, as one can add an integer multiple of the identity matrix. $Y(z)$ is then a univalued holomorphic function, and moreover it is an LN function as long as it is bounded, which we can arrange by choosing an appropriate log-monodromy matrix.

In particular, we take the log-monodromy matrices around $z=0,1$ to be
\begin{equation}
    N_0 = \begin{pmatrix}
        0 & 1 \\ 0 & 0
    \end{pmatrix}, \qquad N_1 =  \begin{pmatrix}
        0 & 0 \\ -1 & 0
    \end{pmatrix}.
\end{equation}
For the singularity around $z=\infty$ we distinguish the monodromy group $\Gamma_1(4)$ from the rest. This is necessary because in this case $z=\infty$ is (in addition to the other singular points) a point of maximal unipotent monodromy, meaning we get log-singular solutions, whereas for $\Gamma_1(n)$ with $n=1,2,3$ we only get fractional powers of $z$.\footnote{This also results in an extra cusp on $D/\Gamma$ for $\Gamma=\Gamma_1(4)$.} We choose the following log-monodromy matrices:
\begin{equation}
    N^{n=1,2,3}_\infty = \log M_\infty, \qquad N_\infty^{n=4} = \log M_\infty - 2\pi i\begin{pmatrix}
        1 & 0 \\ 0 & 1
    \end{pmatrix}.
\end{equation}

\paragraph{LN-format of the holomorphic part.} We can now construct an LN-chain for $Y  = \mathbf{\Pi}\, z^{-N/2\pi i}$ by considering its total derivative:
\begin{equation}
    dY = d\mathbf{\Pi} \,z^{-N/2\pi i} - \frac{1}{2\pi i}\mathbf{\Pi}\, N \,z^{-N/2\pi i} = (AY-YN/2\pi i) dz/z,
\end{equation}
where $A$ is the first matrix in the RHS of equation \eqref{eq:pf-as-connection} and the derivative $d$ acts component-wise on the entries of the matrices. We then define a connection matrix $C(z)$ such that
\begin{equation}
\label{eq:connection-y}
    z dY(z) = C(z) Y(z) dz\, ,
\end{equation}
with the entries of $Y$ now rearranged into a vector $(Y_{11},Y_{12},Y_{21},Y_{22})$.
For the singularities $z=0,1$ these matrices are summarized as
\begin{equation}
        C_0\! =\! \begin{pmatrix}
        0 & 0 & 1 & 0\\
        -\frac{1}{2\pi i} & 0 & 0 & 1\\
         \frac{\alpha_1 \alpha_2\, z}{1-z} & 0 & \frac{z}{1-z} & 0\\
        0 &  \frac{\alpha_1 \alpha_2\, z}{1-z} & -\frac{1}{2\pi i} & \frac{z}{1-z}
    \end{pmatrix},\qquad 
    C_1\! =\! \begin{pmatrix}
        0 & \frac{n}{2\pi i} & 1 & 0\\
        0 & 0 & 0 & 1\\
         \frac{\alpha_1 \alpha_2\,z}{1-z} & 0 & \frac{z}{1-z} & \frac{n}{2\pi i}\\
        0 & \frac{\alpha_1 \alpha_2 \, z}{1-z} & 0 & \frac{z}{1-z}
    \end{pmatrix},
\end{equation}
where for the $z=1$ singularity we made the substitution $z\to 1-z$ so that the singularity is at the origin as required. For $z=\infty$, the matrices are all different and can be found in appendix \ref{app: connection matrices}.

For the singularities around $z=0,1$ we then include the function $\frac{z}{1-z}$ as a link in our LN chain, and we can read the rest of it off of equation \eqref{eq:connection-y}. For the singularity around $z=0$, this leads to an LN chain
\begin{align}
\label{eq:chain-y-lcs}
    z\partial_z F_1 &= -\frac{1}{2\pi i} F_2 + F_3, \qquad &z \partial_z F_2 &= F_4, \nonumber \\
    z \partial_z F_3 &= \alpha_1 \alpha_2 F_5 F_1 + F_5 F_3 -\frac{1}{2\pi i} F_4, \qquad &z \partial_z F_4 &= \alpha_1 \alpha_2 F_5 F_2 + F_5 F_4,\\
    z \partial_z F_5 &= F_5^2+F_5, \nonumber
\end{align}
with solution $(F_1,F_2,F_3,F_4,F_5) = (Y_{11},Y_{12},Y_{21},Y_{22},\frac{z}{1-z})$. The other chains can be found in appendix \ref{app: connection matrices}. Note that this chain mixes all the components of $Y$, and thus we need the full period matrix instead of one of its columns. For the $z=\infty$ case, we instead take $F_5 = \frac{1}{1-z}$ with log-derivative $z \partial_z F_5= F_5^2-F_5$.

We can then fill in formula \eqref{formatLNchain} to find the LN-format of the chain on the domain $D_\circ (r)$. Note that we need to evaluate the suprema of the functions $F_i$ for this, which we find from \eqref{eq:frob-sols}. This format will be dependent on the size $r$ of the cell, but it will also contain a constant part. In other words, schematically we have
\begin{equation}
    \mathcal{F}^{\text{LN}}(Y) = \mathcal{F}^{\rm LN}_{\text{const}}(Y) + \sup_{\substack{i=1, \dots, 5\\ z\in D_\circ(r)}} |F_i (z)|.
\end{equation}
Note that the format of the cell $\mathcal{F}(\mathcal{C}) = \lceil 1+ r\rceil$ is not actually dependent on this size, since $r$ is required to be smaller than 1 in order for the functions in the chain to be bounded.

This constant part $\mathcal{F}_{\text{const}}(Y)$ can be found for the different families of curves in table \ref{tab:formats-elliptic-curves}. An interesting observation here is that the complexity corresponding to the $\Gamma_1(1)$ family of curves is exactly the same at $z=0$ and $z=1$. This can be explained by the observation that the period map actually maps both of these points to $i\infty$ (after accounting for monodromy as explained later in this section). This means that these points are geometrically the same, a property which in physical contexts is known as S-duality.\footnote{To be more precise, the matrix $S=\begin{pmatrix}
    0 & -1\\ 1 & 0
\end{pmatrix}$ relating the points $\tau=0$ and $\tau=i\infty$ by a M\"obius transformation $\tau \to -1/\tau$ is part of the monodromy group, which is not the case for the other monodromy groups considered. This means that $\Gamma_1(1)$ only has a single cusp.}

\begin{table}[h!]
    \centering
    \begin{tabular}{|l|c|c|c|c|}
    \hline
    & $n=1$ & $n=2$ & $n=3$ & $n=4$\\
    \hline
    $\mathcal{F}_{\text{const}}^{\text{LN}}(Y|_{z=0})$ \rule[-.2cm]{0cm}{.6cm} & 24.60 & 24.69 & 24.76 & 24.81\\
    $\mathcal{F}_{\text{const}}^{\text{LN}}(Y|_{z=1})$ \rule[-.2cm]{0cm}{.6cm} & 24.60 & 25.01 & 25.40 & 25.77\\
    $\mathcal{F}_{\text{const}}^{\text{LN}}(Y|_{z=\infty})$ \rule[-.2cm]{0cm}{.6cm} & 25.43 & 26.88 & 29.83 & 28.46\\
    \hline
    \end{tabular}
    \caption{The constant parts of the LN-format of $Y$ near the different singularities. }
    \label{tab:formats-elliptic-curves}
\end{table}

\paragraph{Combining the holomorphic and logarithmic part.} Let us now focus on the expression $z^{N/2\pi i}$. Its entries are of the form $z^\lambda P(\log z)$ with $P$ some polynomial, and $\lambda$ the eigenvalues of the matrix $N/2\pi i$. As shown in section \ref{ch:example-computations}, the complex logarithm is definable in $\mathbb{R}_{\text{LN,PF}}$ with format $\mathcal{F} = 39$, so we can make use of this fact in the following. 

For the singularity around $z=0$, the period matrix is expressed in terms of the entries of $Y(z)$ as
\begin{equation}
    \mathbf{\Pi}(z) = \begin{pmatrix}
        Y_{11}(z) & Y_{12} (z) + Y_{11}(z) \frac{\log(z)}{2\pi i} \\
        Y_{21}(z) & Y_{22}(z) + Y_{21} (z) \frac{\log(z)}{2\pi i}
    \end{pmatrix},
\end{equation}
and the $\tau$ map is given as the ratio of the entries of the first row:
\begin{equation}
    \tau|_0(z) = \frac{Y_{12}(z)}{Y_{11}(z)} + \frac{\log (z)}{2\pi i}.
\end{equation}
Since the holomorphic part $Y(z)$ is an LN-function, it is also an LN,PF-function, but we do need to convert its LN-format into an LN,PF-format. As shown in section \ref{ch:example-computations}, this simply amounts to adding 13 to the format. We can then write $\tau$ as a formula
\begin{equation}\label{eq:splitting-format-tau-0}
   \underbrace{\tilde{x}_1\tau=\tilde{x}_2 + \tilde{x}_1 \tilde{x}_3}_{\text{Algebraic}} \ \wedge \ \underbrace{\tilde{x}_1 = Y_{11}(z) \wedge \tilde{x}_2 = Y_{12}(z)}_{\text{Log-Noetherian}} \ \wedge \ \underbrace{\tilde{x}_3 = \frac{\log z}{2\pi i}}_{\text{Pfaffian}},
\end{equation}
where the subformulas have formats $8,\mathcal{F}(Y),\mathcal{F}(Y),39$ respectively.

We can repeat this calculation for the singularity around $z=1$, and find the expression for the periods around this singularity as
\begin{equation}
    \tau|_1 (z) = \frac{Y_{12}(z)}{Y_{11}(z) - n\,  Y_{12}(z) \frac{\log z}{2\pi i}},
\end{equation}
which we write as a formula
\begin{equation}\label{eq:splitting-format-tau-1}
    ( \underbrace{\tilde{x}_1 + n\tilde{x}_2 \tilde{x}_3 )\tau_1 (z) = \tilde{x}_2}_{\text{Algebraic}} \ \wedge \ \underbrace{\tilde{x}_1 = Y_{11}(z) \ \wedge \ \tilde{x}_2 = Y_{12}(z)}_{\text{Log-Noetherian}} \ \wedge \ \underbrace{\tilde{x}_3  = \frac{\log z }{2\pi i}}_{\text{Pfaffian}},
\end{equation}
with formats $8,\mathcal{F}(Y),\mathcal{F}(Y),39$ respectively.

Finally, around the $z=\infty$ singularity, we see some more differences between the duality groups $\Gamma_1 (n)$ appear. As mentioned, $\Gamma_1(4)$ is the only duality group that results in a point of maximal unipotent monodromy around $z=\infty$, which means that for the other groups we do not get log-singular behaviour. For example, for the monodromy group $\Gamma_1(1) = SL(2,\mathbb{Z})$ we find
\begin{equation}
    \tau|_{\infty}(z) = \frac{Y_{11} (z)(1-z^{1/3}) + \frac{Y_{12}(z)}{\sqrt{3}} \left(e^{5\pi i/6} + e^{\pi i/6}z^{1/3}\right)}{ Y_{12}(z)\left(1-z^{1/3}\right)+\frac{Y_{11}}{\sqrt{3}}\left(e^{\pi i/6} + e^{5\pi i/6} z^{1/3}\right)},
\end{equation}
which can be written as a formula:
\begin{equation}
\begin{split}
    &\underbrace{\tau \left( \tilde{x}_2(1-\tilde{x}_3) + \frac{\tilde{x}_1}{\sqrt{3}} \left(e^{\pi i/6} + e^{5\pi i/6} \tilde{x}_3 \right)\right) = \tilde{x}_1 \left( 1-\tilde{x}_3\right) +\frac{\tilde{x}_2}{\sqrt{3}} \left(e^{5\pi i/6} + e^{\pi i/6} \tilde{x}_3 \right)}_{\text{Algebraic}}\\
    &\wedge \ \underbrace{\tilde{x}_1 = Y_{11}(z) \wedge \tilde{x}_2 = Y_{12}(z)}_{\text{Log-Noetherian}} \ \wedge \ \underbrace{\tilde{x}_3 = z^{1/3}}_{\text{Log-Noetherian}}.
\end{split}
\end{equation}
Here the function $f(z) = z^{1/3}-1$ is a Log-Noetherian function with chain
\begin{equation}
    z \partial_z F_1 = \frac{1}{3} F_1^2,
\end{equation}
and $f=F_1-1$, which has an LN,PF-format $\mathcal{F}(f) = 19$. The algebraic part has a format of 9. The $n=2,3$ curves have similar expressions for $\tau_\infty$ whose algebraic parts also have format 9, and whose fractional powers also have a format of 19. As for the $\Gamma_1(4)$ family of curves, we obtain
\begin{equation}
    \tau|_\infty = \frac{Y_{12} + (Y_{11} - 2Y_{12})\frac{\log z}{2\pi i}}{Y_{11} + 2(Y_{11} - 2 Y_{12}) \frac{\log z}{2\pi i}},
\end{equation}
which as a formula is expressed as
\begin{equation}\label{eq:splitting-format-tau-infty}
    \underbrace{\tau \left(\tilde{x}_1 + 2 \left( \tilde{x}_1 - 2 \tilde{x}_2 \right) \tilde{x}_3 \right) = \tilde{x}_2 + \left( \tilde{x}_1 - 2 \tilde{x}_2 \right) \tilde{x}_3}_{\text{Algebraic}} \wedge \underbrace{\tilde{x}_1 = Y_{11} \wedge \tilde{x}_2 = Y_{12}}_{\text{Log-Noetherian}} \wedge \underbrace{\tilde{x}_3 = \frac{\log z }{2\pi i}}_{\text{Pfaffian}},
\end{equation}
with formats $8,\mathcal{F}(Y),\mathcal{F}(Y),39$.

\paragraph{Format around regular points.} If one wants to find the format of the full period mapping (as opposed to locally, as we have done so far), then one needs to cover the full space $\mathbb{P}^1\setminus\{0,1,\infty\}$. As sketched in figure \ref{fig:riemannsphere-covering}, this requires at least three discs around points that are not singular, since the punctured disc cannot be extended to cover a neighboring singularity. To get the full covering we choose the points $z=-1,\pm i$. Around these points the period map is holomorphic, and we do not need a complicated construction as in the previous paragraph. Instead, the Picard-Fuchs equations directly yield a chain
\begin{equation}
\label{eq:regular-patches-chain}
\begin{aligned}
    r\partial_z F_1 = rF_2, \quad r\partial_zF_2 &= r\alpha_1 \alpha_2 F_1 F_3 + r F_2 F_3 F_4,\\
    r\partial_z F_3 = rF_3^2 F_4, \quad r\partial_z F_4 &= 2r,
\end{aligned}
\end{equation}
solved by $(F_1,F_2,F_3,F_4) = (\Pi_{A,B}, \Pi_{A,B}', \frac{1}{(z+z_0)(1-(z+z_0))},2(z+z_0)-1)$, where the singularity is located at $z_0$, and we took $z \to z+z_0$. Crucially, the periods now each satisfy this LN-chain on their own, whereas in the singular case we really needed the full period matrix, shortening the chain. Thus we find for their LN-formats:
\begin{equation}\label{eq:format-regular-regions}
    \mathcal{F}^{\text{LN}}(Y_{\text{reg}}) = 13 + r(5 + \alpha_1 \alpha_2) + \sup_{\substack{i=1, \dots, 4\\ z\in D(r)}} |F_i (z)|.
\end{equation}

\paragraph{Global complexity -- gluing the patches.}
So far we have discussed how to obtain the format of the periods locally around one of the singularities (or one of the regular points) as a function of its radius. In terms of the Seiberg-Witten example discussed in the next section, this translates to finding the complexity of one of the effective theories around a singularity where certain particles become massless. However, we can also assign a format to the entire moduli space $D/\Gamma$ (seen as the target space of the period map). This means we need to connect the patches of $\mathbb{P}^1\setminus\{0,1,\infty\}$ appropriately:
\begin{itemize}
    \item We need to choose the radii of the discs such that $\mathbb{P}^1\setminus\{0,1,\infty\}$ is fully covered, with the extra constraint that all of the patches contain at most one singularity. Moreover, different parametrizations may yield different formats, so we should choose our covering efficiently.
    \item So far, the target spaces of the period map where taken to be subsets of $\mathbb{H}$. To develop a proper description of the quotiented space, we should perform an explicit mapping to the true moduli space $D/\Gamma$, which we can identify with the fundamental domain corresponding to the monodromy group $\Gamma$. These fundamental domains consist of a finite number of translates of the standard $SL(2,\mathbb{Z})$ fundamental domain, with the number of translates equal to the (projective) index $[SL(2,\mathbb{Z}):\Gamma]$ of the monodromy group.
\end{itemize}

For the first condition, we consider the ansatz in which all the discs around $\pm 1, \pm i$  have the same radius. Letting $r_1, r_2, r_3$ be the radii of the circles around $0,1,\infty$ respectively, we can then find $r_2,r_3$ as\footnote{Note that $r_3$ is associated to the north pole of the Riemann sphere. Thus, we choose the chart $z \to 1/z$ so that the relevant disc has radius $r_3$ in these new coordinates.}
\begin{equation}\label{eq:optimal_radii}
\begin{aligned}
    r_2 &= \sqrt{r_1^2+1-\sqrt{2}\,  r_1},\\
    r_3 &= (2+r_1^2 -2\sqrt{2} r_1)^{-1/2},
\end{aligned}
\end{equation}
as long as $r_1<\sqrt{2}-1$. These discs are found by minimizing the overlap of the discs, since the complexity of the local period map grows with the radius of the disc. The condition on $r_1$ ensures that all radii are smaller than $1$, which is necessary for a finite format.

As mentioned in table \ref{tab:formats-elliptic-curves}, the format of the local periods depends explicitly on the sizes of the domains via the supremum term:
\begin{equation}
    \mathcal{F}(\tau|_{z \in D_\circ (r), D(r)}) \sim \sup_{\substack{i=1,\dots,4\\z \in D_\circ(r), D(r)}} \left|F_i(z)\right|,
\end{equation}
where in the case of a patch around a singularity, $F_i$ are given by the entries of the vector $Y$ as well as the extra function $\frac{z}{1-z}$ or $\frac{1}{1-z}$ (depending on whether the patch is around $z=0,1$ or $z=\infty$ respectively) coming from the connection equation itself. For the patches around regular parts of $\mathbb{P}^1\setminus\{0,1,\infty\}$, they are instead given by the functions solving $\eqref{eq:regular-patches-chain}$. Combining the different patches means taking a union of the local graphs, and therefore by the axioms of effective o-minimality the total format will be given by a repeated use of equation \eqref{eq:formatrules-1}. A detailed analysis in the context of Seiberg-Witten theory will be performed in section \ref{sss: SW global behavior} to optimize the global complexity in terms of the radius $r_1$. For the present analysis, we will apply the result $r_1=0.210$ found there also to the other families of curves.\footnote{Strictly speaking, this may not be the exact optimal radius. However, as also shown in \ref{sss: SW global behavior}, the global format only varies slowly with $r_1$ when near the optimal value.}

\begin{figure}[h]
    \centering
    \includegraphics[width=0.35\linewidth]{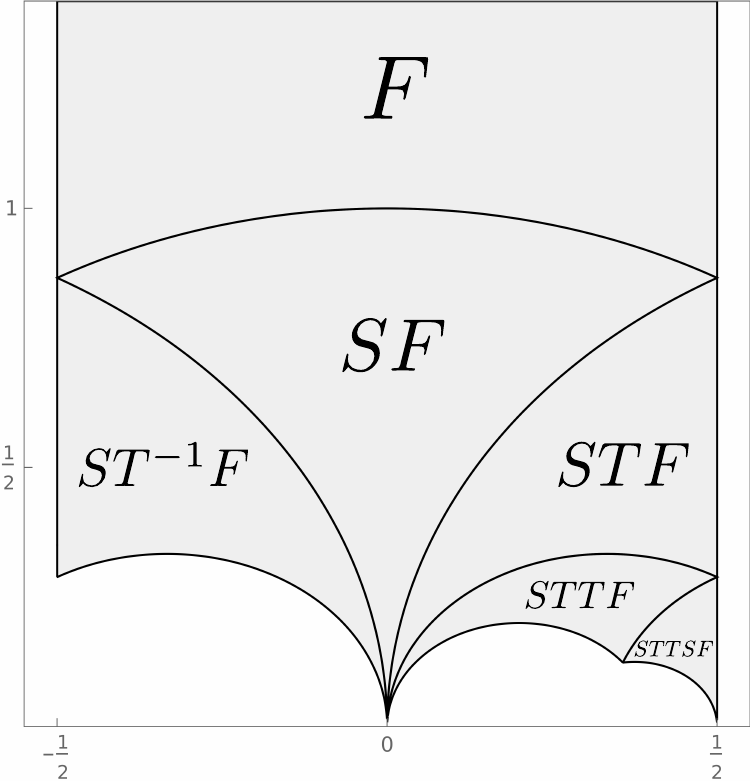}
    \caption{Partial tessellation of the upper half-plane using the images of the fundamental domain $F$ of $SL(2,\mathbb{Z})$ and its images under the generators $T$ and $S$ defined in \eqref{eq: generators of SL(2,Z) on H}.  Since there are (including $F$ itself) 6 translates involved, this is a valid fundamental domain for $\Gamma_1(4)$, which has index 6.}
    \label{fig: tesselation}
\end{figure}

\paragraph{Global complexity -- modding out monodromy.}
We now move on to the mapping of the period image to the fundamental domains $F_n$ corresponding to the families of curves $\Gamma_1 (n)$. These fundamental domains are built up from copies of the standard $SL(2,\mathbb{Z})$ fundamental domain
\begin{equation}
    F = \left\{\tau \in \mathbb{H} \, \middle| \  |\tau| \geq 1, -\frac 1 2 < \text{Re } \tau \leq \frac 1 2 \right\}.
\end{equation}
Translating this fundamental domain amounts to letting one of the generators of $SL(2,\mathbb{Z})$ act on it. These are given by
\begin{equation}
    T: \tau \to \tau + 1, \qquad S: \tau \to -1/\tau,
    \label{eq: generators of SL(2,Z) on H}
\end{equation}
and generate a tessellation of the upper half-plane. The first images of the fundamental domain of $SL(2,\mathbb{Z})$ under the generators of the group are shown in figure \ref{fig: tesselation}.

\begin{figure}[h!]
    \centering
    \begin{subfigure}{0.3\textwidth}
        \includegraphics[width=\textwidth]{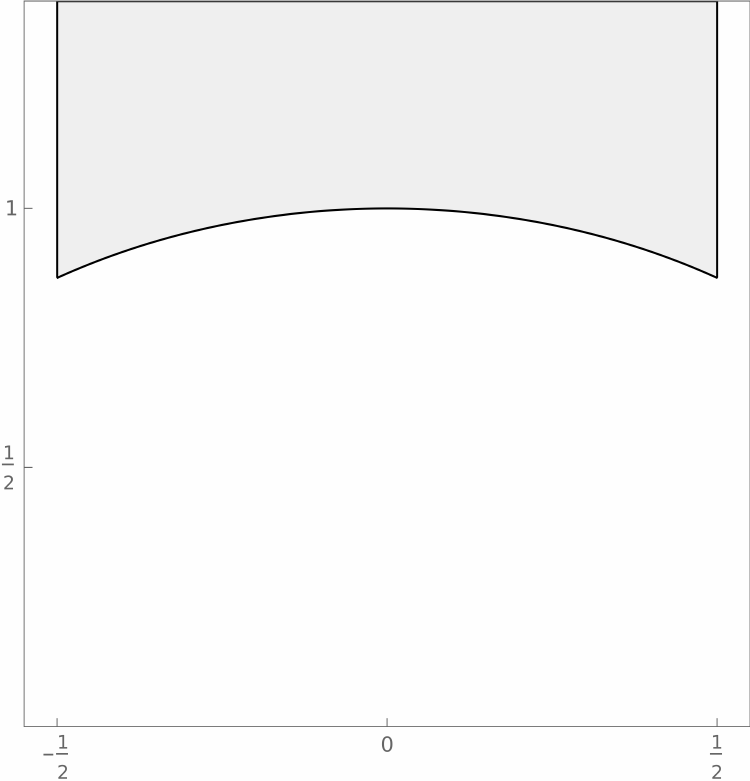}
        \caption{$\Gamma_1(1)$, index $1$.}
    \end{subfigure}\hspace{2cm}
    \begin{subfigure}{0.3\textwidth}
        \includegraphics[width=\textwidth]{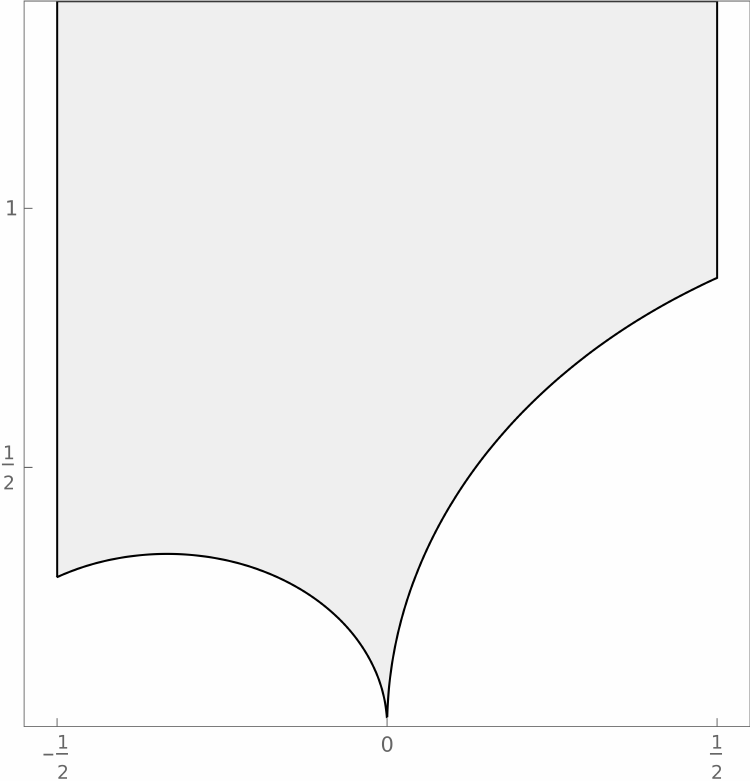}
        \caption{$\Gamma_1(2)$, index $3$.}
    \end{subfigure}\\[.2cm]
    \begin{subfigure}{0.3\textwidth}
        \includegraphics[width=\textwidth]{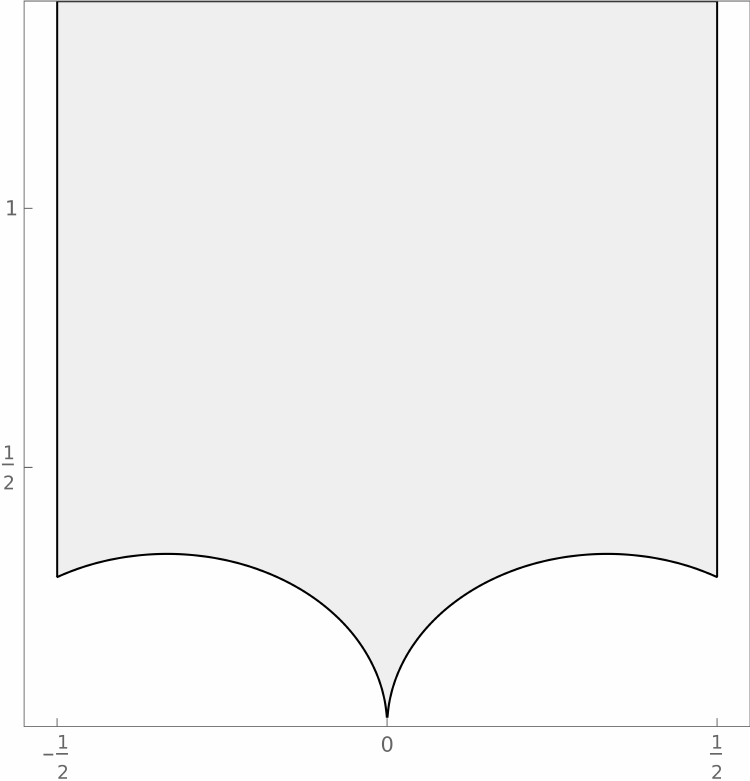}
        \caption{$\Gamma_1(3)$, index $4$.}
    \end{subfigure}\hspace{2cm}
    \begin{subfigure}{0.3\textwidth}
        \includegraphics[width=\textwidth]{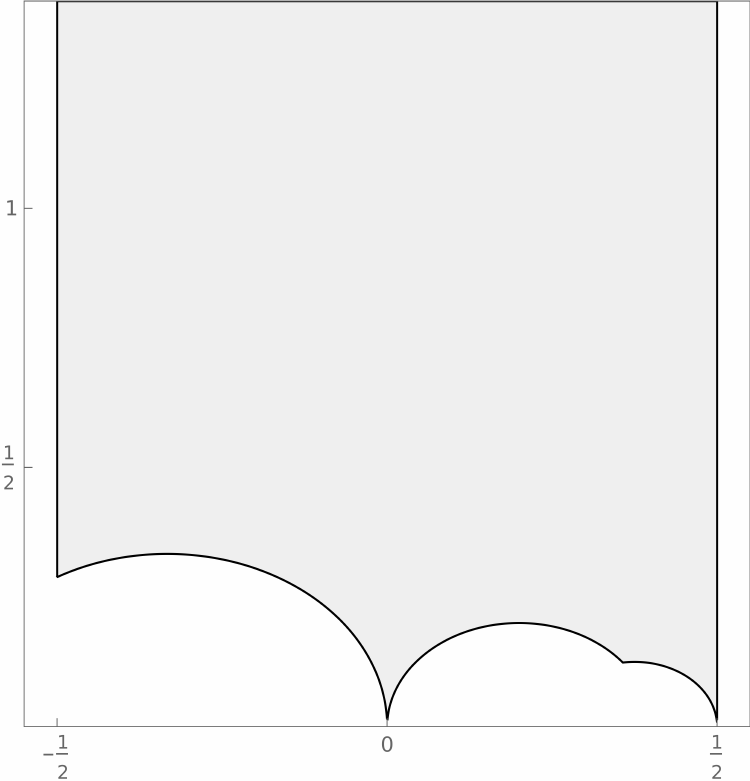}
        \caption{$\Gamma_1(4)$, index $6$.}
    \end{subfigure}
    \caption{Fundamental domains for each of the elliptic curves. }
    \label{fig:fundoms}
\end{figure}

The fundamental domain $F_n$ corresponding to the group $\Gamma_1 (n)$ is then given by a union of translates of $F$. The number of tiles required is equal to the index $[SL(2,\mathbb{Z}):\Gamma]$, which for $n=1,2,3,4$ gives $1,3,4,6$, respectively. These fundamental domains can be found in figure \ref{fig:fundoms}.\footnote{In finding which combination of the generators to let act on $F$ such that the result is connected, we used the program by Verrill \cite{verrill2001algorithm}.}

\begin{figure}[h!]
    \centering
    \begin{subfigure}{0.35\textwidth}
    \includegraphics[width=\textwidth]{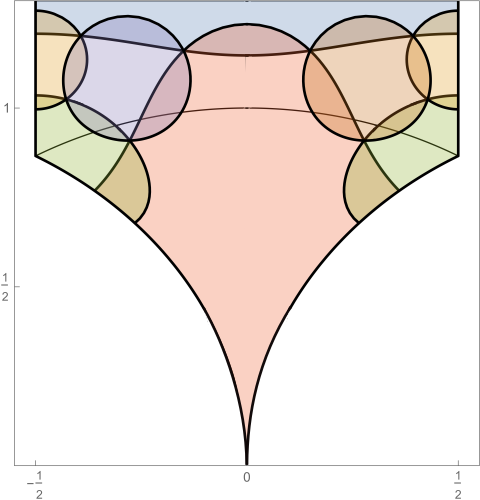}
    \caption{$\Gamma_1(1)$}
    \end{subfigure}\hspace{1cm}
     \begin{subfigure}{0.35\textwidth}
    \includegraphics[width=\textwidth]{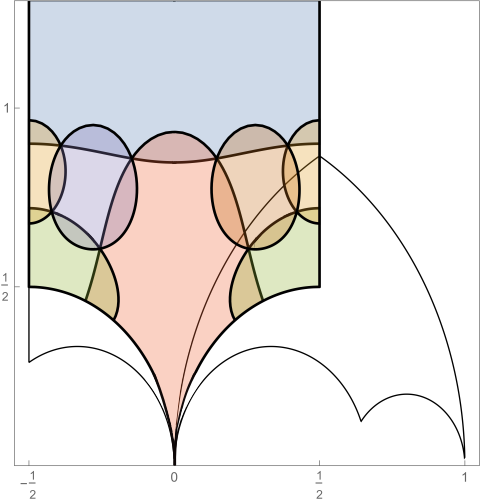}
    \caption{$\Gamma_1(2)$}
    \end{subfigure}\\
     \begin{subfigure}{0.35\textwidth}
    \includegraphics[width=\textwidth]{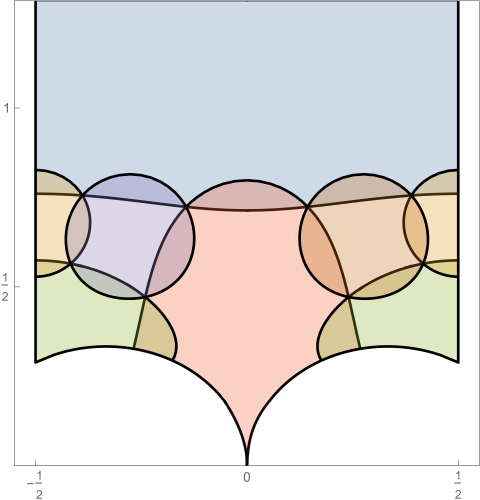}
    \caption{$\Gamma_1(3)$}
    \end{subfigure}\hspace{1cm}
     \begin{subfigure}{0.35\textwidth}
    \includegraphics[width=\textwidth]{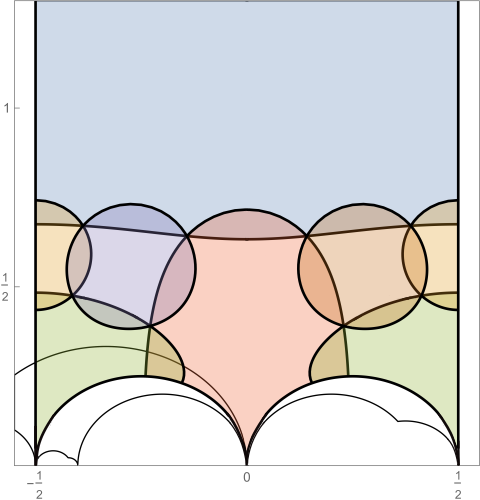}
    \caption{$\Gamma_1(4)$}
    \label{fig: image-of-balls G_1(4)}
    \end{subfigure}
    \caption{Images in the upper-half plane of the patches covering the Riemann sphere (the colors of the patches match figure \ref{fig:riemannsphere-covering}), plotted over the respective fundamental domains along with a single translate of these fundamental domains. In the case of $\Gamma_1(3)$, no translates are needed.}
    \label{fig:image-of-balls}
\end{figure}

In their proof of the tameness of the period map \cite{bakker2020tametopologyarithmeticquotients}, Bakker, Klingler and Tsimerman show that the image of any ball in the defining space (in our case, $\mathbb{P}^1\setminus\{0,1,\infty\}$) meets only finitely many translates of this fundamental domain.
Thus, we only need to consider a mapping that takes a portion of $\mathbb{H}$ that is not within $F$ to $F$. In figure \ref{fig:image-of-balls}, the images of the different discs under the period map can be seen. In particular, we see that for $\Gamma_1(1)$, $\Gamma_1(2)$, $\Gamma_1(4)$, these images cover their fundamental domains along with a single translate. For $\Gamma_1(3)$, the images of the discs stay within the fundamental domain.
Thus, for $\Gamma_1(n)$ with $n=1,2,4$, we can construct a map that takes any point within the translate of the fundamental domain, and maps it onto the fundamental domain by means of a monodromy transformation (which acts as a M\"obius transformation on the upper-half plane).

Explicitly, for the case $\Gamma_1(1)$ we construct a map $\sigma$, given by
\begin{equation}\label{def:sigma_map}
    \sigma(\tau) = \begin{cases}
        -\frac{1}{\tau} & \text{if} \ |\tau|<1,\\
        \tau & \text{otherwise} .
    \end{cases}
\end{equation}
In order to find its format, we write it as a formula:
\begin{equation}\label{eq:comp_sigma_map}
    (\sigma=\tau \ \wedge \ |\tau|\geq 1) \ \lor \ (\sigma \tau = -1 \ \wedge \ |\tau|<1),
\end{equation}
which can be checked to have format 8. If we then compose $\sigma(\tau(z))$, we find that this quotient simply increases the format by 1, since the format of $\tau$ is always higher. This will also be the case for $\Gamma_1(2)$ and $ \Gamma_1(4)$, so the result is that modding out the monodromy in these cases amounts to adding 1 to the format.\footnote{Note that it was essential that we only hit finitely many translates of the fundamental domain. If we needed infinitely many, then $\sigma$ would have had infinite complexity as it would be constructed from infinitely many formulas. Moreover, if we needed a large number of translates then the map $\sigma$ might end up having higher format than $\tau$, and end up increasing the total complexity.}

\begin{table}[h!]
    \centering
    \begin{tabular}{|l||c|c|c|c||c|}
    \hline
         \rule[-.2cm]{0cm}{.6cm} & $\Gamma_1(1)$& $\Gamma_1(2)$ & $\Gamma_1(3)$ & $\Gamma_1(4)$ & Cell\\
         \hhline{|=||=|=|=|=||=|}
         $\mathcal{F}(\tau|_{z=0})$ \rule[-.2cm]{0cm}{.6cm} & 44 & 44 & 43 & 44 & $D_\circ (r_1)$\\
         $\mathcal{F}(\tau|_{z=1})$ \rule[-.2cm]{0cm}{.6cm} & 49 & 50 & 49 & 51& $D_\circ (r_2)$\\
         $\mathcal{F}(\tau|_{z=\infty})$ \rule[-.2cm]{0cm}{.6cm} & 50 & 51 & 53 & 53& $D_\circ (r_3)$\\
         $\mathcal{F}(\tau|_{z=-1})$ \rule[-.2cm]{0cm}{.6cm} & 45 & 45 & 44 & 45& $D (r_2)$\\
         $\mathcal{F}(\tau|_{z=\pm i})$ \rule[-.2cm]{0cm}{.6cm} & 39 & 39 & 38 & 39& $D (r_2)$\\
         \hline
         $\mathcal{F}(\tau)$ \rule[-.2cm]{0cm}{.6cm} & 51 & 52 & 54 & 54& \\
         \hline
    \end{tabular}
    \caption{The formats of the period mapping on local patches, as well as for the particular choice of cell radius $r_1=0.210$, with $r_2$ and $r_3$ as in equations \eqref{eq:optimal_radii} (this choice is motivated by the optimization analysis of section \ref{sss: SW global behavior}). Here we accounted for the mapping back into the fundamental domains, and we also added 1 to account for all the quantifiers used.}
    \label{tab:formats-patches-and-global}
\end{table}

The final outcomes of the complexity analysis, both local and global formats, can be found in table \ref{tab:formats-patches-and-global}. While the complexities are mostly in the same range, it can be seen that the period maps corresponding to smaller monodromy groups have slightly higher formats. Also of note is that despite the fact that $\Gamma_1(3)$ did not require modding out any monodromy, which would otherwise have added 1 to its format, the $\Gamma_1(4)$ period map still has a lower format, possibly due to the fact that it has three points of maximally unipotent monodromy (cusps).

We also provide some comments here on the complexities of the local expressions. First, it is important to note that the formats in table \ref{tab:formats-patches-and-global} are evaluated at different cell sizes given by $r_1,r_2,r_3$ as discussed, meaning that we cannot directly compare them. We can instead refer to table \ref{tab:formats-elliptic-curves} for the `inherent' complexity which is independent of the radius, which carries over to the format of the full period mapping save for the caveat about $\Gamma_1(3)$ not needing the $\sigma$ map. Even still, when choosing the monodromy matrices in equation \eqref{eq:monodromies-pf}, we already demanded a global consistency condition, namely that the product of the three monodromy matrices should equal unity. This means that we did not necessarily choose the least complex representation for the monodromy, particularly around $z=\infty$: we could have chosen a basis in which the monodromy around this point is diagonal in the case of $\Gamma_1(n)$ with $n=1,2,3$ (or else gone to a finite cover to trivialize the monodromy entirely), but we did not do so due to global considerations. In short, there is a subtle interplay between global and local perspectives which has consequences on the complexity that we find, and the present computation has been carried out with the intention of finding a global complexity in the end.

Another interesting conclusion closely related to the previous insight is the fact that in order to find a finite global format for the period map, we had to consider different patches. In all of the chains around singular points, functions appear which grow to infinity as the radius of the corresponding punctured disc approaches 1. Since the format depends on the supremum of these functions, it too will become infinite for a disc of radius 1, and therefore it is vital that we can consider different patches (with different monodromy) which have different areas of convergence. Physically, these patches around different singularities will represent different effective theories, and transformations between them are dualities. Thus, we come to the remarkable conclusion that in order for the complexity to be finite globally, there must be a number of different dual descriptions. This conclusion will be made more precise in the next section, where we apply the methods developed here to the example of Seiberg-Witten theory.

\section{Seiberg-Witten theory and its complexity}
\label{complex_SW}

A very direct application of the effective o-minimal framework displayed in section \ref{effectiveformatofellipticcurves} is Seiberg-Witten theory \cite{Seiberg:1994rs}, a four-dimensional $\mathcal{N}=2$ supersymmetric gauge theory with an exact low-energy effective action (we also refer to \cite{Seiberg:1994aj, Klemm:1995wp, Lerche:1996xu} for a detailed discussion). For simplicity, we will stick to the case of an $SU(2)$ gauge group and focus on the vector multiplet field content. Some comments on the $SU(N)$ extension will be made in the outlook in section \ref{sec:conclusions_and_outlook}.

Recall that the $\mathcal{N}=2$ vector multiplet consists of a spin 1 vector field $A_{\mu}$, two spin $1/2$ Weyl fermions $\psi,\lambda$ and a spinless scalar field $\phi$, all of them living in the adjoint representation of $SU(2)$. In the language of $\mathcal{N}=1$ supersymmetry, the $\mathcal{N}=2$ vector multiplet decomposes into a $\mathcal{N}=1$ chiral multiplet, described by a chiral superfield $\Phi$ that encompasses $\phi$ and $\psi$, and an $\mathcal{N}=1$ vector multiplet, described by a vector superfield that contains $A_{\mu}$ and the corresponding gaugino $\lambda$. The field $\phi$ is subjected to the scalar potential \cite{Seiberg:1994rs}
\begin{equation}\label{scalarpotential}
    V(\phi)=\text{Tr}[\phi,\phi^{\dagger}]^2.
\end{equation}
Such potential is a non-negative function exhibiting flat directions\footnote{As an aside, note that such flat directions are present not only in the classical theory but also in the full quantum corrected theory, since any superpotential contribution required for their lifting is forbidden by symmetry arguments.} that can be  parame-trized after fixing the $SU(2)$ gauge by
\begin{equation}\label{phiinthegauge}
    \phi=a\begin{pmatrix}
    1&&0\\
    0&&-1\\
 \end{pmatrix}=a\sigma^{3},
\end{equation}
where $a$ is a complex number that labels each vacuum. However, the choice in \eqref{phiinthegauge} doesn't fully fix the gauge redundancy, due to the fact that $\phi$ and $-\phi$ or, equivalently, $a$ and $-a$ lie in the same gauge orbit and hence remnant $\mathbb{Z}_{2}$ redundancy persists (such discrete group is precisely the Weyl group of $SU(2)$). Therefore, the natural gauge invariant coordinate that should be used to parametrize different vacua is
\begin{equation}\label{relationuanda}
    u=\text{Tr}(\phi^2)=2a^2.
\end{equation}
The moduli space of vacua $\mathcal{M}$ is then a one-dimensional complex manifold (Riemann surface) described by the complex coordinate $u$. Alternatively, one may formulate everything in terms of the double cover $\mathcal{M}_{a}$, parametrized by $a$, and related with $\mathcal{M}$ via $\mathcal{M}=\mathcal{M}_{a}/\mathbb{Z}_{2}$.

For a given non-trivial value of $a$, there is a subset of $SU(2)$ transformations under which the vacuum remains invariant. These are the ones generated by the matrix $\sigma^{3}$, which give rise to a subgroup $U(1)\subset$ $SU(2)$. In other words, the $SU(2)$ gauge symmetry is broken to $U(1)$, and so two massless gauge bosons become massive. Note that in a low energy effective description, these heavy modes have to be integrated out. The corresponding effective theory in $\mathcal{N}=1$ superspace is given by \cite{Seiberg:1994rs}
\begin{equation}\label{efflagrangian}
    \mathcal{L}_{\text{eff}}=\frac{1}{4\pi}\text{Im}\left[\int d^{4}\theta\hspace{1mm}A^{\dagger}\frac{\partial F}{\partial A}+\frac{1}{2}\int d^{2}\theta \frac{\partial^{2}F}{\partial A^2}W_{\alpha}W^{\alpha}\right],
\end{equation}
where $A$ is the ``effective"\footnote{The UV chiral superfield $\Phi$ which transforms under $SU(2)$ is replaced by an IR chiral superfield $A$ which undergoes $U(1)$ gauge transformations.} chiral superfield whose scalar component is the vacuum expectation value $\langle\phi\rangle=a$, and $W_{\alpha}$ is the field strength of the ``effective" vector superfield $V$ whose spin 1 component is the massless mode of $A_{\mu}$ that survived the breaking. Importantly, we remark that the low-energy theory (gauge couplings and Kähler potential) is described in terms of a holomorphic function $F(A)$ known as the \textit{prepotential}. Clasically (at tree level), it takes the form
\begin{equation}\label{prepotentialtreelevel}
    F(A)=\frac{1}{2}\tau_{\text{cl}}A^2,\quad\text{with}\quad\tau_{\text{cl}}=\frac{\theta_{0}}{\pi}+i\frac{8\pi}{g_{0}^2},
\end{equation}
with $\theta_{0}$ being the Yang-Mills theta angle and $g_{0}$ is the gauge coupling constant at tree level. However, it is clear that, as it underlies the description of an effective theory, it must encode quantum corrections. A full-fledged $F$ involves not only perturbative corrections, with contributions up to one-loop, but also non-perturbative effects \cite{Seiberg:1988ur,Seiberg:1994rs}:
\begin{equation}\label{fullprepotential}
    F(A,\Lambda)=\underbrace{\frac{i}{\pi}A^2\text{ln}\frac{A^2}{\Lambda^2}}_{\text{up to one-loop}}+\underbrace{\frac{1}{2\pi i}\sum_{k=1}^{\infty}F_{k}\left(\frac{\Lambda}{A}\right)^{4k}A^2}_{\text{instantonic corrections}},
\end{equation}
where $\Lambda$ is a dynamical energy scale and $F_k$ are the coefficients in the instanton expansion. The instantonic sum leads in principle to infinitely many terms which will be relevant or not depending on the region of $\mathcal{M}$ that one aims to explore. An example of their suppression happens in the asymptotic limit $a\rightarrow\infty$, where the effective theory \eqref{efflagrangian} features asymptotically free behavior, i.e. $F$ takes the form in \eqref{prepotentialtreelevel}.

\subsection{Structure of the moduli space}\label{structureofthemodulispace}

The aim of this section is to delve into the mathematical structure of the quantum moduli space of vacua $\mathcal{M}$, which we will later connect with the complexity results obtained in section \ref{effectiveformatofellipticcurves}. In four-dimensional supersymmetric field theories, it is well known that the parameter space of the scalar components of the different chiral superfields has the structure of a complex Kähler manifold. In the case at hand, $\mathcal{M}$ is simply a one-dimensional complex Kähler manifold whose Kähler potential can be read off straight away from \eqref{efflagrangian}, 
yielding
\begin{equation}
    K(a,\bar{a})=\text{Im}\left(\bar{a}\frac{\partial F(a)}{\partial a}\right)\ .
\end{equation}
It then follows that the non-trivial component of the Kähler metric is given by
\begin{equation}
    K_{a\bar{a}}=\partial_{a}\partial_{\bar{a}}K=\text{Im}\left(\frac{\partial^2F}{\partial a^2}\right).
\end{equation}
In other words, the line element takes the form
\begin{equation}\label{metricmodulispace}
    ds^2=\text{Im}\hspace{0.5mm}\tau(a)\hspace{1mm}da \,d\bar{a},
\end{equation}
where we have defined
\begin{equation}\label{taugeneral}
    \tau(a)=\frac{\partial^2F}{\partial a^2}=\frac{\theta(a)}{\pi}+i\frac{8\pi}{g^{2}(a)}\,,
\end{equation}
as the function controlling the effective couplings $g(a)$ and $\theta(a)$ for the different vacua. Clearly, \eqref{taugeneral} reduces to $\tau_{\text{cl}}$ at tree level \eqref{prepotentialtreelevel}. A crucial point to make at this stage is that the metric in \eqref{metricmodulispace} is only valid locally, i.e. suitable to describe a certain region of $\mathcal{M}$. This essentially happens because the global positivity condition of Im$(\tau)$ required to preserve unitarity of the theory is incompatible with the maximum principle satisfied by Im$(\tau)$ as a harmonic function.\footnote{Both real and imaginary parts of a holomorphic function $\tau$ are harmonic.} In particular, the coordinate $a$ together with the metric \eqref{metricmodulispace} are useful to describe the asymptotic region of $\mathcal{M}$ with large $\abs{a}$, where Im$(\tau)$ can be proven to be safely positive. New coordinates will therefore be needed to describe the strongly coupled region of $\mathcal{M}$. We will come back to this in section \ref{patchingtheeffectivetheory} after discussing the presence of monodromies.

\subsection{Monodromy behavior}\label{sec:monodromy_behavior}
As we have just explained, a global picture of the moduli space of vacua necessarily requires the introduction of new coordinates. A customary way to proceed is by defining the so-called magnetic multiplet $(A_{D},V_{D})$, representing a `dual' version of the electric multiplet $(A,D)$ describing the theory \eqref{efflagrangian}. That is, the electric photon $A_{\mu}^{3}$ has a dual magnetic photon $(A_{\mu}^{3})_{D}$, as well as the electric Higgs field $a$ in $A$ has a magnetic version $a_{D}$ in $A_{D}$ which is defined as
\begin{equation}\label{eq:magnetic_coordinate}
    a_{D}=\frac{\partial F}{\partial a}.
\end{equation}
The exchange $(A,V)\leftrightarrow (A_{D},V_{D})$ corresponds to the physical phenomenon of electric-magnetic duality.\footnote{It is  important to remark that the magnetic multiplet should not be seen as a new vector multiplet arising from a massless representation of the $\mathcal{N}=2$ algebra but rather, as an equivalent version of the electric multiplet $(A,V)$.} In terms of both $a$ and $a_{D}$, the metric \eqref{metricmodulispace} reads
\begin{equation}\label{metricintermsofaandad}
    ds^2=\text{Im}\hspace{0.5mm}da_{D} \, d\bar{a}\ .
\end{equation}
Remarkably, this expression shows that $a$ and $a_{D}$ enter on equal footing in the metric, meaning that they could both be used as equivalent coordinate systems to describe the same underlying geometry. In fact, it is interesting to look at the different coordinate systems that preserve the metric structure \eqref{metricmodulispace}. To this end, let us assume that $a$ and $a_{D}$ are coordinates of a two-dimensional complex manifold $X$ which depend on a local holomorphic coordinate $u$ parametrizing $\mathcal{M}$ (a very physical choice is given in \eqref{relationuanda}, but for now we keep it arbitrary). One then has a map
\begin{equation}\label{functionf}
\begin{aligned}
        f: \mathcal{M} &\to X\cong\mathbb{C}^{2} \\
        u &\mapsto (a^{\alpha}(u)),
    \end{aligned}
\end{equation}
where $\alpha=1,2$ labels the two complex coordinates, i.e. $a^{\alpha}=(a_{D},a)$. Using this somewhat generic approach, the metric in \eqref{metricintermsofaandad} adopts the form
\begin{equation}\label{metricintermsofu}
    ds^2=-\frac{i}{2}\epsilon_{\alpha\beta}\frac{da^{\alpha}}{du}\frac{d\bar{a}^{\beta}}{d\bar{u}}du \, d\bar{u},
\end{equation}
with $\epsilon$ the two-dimensional Levi-Civita tensor. Now, in order to leave the metric \eqref{metricintermsofu} invariant, we need to look for transformations of $a^{\alpha}$ that preserve the symplectic structure driven by $\epsilon$. This is achieved by the linear transformations
\begin{equation}
    a^{\alpha}\rightarrow N\indices{^\alpha_\beta}a^{\beta}+c^{\alpha},
\end{equation}
where $N\in$ $Sp(2,\mathbb{R})=$ $SL(2,\mathbb{R})$ and $c$ is a constant vector. Furthermore, when magnetic monopoles are included in the theory, the classical $SL(2,\mathbb{R})$ is broken into the discrete quantum $SL(2,\mathbb{Z})$. This is very important, as for every fixed value of $u$, we now have an entire group of `physically' equivalent transformations of $f(u)$. Said differently, the quantum moduli space $\mathcal{M}$ comes equipped with a monodromy bundle $V$, whose fibers are isomorphic to (a subgroup) of $SL(2,\mathbb{Z})$. From this viewpoint, the function $f(\cdot)=(a_{D}(\cdot),a(\cdot))$ is nothing but a smooth holomorphic section of the complexified bundle $V\otimes\mathbb{C}$. 

To get ourselves familiar with the presence of monodromies, let us start by looking at the behavior of $a$ and $a_{D}$ as we go around a loop in the $u$-plane, i.e. under $u\rightarrow ue^{2\pi i}$. When $a\gg 1$, the instantonic corrections in \eqref{fullprepotential} can be neglected, so that $a_{D}$ can be approximated by
\begin{equation}
    a_{D}\simeq \frac{2ia}{\pi}\left(1+2\log\left({\frac{a}{\Lambda}}\right)\right).
\end{equation}
Recalling that the physical parameter $u$ is related with $a$ via \eqref{relationuanda}, one gets that under the aforementioned $u$-loop, the coordinates $a$ and $a_{D}$ undergo the following transformations
\begin{equation}\label{transfaandadaroundinfinity}
    \begin{split}
    a_{D} &\rightarrow -a_{D}+4a, \\
    a &\rightarrow -a.
    \end{split}
\end{equation}
Such behavior is captured by the monodromy matrix $M_{\infty}$ around the point at infinity:
\begin{equation}\label{monodromyinfinity}
     M_{\infty}=\begin{pmatrix}
    -1&&4\\
    0&&-1\\
 \end{pmatrix},
\end{equation}
The appearance of this monodromy indicates the presence of additional singular points in $\mathcal{M}$ (apart from $u=\infty$). In fact, there must be at least two more singular points, which are located at $u=\pm\Lambda^{2}$ and whose monodromies are given by \cite{Lerche:1996xu,Klemm:1995wp}
\begin{equation}\label{monodromyaround1and-1}
   M_{\Lambda^2}=\begin{pmatrix}
    1&&0\\
    -1&&1\\
 \end{pmatrix},\quad M_{-\Lambda^2}=\begin{pmatrix}
    -1&&4\\
    -1&&3\\
 \end{pmatrix}.
\end{equation}

Note that, as a consistency condition, these matrices must satisfy $M_{\Lambda^2}M_{-\Lambda^2}=M_{\infty}$. The set $\lbrace M_{\infty},M_{\Lambda^2},M_{-\Lambda^2}\rbrace$ generates the subgroup $\Gamma_{0}(4)$ of $SL(2,\mathbb{Z})$ defined as
\begin{equation}
    \Gamma_0 (4) = \left\{ \begin{pmatrix}
        a & b \\ c & d
    \end{pmatrix} \in SL(2,\mathbb{Z})\  \middle|  \  b = 0  \ \text{mod } 4 \right\}\,.
\end{equation}
The appearance of $\Gamma_{0}(4)$ monodromies over the punctures on $\mathcal{M}$ together with the positivity requirement of Im$\,\tau$ suggests a geometrization of the setup in terms of the so-called Seiberg-Witten curves. Attached to every point $u\in\mathcal{M}$ we consider a Riemann surface $E_{u}$ given by the elliptic curve
\begin{equation}\label{family_of_SW_curves}
    E_{u}=\lbrace [x:y:1]\in\mathbb{P}^{2}\hspace{0.5mm}\vert\hspace{0.5mm}y^2=(x^2-u)^2-\Lambda^4\rbrace.
\end{equation}
The set of all curves, which we denote by $E=\lbrace E_{u}\rbrace_{u\in\mathcal{M}}$, is seen as a fiber bundle over the moduli space $\mathcal{M}$. The fiber bundle $E$ automatically induces a Hodge bundle $\mathcal{H}^{1}_{\mathbb{C}}=\lbrace H^{1}(E_{u},\mathbb{C})\rbrace_{u\in\mathcal{M}}$, whose fibers are the primitive\footnote{For the case of a elliptic curve, due to dimensionality arguments, the primitive middle cohomology coincides with the middle cohomology classes.} middle cohomology spaces of every $E_{u}$. The monodromy group $\Gamma_{0}(4)$ contains all the information about how the Hodge structures (Hodge decompositions) on each fiber change after parallel transport around loops in the base manifold $\mathcal{M}$.

Based on this purely geometrical idea, the effective gauge coupling function $\tau(u)$ corresponds to the period map \eqref{eq:periodmap} that parametrizes the elliptic curves $E_{u}$ and shares the desirable property Im$\,\tau>0$. This is a very important identification, as we can now apply all the machinery developed in section \ref{effectiveformatofellipticcurves} to compute the complexity of the effective gauge coupling function in different patches of $\mathcal{M}$.   More specifically, the group $\Gamma_{0}(4)$  can be connected with $\Gamma_{1}(4)$ upon noticing a pretty subtle point: for the study of the period map we are interested in the action of $SL(2,\mathbb{Z})$ (and its finite index subgroups) over $\mathbb{H}$. The action of these groups on the upper half-plane is obtained by mapping the elements of $SL(2,\mathbb{Z})$ to Möbius transformations through a surjective homomorphism that takes the standard antisymmetric order $4$ generator $S$ of $SL(2,\mathbb{Z})$ to the order $2$ transformation $\tau\rightarrow -1/\tau$.  This is a manifestation of the fact that the action of $SL(2,\mathbb{Z})$ over $\mathbb{H}$ is free modulo $\mathbb{Z}_2$. Once such quotient by $\mathbb{Z}_{2}$ is taken, generators of $\Gamma_{0}(4)$ and $\Gamma_{1}(4)$ coincide and consequently their fundamental domains also do. Therefore, in the context of the evaluation of the complexity of the period map, one can work with either of the groups. In fact, as we will see in section \ref{patchingtheeffectivetheory}, the PF equation for the periods with monodromies in $\Gamma_{0}(4)$ can be mapped into the PF of Legendre type \eqref{eq:picard-fuchs-legendre} through a linear change of variables, reinforcing the connection.

\subsection{Patching the effective theory along $\mathcal{M}$}\label{patchingtheeffectivetheory}

As we just have motivated, realizing the effective gauge coupling constant $\tau$ as the parameter describing the complex structure of Riemann surfaces gives us a chance to compute its complexity over different patches in $\mathcal{M}$. This is a significant achievement, since the physical prepotential $F$ is directly linked to $\tau$ via \eqref{taugeneral}. In other words, the complexity of the EFT described by a prepotential $F$ is directly related with that of the period map $\tau$. Before commenting on the different complexities, let us have a small discussion on the singular structure of $\mathcal{M}$ that will help us set the correct notation compatible with the calculations in  section \ref{effectiveformatofellipticcurves}.

The point $u=0$ is a singularity of the classical moduli space, since there we restore the $SU(2)$ symmetry and consequently extra massless gauge fields appear in the spectrum. Consequently, the dynamics of these new fields should be explicitly present in the EFT description. 
This should be contrasted with the description in any neighborhood of $u=0$, where such fields have been integrated out as they became massive. However, as pointed out in \cite{Lerche:1996xu}, this singularity is `resolved' at the quantum level and instead, two quantum singularities appear at finite distance $u=\pm\Lambda^2$, as we motivated with the introduction of the two monodromies \eqref{monodromyaround1and-1}. In the same way additional massless gauge bosons appeared at $u=0$, one now has that certain 't Hooft-Polyakov monopoles become massless near $u=\pm\Lambda^2$. These monopoles are BPS states living in hypermultiplets of the $\mathcal{N}=2$ algebra of central charge
\begin{equation}
    Z=q_{e}a+q_{m}a_{D},
\end{equation}
where $a_{D}$ is the magnetic coordinate introduced in \eqref{eq:magnetic_coordinate} and $(q_{m},q_{e})$ are the respective magnetic and electric charges of the associated state. At the singularity $u=\Lambda^2$, one has that $a_{D}=0$, which means that a magnetic monopole with charges $(1,0)$ becomes massless. In the case of the singular point $u=-\Lambda^2$, it is a dyon with charges $(1,-2)$ the one becoming massless. It is always possible to read off the electric and magnetic charges of the massless BPS states by looking at the monodromies around the singularities. In fact, the condition $M_{\Lambda^2}M_{-\Lambda^2}=M_{\infty}$ does set stringent constraints on the respective choices. Nonetheless, it is important to keep in mind that the charges $(q_{m},q_{e})$ are not unique and are always defined up to conjugacy of the associated monodromies.

We now return to the discussion of the effective coupling constant $\tau$. In order to relate the periods of the elliptic curve\footnote{This choice to parametrize the elliptic curve has been considered before in the literature \cite{Klemm:1995wp}, \cite{Seiberg:1994aj} and turns out to be convenient when incorporating matter content in the theory and also from the perspective of electrical charge normalization. Another parametrization $y^2=(x-\Lambda^2)(x+\Lambda^2)(x-\tilde{u})$, was considered, for example, in \cite{Seiberg:1994rs} and has the monodromy group $\Gamma(2)$ defined in \eqref{definitionGamma2}. Both representations lead to the same physics, and their complex structure parameters $u$ and $\tilde{u}$ are related via the duality transformation $u= \tilde{u}\Lambda^2\, /\sqrt{\tilde{u}^2-\Lambda^4}$.
This map interchanges electric and magnetic singularities, but preserves the correct structure of the physical gauge and dual gauge coupling constants~\cite{Klemm:1995wp}.}
\begin{equation}\label{SW_curve}
    y^2=(x^2-u)^2-\Lambda^4,
\end{equation}
defining the family \eqref{family_of_SW_curves} with those of the Legendre curve \eqref{eq:legendre-family}, one has to find a transformation mapping their PF equations onto one another. The PF operator of the curve \eqref{SW_curve} gives rise to the following differential equation for the periods \cite{Klemm:1995wp}
\begin{equation}\label{PFeqinutilde}
    (\Lambda^4-u^2)\ddot{\varpi}-2u\dot{\varpi}-\frac{1}{4}\varpi=0,
\end{equation}
where $\varpi(u)$ are the periods of the curve \eqref{SW_curve}, $\dot\varpi=\partial_{u}\varpi$ and $\ddot\varpi=\partial_{u}^2\varpi$. A simple change of variables of the form
\begin{equation}\label{eq:change_from_z_to_u}
    z=\frac{u+\Lambda^2}{2\Lambda^2}\,,
\end{equation}
brings the equation \eqref{PFeqinutilde} back to the Legendre form \eqref{eq:picard-fuchs-legendre}
\begin{equation}\label{eq:legendretypecurve}
    z(1-z)\varpi''+(1-2z)\varpi'-\frac{1}{4}\varpi=0,
\end{equation}
where now $\varpi '=\partial_{z}\varpi$ and $\varpi''=\partial_{z}^2\varpi$. Solutions to \eqref{PFeqinutilde} are then given by those of \eqref{eq:legendretypecurve} together with the rescaling \eqref{eq:change_from_z_to_u} and thus, we can directly apply the techniques and results from the previous section to evaluate the complexity of the period map. 

To clarify some notation we would like to stress that, from now on, we will say that $u=-\Lambda^2$ is the dyonic point, $u=\Lambda^2$ the magnetic point and $u=\infty$ the electric point. We proceed by analyzing the local solutions to \eqref{PFeqinutilde} both around singular and regular patches of $\mathcal{M}$.

\subsubsection{Singular regions}

Let us start by talking about the semi-classical patch of $\mathcal{M}$, i.e.~the weakly-coupled asymptotic region which involves the electric singularity $u\rightarrow\infty$. Along this patch, we can use the coordinate $a$ as a trustworthy parameter and the EFT comes totally described by the prepotential \eqref{fullprepotential}
\begin{equation}\label{fullprepotential2}
    F(a,\Lambda)=\frac{i}{\pi}a^2\text{ln}\frac{a^2}{\Lambda^2}+\frac{a^2}{2\pi i}\sum_{k=1}^{\infty}F_{k}\left(\frac{\Lambda}{a}\right)^{4k}\,.
\end{equation}
Clearly, as we move towards the interior of $\mathcal{M}$, we will reach a point where we leave the domain of convergence of the instantonic sum and therefore the EFT will no longer be described by \eqref{fullprepotential2}. This happens approximately when $u\sim \pm\Lambda^2$, moment at which the EFT stops being valid as we hit the next singular points. Such observation goes in accordance with the fact that the coordinate $a$ stops being useful to describe regions of $\mathcal{M}$ outside the semi-classical patch, as we explained in the last paragraph of section \ref{structureofthemodulispace}. Therefore, the idea is to change the coordinate representation employed and resum the instantonic contribution in terms of the new variables such that the EFT remains finite. 

From the complexity perspective, recall from the covering discussion of section \ref{section:format_of_period_maps} that for the singularity $u=\infty$ we choose to work on a patch parametrized by the coordinate $\hat{u}=1/u$ instead of $u$. That is, in terms of the new local coordinate $\hat{u}$, the LN cell describing the semi-classical patch is given by the punctured disc
\begin{equation}\label{eq:cell_around_infinity}
    \mathcal{C}_{\infty}=D_{\circ}^{-1/\Lambda^2}(1/R_{\infty})\,,
\end{equation}
where the extra label in the exponent represents the center of the disc, which has been chosen in a suitable way to preserve the structure of the covering displayed at the right in figure \ref{fig:riemannsphere-covering}.\footnote{Note that the disc of radius $1/R_{\infty}$ centered at $\hat{u}=-1/\Lambda^2$ corresponds, in the $u$-plane, to an annulus of inner radius $R_{\infty}$ and infinite outer radius centered at $u=-\Lambda^2$.} A local analysis of the periods around $\mathcal{C}_{\infty}$ using the procedure of section \ref{section:format_of_period_maps} leads to the following value of the format
\begin{equation}\label{eq:format_around_infinity}
    \mathcal{F}\left(\tau\vert_{\mathcal{C}_{\infty}}\right)=\left\lceil46.46+\frac{1}{1-\frac{2\Lambda^2}{R_{\infty}}}\right\rceil,
\end{equation}
where we want to emphasize that, here, $\mathcal{C}_{\infty}$ should be interpreted as the associated domain to \eqref{eq:cell_around_infinity} in the $u$-plane. Let us stress what is the origin of the two pieces in \eqref{eq:format_around_infinity}: the LN format of $Y$ around the electric point found in table \ref{tab:formats-elliptic-curves} (recall that this value does not contain the supremum term) acquires an additional contribution of $13$ coming from the appropriate embedding into the bigger structure $\mathbb{R}_{\text{LN,PF}}$, see equation \eqref{eq:embeddingintoRLNPF}. In total, we have $28.46+13=41.46$, which surpasses the format $39$ coming from the Pfaffian piece in \eqref{eq:splitting-format-tau-infty}. It therefore follows that the overall format of the formula \eqref{eq:splitting-format-tau-infty} is given by $41.46+3+1+1=46.46$, where the $3$ is just a consequence of having $3$ conjunctions, $1$ stems from the composition of $\tau$ with the $\sigma$ map (as explained below equation \eqref{eq:comp_sigma_map}) and the last $1$ stands for any possibly existent quantifier in the final formula. The second term in \eqref{eq:format_around_infinity} is precisely the supremum term in expression \eqref{formatLNchain} that we left unattended in table \ref{tab:formats-elliptic-curves}. Such supremum is attained by the function $\frac{1}{1-z}$ introduced in section \ref{section:format_of_period_maps} and will be finite or infinite depending on the specific value of $R_{\infty}$ in terms of $\Lambda$. In particular, as we approach the magnetic singularity $u=\Lambda^2$, the complexity increases more and more, blowing up when $R_{\infty}$ finally reaches $2\Lambda^2$ from above. Such blow up signals a pathological behavior of the physical EFT, as expected from the prepotential discussion above. To resolve this unphysical outcome, one has to change to a dual representation of the coordinates, as we now show for the strongly coupled regions surrounding the points $u=\pm\Lambda^2$. 

Due to their similarities, let us treat both singularities in the strongly coupled regime simultaneously. Their LN domains consist of punctured discs centered at $u=\pm\Lambda^2$ with radii $R_{\text{mag}}$ and $R_{\text{dyon}}$, denoted as
\begin{equation}
    \mathcal{C}_{\text{mag}}=D^{\Lambda^2}_{\circ}(R_{\text{mag}}),\quad \mathcal{C}_{\text{dyon}}=D^{-\Lambda^2}_{\circ}(R_{\text{dyon}}).
\end{equation}
Around the magnetic monopole region, the magnetic variable $a_{D}$ turns into the appropriate coordinate to underlie the EFT description. Recall that $a_{D}$ vanishes at the point $u=\Lambda^2$, so it makes sense to use it as a parameter on which to expand physical quantities.\footnote{A subtlety follows from this last comment: the theory is weakly coupled with respect to $a_{D}$ but it keeps being strongly coupled with respect to $a$.} In particular, the prepotential in \eqref{fullprepotential2} admits a dual description of the form
\begin{equation}\label{eq:magentic_prepotential}
    F_{D}(a_{D})=\frac{a_{D}^2}{4\pi i}\hspace{0.5mm}\text{ln}\frac{a_{D}}{\Lambda}-\frac{\Lambda^2}{2\pi i}\sum_{k=0}^{\infty}F_{k}^{D}\left(\frac{ia_{D}}{\Lambda}\right)^{k},
\end{equation}
where $F_{k}^{D}$ are the dual version of the parameters $F_{k}$ driving the instantonic sum. Note that as $a_{D}\rightarrow 0$, the infinite sum converges and the prepotential remains finite. When coming to the dyonic point, no new physics is expected to emerge. This is essentially because $u=-\Lambda^2$ and $u=\Lambda^2$ are connected via the $\mathbb{Z}_{2}$ symmetry $u\rightarrow -u$ present in the quantum moduli space $\mathcal{M}$.\footnote{The $\mathbb{Z}_{2}$ symmetry in the quantum $u$-plane arises after quantizing the classical (global) $U(1)$ $\mathcal{R}$-symmetry of the superalgebra.} More concretely, the prepotential around the dyonic singularity, denoted by $\tilde{F}_{D}$, is exactly the same as the magnetic version \eqref{eq:magentic_prepotential} with a replacement of $a_{D}$ by $a-2a_{D}$, yielding
\begin{equation}\label{eq:prepotential_dyonic}
    \tilde{F}_{D}=F_{D}(a-2a_{D}).
\end{equation}
The similarity of the regions $\mathcal{C}_{\text{mag}}$ and $\mathcal{C}_{\text{dyon}}$ is also reflected at the level of the local format of $\tau$. In particular, one obtains the following complexities around the cells 
\begin{equation}\label{eq:formatLCSandC}
\mathcal{F}\left(\tau\vert_{\mathcal{C}_{\text{dyon}}}\right)= \left \lceil42.81+\frac{\frac{R_{\text{dyon}}}{2\Lambda^2}}{1-\frac{R_{\text{dyon}}}{2\Lambda^2}}\right\rceil,\quad\mathcal{F}\left(\tau\vert_{\mathcal{C}_{\text{mag}}}\right)= \left \lceil 43.77+\frac{\frac{R_{\text{mag}}}{2\Lambda^2}}{1-\frac{R_{\text{mag}}}{2\Lambda^2}}\right\rceil.
\end{equation}
The first piece of both formats can be explained in a very similar way we did for the point at infinity below equation \eqref{eq:format_around_infinity}: one more time, the LN formats of $Y$ for the dyonic and the magnetic singularities found in table \ref{tab:formats-elliptic-curves} need to be inflated by 13 units due to the embedding into $\mathbb{R}_{\text{PF,LN}}$, giving $24.81+13=37.81$ and $25.77+13=38.77$, respectively. These numbers are lower than the format of  $39$ coming from the Pfaffian terms in \eqref{eq:splitting-format-tau-0} and $\eqref{eq:splitting-format-tau-1}$. However, when factoring in the contribution from the supremum term, $\mathcal{F}(Y)$ gets bigger than $39$, making $37.81$ and $38.77$ crucial in controlling the finite part of the complexity. On top of them, we need to add extra information from the formulas \eqref{eq:splitting-format-tau-0} and $\eqref{eq:splitting-format-tau-1}$, resulting in the finite complexities $42.81=37.81+3+1+1$ and $43.77=38.77+3+1+1$, where the $3$ comes from the $3$ conjunctions, the $1$ from the $\sigma$ map, and the last $1$ from possible quantifiers in the $\tau$ formulas \eqref{eq:splitting-format-tau-0} and \eqref{eq:splitting-format-tau-1}. Concerning the second piece, it is now the function $\frac{z}{1-z}$, introduced above equation \eqref{eq:chain-y-lcs}, the one contributing to the supremum term. Again, this second piece might be infinite. Indeed, the formats blow up when the corresponding cell hits the next singularity, i.e. when $R_{\text{dyon}},R_{\text{mag}}\rightarrow 2\Lambda^2$. Before reaching the next singularity one therefore has to change the coordinate representation and describe the new EFT based on either \eqref{eq:magentic_prepotential} or \eqref{eq:prepotential_dyonic}.

\subsubsection{Regular regions}
As we mentioned in section \ref{section:format_of_period_maps} we need 6 cells to entirely cover the moduli space. In addition to the three singular cells just described above we consider domains of the disc-type domains
\begin{equation}
    \mathcal{C}_{i}=D^{u_{i}^{\text{reg}}}(R_{i}),\quad i=1,2,3,
\end{equation}
centered around the regular points\footnote{The choice of regular points is not unique, but this selection fits precisely with the image in figure~\ref{fig:riemannsphere-covering}.}
\begin{equation}
     u^{\text{reg}}_{1}=-\Lambda^2-2i\Lambda^2,\quad u^{\text{reg}}_{2}=-3\Lambda^2,\quad u^{\text{reg}}_{3}=-\Lambda^2+2i\Lambda^2. 
\end{equation}
The corresponding format of $\tau$ along these regions is given by the following expressions
\begin{equation}\label{eq:format_regular_balls}
\begin{aligned}
    &\mathcal{F}\left(\tau\vert_{\mathcal{C}_{1}}\right)=27+\left\lceil\frac{21}{4}\frac{R_{1}}{2\Lambda^2}+\frac{1}{\abs{\left(1-(\frac{iR_{1}}{2\Lambda^2}-i)\right)(\frac{iR_{1}}{2\Lambda^2}-i)}}\right\rceil,\\
    &\mathcal{F}\left(\tau\vert_{\mathcal{C}_{2}}\right)=27+\left\lceil\frac{21}{4}\frac{R_{2}}{2\Lambda^2}+\frac{1}{\abs{\left(1-(\frac{R_{2}}{2\Lambda^2}-1)\right)(\frac{R_{2}}{2\Lambda^2}-1)}}\right\rceil,\\
    &\mathcal{F}\left(\tau\vert_{\mathcal{C}_{3}}\right)=27+\left\lceil\frac{21}{4}\frac{R_{3}}{2\Lambda^2}+\frac{1}{\abs{\left(1-(\frac{-iR_{3}}{2\Lambda^2}+i)\right)(\frac{-iR_{3}}{2\Lambda^2}+i)}}\right\rceil.
\end{aligned}
\end{equation}
Here, we have essentially considered expression \eqref{eq:format-regular-regions} and added a contribution of $12$ coming from the embedding into $\mathbb{R}_{\text{LN,PF}}$,\footnote{The reason why we consider $12$ and not $13$, as in \eqref{eq:embeddingintoRLNPF}, is because the LN format of a regular disc $D(r)$ is $4$, contrary to a puntured disc $D_{\circ}(r)$ which has format $5$.} a factor of $1$ arising from the usual $\sigma$ map and a final $1$ from possible quantifiers in the formula of $\tau$. These constant factors get added to the $13$ in the first term of \eqref{eq:format-regular-regions}, yielding $27$. In contrast with the format around the singular discs, we now have a linear term in the radii directly coming from the second term in \eqref{eq:format-regular-regions} particularized to the $n=4$ case $(\alpha_{1},\alpha_{2})=(\frac{1}{2},\frac{1}{2})$. Finally, the last piece originates from the supremum contribution in the third term of \eqref{eq:format-regular-regions}, which is always driven by the function $\frac{1}{(z+z_0)(1-(z+z_0))}$ appearing in the chain system \eqref{eq:regular-patches-chain}. Clearly, the value of $z$ for which the maximum contribution is achieved depends on the specific cell, being $iR_{1}/2\Lambda^2$ for $\mathcal{C}_{1}$, $R_{2}/2\Lambda^2$ for $\mathcal{C}_{2}$ and $-iR_{3}/2\Lambda^2$ for $\mathcal{C}_{3}$.

We would like to stress one more time that the formats in \eqref{eq:format_regular_balls} reproduce the expected behavior. When $R_{i}\rightarrow 2\Lambda^2$, the last terms in \eqref{eq:format_regular_balls} blows up, which agrees with the fact that we would be hitting the dyonic singularity $u=-\Lambda^2$ (see right side of figure \ref{fig:riemannsphere-covering}).  

\begin{figure}[h!]
    \centering
    \includegraphics[width=0.95\linewidth]{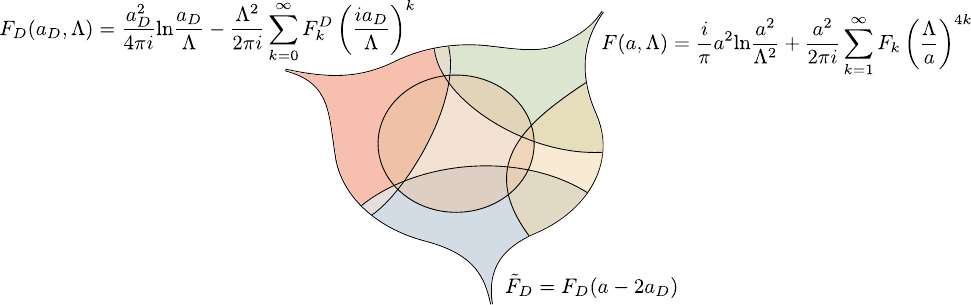}
    \caption{Depiction of the moduli space $\mathcal{M}$ covered by a total of 6 LN cells  centered around 3 singular and 3 regular points (only two of the latter are visible) as in \ref{fig: image-of-balls G_1(4)}. The asymptotic regions correspond to the three punctures of the sphere surrounded by the cells $\mathcal{C}_{\infty}$, $\mathcal{C}_{\text{mag}}$ and $\mathcal{C}_{\text{dyon}}$, on top of which the EFT is described by a different representation of the coordinates and associated prepotential: $a$ and $F$ for $\mathcal{C}_{\infty}$, $a_{D}$ and $F_{D}$ for $\mathcal{C}_{\text{mag}}$ or $a-2a_{D}$ and $\tilde{F}_{D}$ for $\mathcal{C}_{\text{dyon}}$, as discussed in  \eqref{fullprepotential2}, \eqref{eq:magentic_prepotential} and \eqref{eq:prepotential_dyonic}. The cells around the regular points overlap with the singular regions completing the covering in a consistent way that guarantees a finite global complexity.}
    \label{fig: SW moduli space limits}
\end{figure}

It is also worth mentioning that, in general, the finite complexity of $\tau$ around $\mathcal{C}_{i}$ will never exceed the one around the singular cells, as may be seen from the fact that $R_{i}/2\Lambda^2$ is always upper bounded by $1$ in \eqref{eq:format_regular_balls}. Physically speaking, the regular domains represent the transition from the weakly to the strongly coupled regime of the EFT. They create a bridge from $\mathcal{C}_{\infty}$ to $\mathcal{C}_{\text{dyon}}$ (see figure \ref{fig:riemannsphere-covering}) and thus, they cover patches of $\mathcal{M}$ where the EFT can be expressed in terms of any of the prepotentials, \eqref{fullprepotential2} or \eqref{eq:prepotential_dyonic}. The entire covering of $\mathcal{M}$ with the corresponding EFT limits is shown in figure~\ref{fig: SW moduli space limits}. 

\subsubsection{Global behavior}
\label{sss: SW global behavior}
\begin{figure}[h!]
    \centering
     \includegraphics[width=0.6\linewidth]{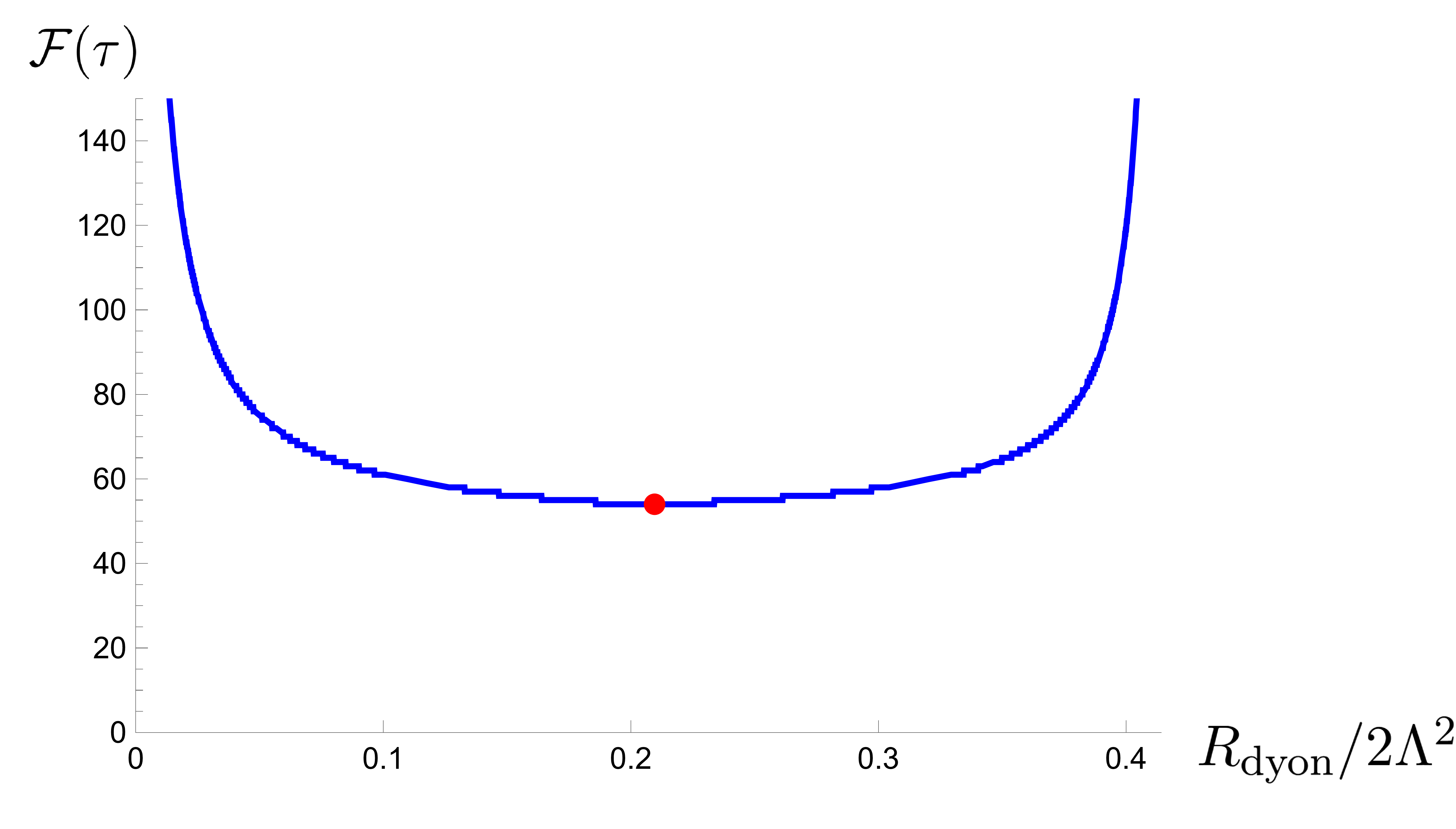}
    \caption{Representation of the $\tau$ format described by equation \eqref{ffunction} for different values of $R_{\text{dyon}}/2\Lambda^2$, where every other radius has been written in terms of $R_{\text{dyon}}/2\Lambda^2$ as explained in \eqref{eq:optimal_radii} and \eqref{eq:identifications}. The red dot corresponds to the minimal complexity configuration, located at $(R_{\text{dyon}}/2\Lambda^2,\mathcal{F})\simeq(0.210,54)$. The admissible range for $R_\text{dyon}$ is $(0,\sqrt{2}-1)$, as discussed below equation \eqref{eq:optimal_radii}. As we clearly see in the picture, the format blows up both when $R_{\text{dyon}}/2\Lambda^2\rightarrow 0$, moment at which $\mathcal{C}_{\text{mag}}$ and $\mathcal{C}_{\text{reg}}$ hit the dyonic singularity, and $R_{\text{dyon}}/2\Lambda^2\rightarrow \sqrt{2}-1$, which is when $\mathcal{C}_{\infty}$ reaches the magnetic singularity.}
    \label{fig:optimization_plot}
\end{figure}

Having examined the local complexity of $\tau$ on the cells covering the moduli space, we can now determine a global complexity bound. Based on the general recipe described in section \ref{section:format_of_period_maps}, the global graph of $\tau$ can be seen as a union of the local graphs around the 6 cells. The overall format is nothing but the format of a conjunction of 6 formulas, each describing $\tau$ locally. Viewed as a function of the radii, we can express the global complexity of $\tau$ as 
\begin{align}\label{ffunction}
& \mathcal{F}(R_{\text{dyon}},  R_{\text{mag}}, R_{\infty}, R_{\text{reg}}) 
:= \nonumber \\ & \max \Bigg\{ 
    \max \Bigg\{ 
        \max \Bigg\{ 
            29+\left\lceil\frac{21}{4}\frac{R_{\text{reg}}}{2\Lambda^2}+\frac{1}{\abs{\left(1-(\frac{R_{\text{reg}}}{2\Lambda^2}-1)\right)(\frac{R_{\text{reg}}}{2\Lambda^2}-1)}}\right\rceil,\hspace{1mm}\nonumber
            \left\lceil 42.81 \phantom{\frac{\frac{R_{\text{dyon}}}{2\Lambda^2}}{1 - \frac{R_{\text{dyon}}}{2\Lambda^2}}}\right. \\
            &+  \left. \frac{\frac{R_{\text{dyon}}}{2\Lambda^2}}{1 - \frac{R_{\text{dyon}}}{2\Lambda^2}} \right\rceil 
        \Bigg\} + 1, \hspace{1mm} \left\lceil 43.77 +\frac{\frac{R_{\text{mag}}}{2\Lambda^2}}{1 - \frac{R_{\text{mag}}}{2\Lambda^2}} \right\rceil 
    \Bigg\} + 1,\ 
    \left\lceil 46.46 + \frac{1}{1 - \frac{2\Lambda^2}{R_{\infty}}} \right\rceil 
\Bigg\} + 1.
\end{align}
Here, motivated by symmetry arguments, we are considering the three regular discs with the same radius, i.e.~$R_{i}=R_{\text{reg}}$ $\forall\hspace{0.5mm}i$. In addition, just for the sake of shortening the expression, we are implicitly assuming that the local format of $\tau$ around $\mathcal{C}_{2}$ is higher than the one over $\mathcal{C}_{1}$ and $\mathcal{C}_{3}$, reason why the number $27$ showing up in \eqref{eq:format_regular_balls} has been replaced by $29$ (this assumption has not been taken into account for the full optimization analysis shown in figure \ref{fig:optimization_plot}). Lastly, observe that every radius enters with the proper rescaling on $\Lambda$ needed to generate dimensionless quantities in the format.\footnote{Recall that $[R]=2$, as it describes the radius of a disc in the $u$-plane, and $[\Lambda]=1$, so dimensionless couplings are of the form $R/\Lambda^2$.}

Our goal is to find the values of the radii that minimize the function $\mathcal{F}$ while making sure we are covering $\mathcal{M}$ entirely. For simplicity, we will additionally assume that $R_{\text{mag}}=R_{\text{reg}}$, so that we can recycle the optimal radii expressions found in equation \eqref{eq:optimal_radii} by means of the identifications
\begin{equation}\label{eq:identifications}
    r_{1}=\frac{R_{\text{dyon}}}{2\Lambda^2}\, ,\qquad r_{2}=\frac{R_{\text{mag}}}{2\Lambda^2}=\frac{R_{\text{reg}}}{2\Lambda^2}\,,\qquad r_{3}=\frac{2\Lambda^2}{R_{\infty}}\,.
\end{equation}

Using the conditions in \eqref{eq:optimal_radii}, one may write $\mathcal{F}$ in \eqref{ffunction} as a function of just $R_{\text{dyon}}$, which attains a minimum for the following values
\begin{equation}\label{eq:optimal_radii_final}
    \frac{R^{*}_{\text{dyon}}}{2\Lambda^2}\simeq 0.210,\qquad \frac{R^{*}_{\text{mag}}}{2\Lambda^2}=\frac{R^{*}_{\text{reg}}}{2\Lambda^2}\simeq 0.865,\qquad \frac{R^{*}_{\infty}}{2\Lambda^2}\simeq 1.205\,.
\end{equation}
Plugging the optimal radii \eqref{eq:optimal_radii} back into \eqref{ffunction}, one gets the final global format
\begin{equation}\label{eq:final_value_format}
    \mathcal{F}(\tau)=54.
\end{equation}

The result is a finite number that quantifies the information encoded in the physical coupling $\tau$, and thus the information required to construct the corresponding EFT (see figure \ref{fig:optimization_plot}). We emphasize again that the resulting format is merely one representation of the complexity and need not be the optimal one. In principle, lower-complexity representations could arise from alternative rearrangements of the local formats in \eqref{ffunction}, or by treating $R_{\text{mag}}$ and $R_{i}$ as independent variables in the optimization. However, such refinements would not significantly shift the value found in \eqref{eq:final_value_format}.
The key point is that the total complexity of the coupling is finite and quantifiable, a fact that relies crucially on dualities that give rise to local descriptions with bounded complexities summing to a finite result.

Let us close by noting that we do not expect to be able to provide a finite complexity description for an arbitrary EFT. On the contrary, we believe this to be a non-trivial constraint that it is only satisfied by a selective class of EFTs. In a companion paper \cite{complexityconj}, we frame this observation in the context of Swampland program and conjecture that any EFT that can be consistently coupled with Quantum Gravity admits a description with finite complexity in the language of o-minimality.   The $SU(2)$ Seiberg-Witten theory considered here is 
a suitable example that reinforces this proposal. In fact, it is known that the Seiberg-Witten curve \eqref{SW_curve} can be embedded into 
a non-compact Calabi-Yau threefold \cite{Klemm:1996bj,Katz:1996fh}, which itself can be viewed as a local limit of a compact Calabi-Yau threefold. 
The finiteness of complexity observed for the Seiberg–Witten curve, and its close connection to dualities across moduli space, suggests a natural generalization to the full Calabi–Yau threefold moduli space. More broadly, we expect such finiteness of complexity to be a general feature in compactifications of string theory, offering a new perspective on the necessity of string dualities. We see this as another manifestation of how tameness arises as a quantum gravity principle, and expect this to be related to the existence of a tame isometric embedding of moduli spaces \cite{Grimm:2025lip}.

\section{Conclusions}\label{sec:conclusions_and_outlook}

In this work we have used quantitative language provided by effective o-minimality to study the complexity of effective field theories, focusing on the Seiberg–Witten solution of four-dimensional $\cN=2$ $SU(2)$ pure super Yang–Mills theory. By viewing the effective gauge coupling $\tau$ as the period map of the Seiberg–Witten elliptic curve fibered over the moduli space of quantum vacua $\mathcal{M}$, we were able to assign to it a precise complexity using the recent formalism for Log-Noetherian and Pfaffian functions \cite{binyamini2024log}. The moduli space is a thrice-punctured sphere with one weakly coupled region $(u=\infty)$ and two strongly coupled regions ($u=\pm \Lambda^2$) where magnetic monopoles and dyons become massless. The complexity computation requires to decompose this space into complex cells constructed using discs and annuli. We computed the complexity of the local period maps and on each cell by covering the moduli space with six discs: three punctured discs around the singular points and three regular discs. We then combined all local complexities into a complexity of the full setting. Our local complexity results can be interpreted as quantifying the complexity of the $U(1)$ effective field theories (EFTs) in these cells with all massive states integrated out, while the total yields the complexity of the space of EFTs.   

A key outcome is that the total complexity of the effective coupling is quantifiable and finite. This statement is highly non-trivial and it is worth emphasizing some of the key points that make this result possible. Most crucial is the existence of a consistent complexity measure, which relies on the deep result that $\bbR_{\rm LN}$ is an effective o-minimal structure by using a complex cell decomposition \cite{binyamini2019complex,binyamini2024log}.\footnote{While $\bbR_{\rm LN}$ is a real structure, the use of complex and hyperbolic geometry is thereby the main tool.} In our explicit example, we were able to find a simple complex cell covering with a tractable optimization of complexity. Roughly speaking, the complexity measure captures the amount of information (when counting real numbers as single units), in the differential equations and domains that are needed to specify the coupling functions of the effective theory. This measure clearly depends on the representation of the theory and it is a non-trivial task to find the optimal representation with minimal complexity. Implementing a concrete optimization, we were able to compute the complexities of the periods for elliptic curves with monodromy groups $\Gamma_1(n)$, $n=1,2,3,4$. We observed that the singularities most strongly influence the total complexity of the theory and induce the dependence of the local and global complexity on $n$.

More important than the precise value for the total complexity of $\tau$ and the EFTs is the mechanism behind its finiteness. 
The local complexity on a given cell grows as the cell is extended toward another singularity and would diverge if we insisted on a single global description. Finiteness is restored only when we allow changes of duality frame and pass to new local effective theories adapted to the region. From a mathematical point of view the domain dependence of the local complexity is unavoidable if one aims to define a complexity for a rich o-minimal structure such as $\bbR_{\rm LN}$. It thus appears to be crucial when trying to assign a meaningful notion of complexity to period integrals. This teaches us the important lesson: any attempt to define a global notion of complexity along a full quantum moduli space, the finite information in the local differential equations gets completed globally with the help of dualities. In this sense, dualities are not merely a useful organizing principle but a structural requirement for having a tame, finite-complexity description of the physical theory on the full moduli space. 

A key step in our analysis was to determine the local complexities in cells around the singular points. These are the boundaries of the moduli space $\cM$, which are responsible for the non-trivial monodromy group $\pi_1(\cM)$. A central fact of asymptotic Hodge theory is that the local monodromy is crucial in describing the local period expansions near these boundaries \cite{schmid}. This matches the observation that the local complexities depend on the local monodromy group, leading to a pattern of complexities displayed in table~\ref{tab:formats-patches-and-global}. This fact suggests the possibility to turn the classification of asymptotic limits (see, e.g.~\cite{Kerr2017,Grimm:2018ohb,Grimm:2018cpv,Corvilain:2018lgw,Grimm:2019ixq,Lee:2021qkx,Hassfeld:2025uoy}) into a split into complexity classes. It would be interesting to carry this out explicitly and check how far one can distinguish limits in moduli space by this measure. The structure $\bbR_{\rm LN,PF}$ was conjectured to be sharply o-minimal in \cite{binyamini2024log}, which would provide proper notion of sharp complexity for period integrals, given by two integers $(F,D)$. We expect that with that refined description, all boundaries in the classification of \cite{Kerr2017} will have distinct patterns. We leave a further study of this interesting connection to future work.

There are a number of immediate generalizations of our analysis that are of interest. 
A natural step is to consider $\cN=2$ super-Yang-Mills theory with higher-rank gauge groups $G$ and to include the coupling to charged hypermultiplet matter fields. Focusing on the pure super-Yang-Mills setting, e.g.~with gauge group $G=SU(N)$, we have new parameter $N = \text{rank}(G)$ entering the complexity. These theories have a quantum vacuum moduli space that arises form a Seiberg-Witten curve of higher 
genus \cite{Klemm:1994qs,Argyres:1994xh, Klemm:1995wp}. It is clear that the complexity grows with $N$, since both the 
dimension of the moduli space and the number of algebraically independent period integrals grows with $N$. With the effective 
o-minimality framework at hand, it is possible to quantify this growth and we expect it to be at least $\sim N^2$. In is then interesting to 
contrast this complexity growth with the claimed complexity bounds of \cite{complexityconj}. This discussion has a strong connection with ideas about the boundedness of the rank of gauge groups appearing in EFTs consistent with quantum gravity \cite{Vafa:2005ui}. The sharpest bounds in this context are provided for configurations with high supersymmetry. For instance, in a four-dimensional $\cN=4$ supersymmetric theory with gravity the rank of the gauge group is claimed to be bounded by $N\leq 22$ \cite{Kim:2019ths}. 
In settings with lower amounts of supersymmetry specialized finite-rank claims have been made in \cite{Kumar:2010ru,Morrison:2011mb,Grimm:2012yq}  and more  recently \cite{Lee:2019skh,Kim:2019vuc,Katz:2020ewz,Tarazi:2021duw,Martucci:2022krl}. Supporting evidence for the rank-boundedness in $\cN=2$ comes from the claimed finiteness of compact Calabi-Yau threefolds \cite{reid1987moduli,yau2008survey}, which has recently seen much progress when restricting to elliptic fibrations \cite{Gross1993AFT, MR4939522,MR4801611, Kim:2024eoa,Birkar:2025gvs}. In accordance with these results, we expect that high-complexity (that is higher-genus) Seiberg-Witten curves no longer admit an embedding into a controlled string compactification to four dimensions when requiring a finite Planck mass and fixed finite cut-off scale. 

Let us end by noting that it is desirable to develop a wider generalization of the presented ideas beyond the Seiberg-Witten setting and study the complexity of more general EFTs. For example, one natural next set-up is to consider F-theory and Type IIB string theory compactifications on Calabi-Yau manifolds. In these settings period integrals still determine parts of the EFTs but there are additional coupling functions and quantum corrections that require to extend the tool-set beyond period integrals. To eventually understand such broad classes of EFTs one first needs to establish that tame geometry can be used to assign a quantitative complexity to the full EFT description, following up on the suggestions of \cite{Grimm:2024elq}. Finding the  proper characterization for the complexity of such an abstract object is a very nuanced question. It requires to analyze the functional dependence of the Lagrangian but also control the field content and the cutoff dependence across the quantum moduli space. In an upcoming work \cite{complexityconj} we will investigate how this might be done using the complexity measure provided by sharp o-minimality and will connect the bounds on complexity to the expected properties of the EFTs compatible with the coupling to quantum gravity. More broadly, we believe that analyzing complexities of EFTs has the potential to unify many ideas about the landscape of effective theories and deserves much further study.

\subsubsection*{Acknowledgements}
We would like to thank Gal Binyamini, Johan Commelin, Arno Hoefnagels, Jeroen Monnee, Thorsten Schimannek, Cumrun Vafa, Stefan Vandoren, and  Mick van Vliet for valuable discussions and comments. Furthermore, TG and DP are grateful to the CMSA for support and hospitality during the final stages of this work. The research of TG and DP is supported, in part, by the Dutch Research Council (NWO) via a Vici grant. The work of MC is supported by an STFC Consolidated Grant.


\appendix
\section{Effective o-minimality}
\label{Effective_appendix}

This appendix will put the relatively informal discussion on tameness and complexity as presented in section \ref{ch:tameness-eff-o-minimality} on a more rigorous footing. We define o-minimal structures in appendix \ref{o-min_app}, while effective o-minimality is introduced in appendix \ref{eff_app}. $\bbR_{\rm LN}$ and $\bbR_{\rm LN,PF}$ are the relevant examples of effective o-minimal structures. We review the constructions of cells in $\bbR_{\rm LN}$ in appendix~\ref{ch:app-lncells} and introduce the associated complexity notion in appendix~\ref{LN_Comp_app}. The connection into mathematical logic will be explained in appendix~\ref{Logic-semialg}. We close by discussing an algorithmic way to determine the complexity of conjunctions in appendix~\ref{Conj_app}. 

\subsection{Tameness via o-minimality} \label{o-min_app}

Tameness is a generalized notion of finiteness, on which our idea of complexity is based. By demanding that the functions and sets we study satisfy certain properties, we exclude many pathological functions that in some sense have infinite complexity. 

The precise statement of tameness comes from considering \textit{o-minimal structures}, which are collections of subsets of $\mathbb{R}^n$ satisfying some finite properties as follows:
\begin{definition}
    \label{def:ominstructure}
An o-minimal structure on $\mathbb{R}$ is a sequence of collections of subsets $S = 
(S_m \subset \mathbb{R}^m)_{m \in \mathbb{N}}$ such that:
\begin{enumerate}
    \item $S_m$ is closed under finite unions, intersections and complements with respect to $\mathbb{R}^m$.
    \item If $A \in S_m$ and $B \in S_n$ then $A \times B \in S_{m+n}$.
    \item If $P(x_1, \dots , x_m)$ is a polynomial then $\{P(x_1, \dots, x_m) = 0\} \in S_m$.
    \item If $A \in S_m$ then $\pi(A) \in S_{m-1}$ where $\pi: \mathbb{R}^m \to \mathbb{R}^{m-1}$ is a linear projection.
    \item $S_1$ consists of finite unions of points and intervals.
\end{enumerate}
\end{definition}
The first four axioms define a \textit{structure}, and the last one enforces the finiteness principle and is therefore known as the o-minimality axiom.

Sets within an o-minimal structure are said to be \textit{tame} or \textit{definable} in that structure. One further defines a function $f: A \to B$ to be tame (or definable) if $A \times B$ is tame and the graph $\text{Gr } f$ is a tame subset of $A \times B$. These tame functions have a number of attractive features. If $f: A \to \mathbb{R}$ is a tame function, then both the domain $A$ and the image $f(A)$ are tame within the same o-minimal structure. Furthermore, $f^{-1}$ is tame and crucially, the sum, product and composition of two tame functions is tame. Tameness also guarantees that $f$ is continuous and differentiable except at finitely many points.

Let us take a look at some examples of o-minimal structures. A way of constructing them is by specifying a collection of sets $ A = (A_n \subset \mathbb{R}^n )_{n \in \mathbb{N}}$ that should be in the o-minimal structure, and then considering the smallest o-minimal structure $S$ containing them. We say that $S$ is generated by $A$. The simplest example of such a construction is the structure generated by semi-algebraic sets (i.e. sets defined by (in)equalities of polynomials), and is known as $\mathbb{R}_{\text{alg}}$. This is the smallest o-minimal structure, and by axiom 3 any other o-minimal structure will contain $\mathbb{R}_{\text{alg}}$.

A slightly more elaborate example is the structure $\mathbb{R}_{\text{exp}}$, which is the o-minimal structure generated by the (graph of the) exponential function. One could then also further add graphs of analytic functions, restricted to compact domains. In this way, the o-minimal structure $\mathbb{R}_{\text{an,exp}}$ is generated. Note that this restriction to compact domains is crucial, because analytic functions on infinite domains can contain infinitely many zeroes, and hence $S_1$ should contain an infinite union of points (because the intersection of the graph of a function of one variable and the $x$-axis must be definable), in contradiction with the o-minimality axiom.

While the structure $\mathbb{R}_{\text{an,exp}}$ is large enough to contain many physically interesting functions, it is in fact too large to assign a complexity to every function or set within it. Instead, we focus on a smaller structure known as $\mathbb{R}_{\text{LN,PF}}$, analyzed in \cite{binyamini2024log}, which is large enough to contain all period mappings of algebraic varieties, meaning it contains many functions found in physics still.

\subsection{Effective o-minimality} \label{eff_app}

In order to make our notion of complexity precise, we define effective o-minimal structures. These structures, introduced by Binyamini in \cite{binyamini2024log}, are a weaker alternative to sharp o-minimality \cite{binyamini2022sharplyominimalstructuressharp}, which classifies complexity by two integers $(F,D)$. In effective o-minimality, we instead consider only a single integer, known as the format $\mathcal{F}$. More precisely, we define:
\begin{definition}
\label{def:eff-o-minimality}
    An \textit{effective o-minimal } structure consists of an o-minimal structure $\mathcal{S}$ along with a so-called \textit{format filtration} $\Omega_{\mathcal{F}}$ and a primitive recursive function $\mathcal{E}: \mathbb{N} \to \mathbb{N}$ such that:
    \begin{enumerate}
        \item $\Omega_{\mathcal{F}} \subset \Omega_{\mathcal{F}+1}$ for any $\mathcal{F}$.
        \item $\bigcup_{\mathcal{F}} \Omega_{\mathcal{F}} = \mathcal{S}$.
        \item If $A, B \subset \mathbb{R}^n$ and $A \in \Omega_{\mathcal{F}(A)}$, $B \in \Omega_{\mathcal{F}(B)}$ then:
        \begin{align}
            A \cup B, \, A \cap B, \, \, A \times B, \, &\in \Omega_{\text{max}\{\mathcal{F}(A),\mathcal{F}(B)\}+1},\\
             \mathbb{R}^n \setminus A, \pi_k(A) &\in \Omega_{\mathcal{F}(A)+1},
        \end{align}
        where $\pi_k$ is a linear projection to the first $k$ coordinates.
        \item If $A \subset \mathbb{R}$ and $A \in \Omega_{\mathcal{F}}$ then $A$ has at most $\mathcal{E}(\mathcal{F})$ connected components.
    \end{enumerate}
\end{definition}
One thing to note in this definition is that since the format $\mathcal{F}$ labels a filtration, a given set has no unique format: if it is contained in $\Omega_{\mathcal{F}}$ it is also contained in any $\Omega_{\mathcal{F}'}$ for $\mathcal{F}' \geq \mathcal{F}$. It does, however, have a unique lowest format, in contrast to sharp o-minimality.

In order to generate interesting structures, we demand that functions satisfy some differential equations. It was realized by Khovanskii in \cite{khovanskiĭfewnomials} that solutions to certain systems of differential equations, known as Pfaffian equations, admit bounds on their number of zeroes. Based on this, one can construct the effective o-minimal structure $\mathbb{R}_{\rm Pfaff}$. This structure was proven to be o-minimal by Wilkie in \cite{Wilkie1999ATO}, and the bound by Khovanskii ensures effective o-minimality. Moreover, one can also consider solutions to more general systems of differential equations, known as (log-)Noetherian systems, restricted to special kinds of domains. This is what was done in \cite{binyamini2024log} to define the structure of Log-Noetherian functions, $\mathbb{R}_{\rm LN}$. Finally, it is known that given any (effective) o-minimal structure, one can take its Pfaffian extension to generate a larger one. This leads to the Pfaffian extension of Log-Noetherian functions $\mathbb{R}_{\rm LN,PF}$, also defined in \cite{binyamini2024log}, which is the structure we work in for this paper.

\subsection{Log-Noetherian cells}\label{ch:app-lncells}
The main structures in this work are $\mathbb{R}_{\rm LN}$ and $\mathbb{R}_{\rm LN,PF}$, the structure of Log-Noetherian functions and its Pfaffian extension. We now focus on the former. As mentioned previously, Log-Noetherian functions are those satisfying certain differential equations on specific (bounded) domains. These domains are known as LN-cells, and are defined recursively:
\begin{definition}\label{definition_LN_cell}
    An LN cell $\mathcal{C}$ of length 0 is a single point $\mathbb{C}^0$. An LN cell $\mathcal{C}_{1,\dots,l+1} \subset \mathbb{C}^{l+1}$ of length $l+1$ is a fibration (which we denote by the symbol $\odot$):
    \begin{equation}
        \mathcal{C}_{1,\dots,l+1} = \mathcal{C}_{1\dots l} \odot F,
    \end{equation}
    where $F$ is one of four types of fibers (see also figure \ref{figblocks}):
    \begin{itemize}
        \item $*=\{0\}$, the trivial fiber.
        \item $D(r)$, a disc around $0$ with radius $|r|$.
        \item $D_\circ (r)$, a punctured disc around $0$ with radius $|r|$.
        \item $A(r_1,r_2)$, an annulus around $0$ with inner radius $|r_1|$ and outer radius $|r_2|$.
    \end{itemize}
    Since this is a fibration, the radii $r, r_1, r_2$ are functions depending on the previous coordinates:
    \begin{equation}
        r,r_1,r_2:\mathcal{C}_{1,\dots,l} \to \mathbb{C}\setminus\{0\},
    \end{equation}
    where we require $|r_1 (z)| < |r_2 (z)|$ for every $z \in \mathcal{C}_{1,\dots,l}$ in the case of the annulus. Note also that these functions are allowed to be complex (hence the absolute value signs). If they are real instead, we speak of a real cell (which still defines a subset of $\mathbb{C}^n$). Finally, in the case of fibers of type $D(r)$, we demand $r$ to be a constant function.
\end{definition}
To illustrate how this notation simplifies complicated cellular domains, consider the following example:
\begin{equation}
\begin{split}
    D(1) \odot D(z_1^2) &=\{(z_1,z_2) \in \mathbb{C}^2 \, | \, |z_1|<1, \, |z_2|<|z_1^2|\}\\
    &= \left\{(x_1,y_1,x_2,y_2) \in \mathbb{R}^4 \, | \, \sqrt{x_1^2+y_1^2} < 1, \, \sqrt{x_2^2+y_2^2} < \sqrt{ x_1^4+2x_1^2y_1^2+y_1^4} \right\}.
\end{split}
\end{equation}
These cellular constructions allow for very general domains. Moreover, one can define maps between cells and pull back functions by such maps in order to construct even more general domains, as explained in detail in \cite{binyamini2024log} and \cite{binyamini2020effectivecylindricalcelldecompositions}.

Sometimes it may be necessary to enlarge every fiber making up a cell simultaneously. This may be done by means of a $\delta$-extension $\mathcal{C}^\delta$ with $\delta \in (0,1)$, in which the fibers are changed as follows:
\begin{itemize}
    \item $*$ remains unchanged.
    \item Fibers of type $D(r)$ become $D(r/\delta)$.
    \item Fibers of type $D_\circ (r)$ become $D_\circ (r/\delta)$.
    \item Fibers of type $A(r_1, r_2)$ become $A(r_1 \delta , r_2/\delta)$.
\end{itemize}
While extending cells in this way is sometimes necessary for technical reasons (e.g. needing to extend the possibly closed domain of a function to an open set) we will not be concerned with these subtleties.

\subsection{Log-Noetherian Complexity} \label{LN_Comp_app}

With their domains of definition set, we are now ready to define Log-Noetherian functions:
\begin{definition}
    Let $\mathcal{C} \subset \mathbb{C}^n$ be an LN cell. Then a \textit{Log-Noetherian (LN) chain} is a collection of holomorphic functions $F_1, \dots, F_n$ such that:
    \begin{equation}
    \label{eq:app-def-ln-func}
        \partial_{z_j}^\mathcal{C} F_i = G_{ij} (F_1, \dots, F_n),
    \end{equation}
    where the derivative is defined as:
    \begin{equation}
    \label{eq:cell-derivative}
        \partial_{z_j}^\mathcal{C} = \begin{cases}
            r\partial_{z_j} \quad \quad  \text{if the $j$'th coordinate has fiber type } \, D(r),\\
            z_j \partial_{z_j} \quad \, \text{if the $j$'th coordinate has fiber type } \, D_\circ (r) \, \text{or}\, A(r_1,r_2) ,\\
            0 \quad \quad \ \ \, \text{if the $j$'th coordinate has fiber type }\, *,
        \end{cases}
    \end{equation} and $G_{ij}$ are polynomials over $\mathbb{C}$. Finally, a \textit{Log-Noetherian function} is then a function $F: D \to \mathbb{C}$ given as a polynomial of the $F_i$'s, i.e. $F = G(F_1, \dots, F_n)$, where $G$ is a polynomial.
\end{definition}
In order to make contact with earlier notions of o-minimal structures, which are always defined over the real numbers, one can simply restrict all of the functions and domains to the real numbers in order to define the o-minimal structure $\mathbb{R}_{\text{LN}}$ generated by LN functions. In fact, it turns out that this structure also contains the graphs of LN functions as complex functions when identifying $\mathbb{C} \cong \mathbb{R}^2$ \cite{binyamini2024log}.

Since this is an effective o-minimal structure, there is a corresponding format filtration. The format of an LN-function will depend on the format of the cell it is defined on, which is constructed recursively:
\begin{itemize}
    \item For a cell of length 0, the format $\mathcal{F}(\mathcal{C})=1$.
    \item The format of a constant function $r:\mathcal{C}\to \mathbb{C}$ is the least integer upper bound for $|r|$.
    \item $\mathcal{F}(\mathcal{C}_{1,\dots,l+1} \odot *) = 1 + \mathcal{F}(\mathcal{C}_{1,\dots,l} )$,
    \item $\mathcal{F}(\mathcal{C}_{1,\dots,l+1} \odot D_\circ (r)) = \mathcal{F}(\mathcal{C}_{1,\dots,l+1} \odot D(r)) = 1 + \mathcal{F}(\mathcal{C}_{1,\dots,l} ) + \mathcal{F}(r)$,
    \item $\mathcal{F}(\mathcal{C}_{1,\dots,l+1} \odot A(r_1,r_2) = 1 + \mathcal{F}(\mathcal{C}_{1,\dots,l} ) + \mathcal{F}(r_1) + \mathcal{F}(r_2)$.
\end{itemize}
In particular, the format of a punctured disc with constant radius $r$ will be $\mathcal{F}(D_\circ (r)) = \lceil 1 + |r|\rceil$. As for LN functions as defined in formula \eqref{eq:app-def-ln-func}, their format is given by:
\begin{align}
    \label{eq:app-format-ln-chain}
    \mathcal{F}(F_1, \dots, F_N) &= \mathcal{F}(\mathcal{C}) + N + \sum_{i,j} \text{deg} \, G_{ij} + \Vert G_{ij} \Vert +\sup_{\substack{i=1, \dots, N\\ z\in \mathcal{C}}} |F_i (z)|, \\
    \label{eq:app-format-ln-func}
    \mathcal{F}(F) &= \mathcal{F}(F_1, \dots, F_N) + \text{deg} \,G + \Vert G  \Vert,
\end{align}
where the norm $\Vert G_{ij} \Vert$ of a polynomial $G_{ij}$ is given by the sum of the absolute values of its coefficients, and it is understood that one takes the least integer upper bound for these expressions. Note that this expression has explicit dependence on both the coefficients appearing in the LN chain, the format (and therefore size) of the LN cell, and the supremum of the functions used in the construction in the chain. The supremum term in particular also ensures that the functions $F_i$ cannot have poles on the LN cell, as this would yield an infinite format. 

The fact that Log-Noetherian functions are defined on finite domains and have to be bounded functions is a problem if we want to define an o-minimal structure containing them, as any o-minimal structure has to include the zero sets of polynomials, which in general are unbounded. Thus, for our construction of the effective o-minimal structure $\mathbb{R}_{\text{LN}}$ we also include all graphs of polynomials on unbounded domains by including the graphs of the functions $+,\cdot:\mathbb{R}^2 \to \mathbb{R}$ (and assigning both a format of 1), which we will discuss in detail in the next section on mathematical logic. 
It is then shown in \cite{binyamini2024log} that there exists a primitive recursive function $\mathcal{E}:\mathbb{N} \to  \mathbb{N}$ that translates a format into a bound on the number of connected components of the corresponding set, so that effective o-minimality is satisfied.

If we want to consider functions that do have poles, we need to extend this framework using what is known as a Pfaffian closure. This Pfaffian closure can be defined for any effective o-minimal structure:
\begin{definition}
    Let $\mathcal{S}$ be an effective o-minimal structure. Then the \textit{Pfaffian closure} over $\mathcal{S}$ is given by the collection of functions defined as follows. Let $G=I_1 \odot I_2 \odot \cdots \odot I_n \subset \mathbb{R}^n$ be a series of intervals fibered over each other (similarly to the LN cells), i.e. each interval is of the form:
    \begin{equation}
        I_k = (a_k, b_k), \quad a_k,b_k: I_1 \odot \cdots \odot I_{k-1} \to \mathbb{R}\cup \{\pm \infty\},
    \end{equation}
    such that $a_k<b_k$ everywhere. Now let $\zeta_1,\dots,\zeta_N$ be a collection of functions that satisfy a triangular system of differential equations:
    \begin{equation}
        \frac{\partial \zeta_i}{\partial x_j} = P_{ij} (\zeta_1, \dots \zeta_i),
    \end{equation}
    where now the $P_{ij}:X_i \to \mathbb{R}$, instead of being polynomials, are functions belonging to $\mathcal{S}$, and are also required to be real analytic. Their domains are defined in triangular way such that they contain the images of $(\zeta_1,\dots,\zeta_N)$:
    \begin{equation}
    \begin{split}
        \zeta_1 (G) &\subset X_1,\\
        \zeta_1(G) \times \zeta_2 (G) &\subset X_2,\\
        &\vdots\\
        \zeta_1(G) \times \cdots \times \zeta_N (G) &\subset X_N \equiv X.
    \end{split}
    \end{equation}
    A function $f$ in this collection is then of the form $f = P(\zeta_1,\dots,\zeta_N)$, where $P:X \to \mathbb{R}$ again belongs to $\mathcal{S}$ and $\zeta_i$ are in the Pfaffian chain. In our case, we choose $\mathcal{S} = \mathbb{R}_{\text{LN}}$, and call the resulting structure $\mathbb{R}_{\text{LN,PF}}$.
\end{definition}

The new structure $\mathbb{R}_{\text{LN,PF}}$ is again effective o-minimal, with format filtration constructed as follows. Let $f=P(\zeta_1, \dots, \zeta_N)$ be a function in $\mathbb{R}_{\text{LN,PF}}$ as before. Then the format of the corresponding Pfaffian chain is given by:
\begin{equation}
    \label{eq:app-lnpfchainformat}
    \mathcal{F}^{\text{LN,PF}} (\zeta_1, \dots, \zeta_N) = N + \mathcal{F}^{\text{LN}} (G) + \sum_{i,j} \mathcal{F}^{\text{LN}} (P_{ij}),
\end{equation}
where $\mathcal{F}^{\text{LN}}$ denote the LN-formats, and the format of $f$ is given as:
\begin{equation}
    \label{eq:app-lnpffuncformat}
    \mathcal{F}^{\text{LN,PF}} (f) = \mathcal{F}^{\text{LN,PF}} (\zeta_1, \dots, \zeta_N) + \mathcal{F}^{\text{LN}} (P).
\end{equation}
In this case, since polynomials are definable in $\mathbb{R}_{\text{LN,PF}}$ already, we technically do not need to add them by hand, but adding them in the same way as we will do for $\mathbb{R}_{\text{LN}}$ simplifies calculations greatly while retaining effective o-minimality. 
Note also that since polynomials are definable within $\mathbb{R}_{\text{LN}}$, the structure $\mathbb{R}_{\text{LN,PF}}$ contains the o-minimal structure $\mathbb{R}_{\text{Pfaff}}$ as well.

\subsection{Mathematical logic and formats of semi-algebraic sets} \label{Logic-semialg}

In order to properly assign a format to semi-algebraic sets in $\mathbb{R}_{\text{LN}}$, we first discuss some basic mathematical logic. While not necessary in principle, logical formulas are a much more convenient way to combine functions, and moreover the primitive recursive function $\mathcal{E}$ found in \cite{binyamini2024log} actually converts the format of such logical formats into bounds on the sets they define.

A logical formula is made using a \textit{language} or \textit{signature}, which contains symbols that are grouped by type:
\begin{definition}
    A language is a collection of symbols, which form the building blocks of logical statements. These symbols consist of the following:
    \begin{itemize}
        \item \textbf{Relation symbols} or \textbf{predicates}, which represent properties or relations. Examples include $=$ and $<$, which describe relations between variables.
        \item \textbf{Function symbols}, which represent functions (mappings) between symbols. For example, $+$ or $\cdot$ both take two variables as input, and have a single variable as output.
        \item \textbf{Constant symbols} are function symbols that do not accept any input.
    \end{itemize}
\end{definition}
These symbols are used to form \textit{terms}, which are defined recursively as:
\begin{definition}
\label{def:term}
    A \textit{term} is the most basic building block of a formula, and can consist of the following types of objects (defined recursively):
    \begin{itemize}
        \item A variable $x$,
        \item A constant $a$,
        \item A function symbol $f(t_1,\dots,t_n)$, where $t_i$ are again terms.
    \end{itemize}
    These can then be combined to form \textit{atomic predicates}, which are either of the form $P(t_1,\dots,t_n)$ where $P$ is a relation symbol and $t_1,\dots,t_n$ are terms, or of the form $t_1=t_2$ where $t_1,t_2$ are terms.
\end{definition}
In addition to these symbols, first-order logical statements also require some universal ingredients:
\begin{itemize}
    \item \textbf{Quantifiers}, namely the existential quantifier $\exists$ and the universal quantifier $\forall$.
    \item \textbf{Logical connectives}, namely the conjunction $\wedge$ (`and'), the disjunction $\lor$ (`or') and the negation $\neg$ (`not').
\end{itemize}
In order to assign a format to a formula in a language, all of the symbols in the language should be assigned a format. The quantifiers and logical connectives are fixed by effective o-minimality as can be seen in table \ref{tab:logical_operations}.
\begin{table}[h!]
    \begin{tabular}{c|l|l}
         Set operation & Corresponding formula & Format from eff. o-minimality\\
         \hline
         $A$ & $\phi(x) = x  \in A$ & $\mathcal{F}(\phi) = \mathcal{F}(A)$\\
         $A \cup B$ & $\phi(x) = \phi_1 (x) \lor  \phi_2(x)$ & $\mathcal{F}(\phi) =  \max\{\mathcal{F}(\phi_1), \mathcal{F}(\phi_2)\}+ 1$\\
         $A \cap B$ & $\phi(x) = \phi_1(x) \wedge  \phi_2(x) $ & $\mathcal{F}(\phi) = \max\{\mathcal{F}(\phi_1), \mathcal{F}(\phi_2)\} + 1$\\
         $\pi_n (A)$ & $\phi(x) = \exists y_1\dots y_n:(\phi'(x,y_1,\dots,y_n))$ & $\mathcal{F}(\phi) = \mathcal{F}(\phi')+ 1$\\
         $\mathbb{R}^n\setminus A$ & $\phi(x) = \neg \phi'(x)$ & $\mathcal{F}(\phi) = \mathcal{F}(\phi') + 1$\\
         $A \times B$ & $\phi(x,y) = \phi_1(x) \wedge \phi_2(y)$ & $\mathcal{F}(\phi) = \max\{\mathcal{F}(\phi_1), \mathcal{F}(\phi_2)\} + 1$
    \end{tabular}
     \caption{Relation between formulas of logical symbols and their formats, as well as their corresponding set operation. Note that the universal quantifier $\forall$ is missing from the list because $\forall x: \phi \iff \neg \exists x: \neg \phi$.}
    \label{tab:logical_operations}
\end{table}
Furthermore, we declare that the format of an atomic predicate is the sum of the formats of its components (with variables having format 0), and that the format of a set defined by a formula is equal to the format of that formula. 

With these definitions, we now define the language corresponding to the structure $\mathbb{R}_{\text{LN}}$ as:
\begin{equation}
    \mathcal{L}_{\text{LN}} = \{=,>,+,\cdot, \text{Gr }F, a \in \mathbb{R} \},
\end{equation}
where $=,>,+,\cdot$ are relation symbols (for $+,\cdot$ this means that they accept three variables/constants $x,y,z$ and return True if $x+y=z$ or $x \cdot y =z$ respectively), $a \in \mathbb{R}$ implies that we add a constant symbol for every real number, and $\text{Gr } F$ means that we add the graphs of LN functions, seen as subsets of $\mathbb{R}^{2l} \cong \mathbb{C}^l$. This represents a slight departure from the definition of in \cite{binyamini2024log}, where instead the LN function has to be real and is restricted to the real part of a cell. Since it is subsequently shown in \cite{binyamini2024log} (proposition 48) that the graphs of complex LN functions also belong to the structure $\mathbb{R}_{\text{LN}}$ and their format is bounded by a primitive recursive function of their format, we absorb this primitive recursive function into $\mathcal{E}$ from axiom 4 in definition \ref{def:eff-o-minimality}. In principle this means that for real functions we will be overestimating the bounds obtained from $\mathcal{E}$, however we do not have an explicit expression for $\mathcal{E}$ at the moment anyway. In Binyamini's construction, there is also an explicit dependence on the $\delta$-extension that the cell admits which we omit here.

The formats of the symbols in the language are given as:
\begin{itemize}
    \item $\mathcal{F}(=) = \mathcal{F}(>) = \mathcal{F}(+) = \mathcal{F}(\cdot) = \mathcal{F}(a) = 1$,
    \item $\mathcal{F}(\text{Gr } F) = \mathcal{F} (F) $ as in equation \eqref{eq:app-format-ln-func},
\end{itemize}
and we define the format of an atomic formula constructed using these symbols to be the sum of the formats of the symbols used (with variables contributing a format of 0).
This allows us to in principle find the format of any definable set in $\mathbb{R}_{\text{LN}}$, including the semi-algebraic sets. For the latter, we decompose the corresponding semi-algebraic formula into atomic predicates. Some examples were shown in the main text in section \ref{ch:example-computations}.
In general, finding the lowest possible format for a polynomial is a difficult task, because it is often possible to find clever ways to decompose them into atomic formulae. For example, the monomial $y=x^{60}$ can be written as $y=((((x)^2)^2)^3)^5$ which means one only needs a small number of formulae, as opposed to a more naive approach where one simply uses repeated multiplication. The most efficient way to compose polynomials into repeated addition and multiplication is a well-known problem in computer science, and is extensively treated in chapter 4.6 of \cite{knuth97}.
Once we have found the decomposition of a complicated formula in terms of a conjunction of atomic formulas, we have to evaluate the format of this conjunction, which involves repeated application of the $\text{max} +1$ rule. We explain how to find such formats in the next section.

For the Pfaffian extension $\mathbb{R}_{\text{LN,PF}}$, we take as a language
\begin{equation}
    \mathcal{L}_{\text{LN,PF}} = \{=,>,+,\cdot, \text{Gr f}, a \in \mathbb{R}\},
\end{equation}
where now $f$ is an LN,PF-function instead. This again represents a slight departure from Binyamini's construction, since we are explicitly including the symbols $=,>,+,\cdot$ in our language, despite the fact that $+,\cdot$ are already LN,PF-functions by themselves (and moreover atomic formulas can always contain $=$, and $>$ can be constructed using projections, meaning they are not strictly necessary either). This is done to keep computations of formats of semi-algebraic sets tractable, and does not break effective o-minimality: after all, we already have bounds on sets constructed this way as in $\mathbb{R}_{\text{LN}}$. The symbols $=,>,+,\cdot,a\in \mathbb{R}$ are all assigned a format of 1 again, and the format of $\text{Gr } f$ is given by equation \eqref{eq:app-lnpffuncformat}.

\subsection{Conjunctions of formulas} \label{Conj_app}
We now present a way to find the format of a conjunction of formulas, which makes it easier to find the formats for complicated semi-algebraic sets.
First, we note that very often, it does not matter how we place parentheses. For example, the formula
\begin{equation}
    \phi_1 \equiv \left(y = a \cdot t_1 \ \wedge \ t_1 = t_2 \cdot x \right)\ \wedge \ t_2 = y + z \,,
\end{equation}
defines the same set as
\begin{equation}
    \phi_2 \equiv y =  a \cdot t_1 \ \wedge \ (t_1 = t_2 \cdot x \ \wedge \ t_2 = y + z) \,.
\end{equation}
However, the formats are respectively given by
\begin{align}
    \mathcal{F}(\phi_1) &= \max\{\max\{2,1\}+1,1\}=4\\
    \mathcal{F}(\phi_2) &= \max\{\max\{1,1\}+1,2\}=3.
\end{align}
This means that we should be clever in ordering our subformulas when we are free to switch them around. We will now develop a method that yields the lowest format for such a conjunction.
In order to gain some intuition, let us  consider a formula $\phi$, made up of a conjunction of of 9 subformulas, which each have format $\mathcal{F}=3$. So schematically, we have
\begin{equation}
    \phi = 3 \wedge 3 \wedge 3 \wedge 3 \wedge 3 \wedge 3\wedge 3 \wedge 3 \wedge 3 \,,
\end{equation}
where instead of the formulas themselves, we simply note their format.

We can perform the conjunction by making pairs of these formulas, which will increase the format by 1 every time. This schematically looks like:
\begin{equation}
\begin{gathered}
        \wick{\c1 3 \wedge \c1 3 \wedge \c1 3 \wedge \c1 3 \wedge \c1 3 \wedge \c1 3 \wedge \c1 3 \wedge \c1 3  \wedge 3}\\
        \Downarrow \\
        \wick{ 4 \wedge \c1 4 \wedge \c1 4 \wedge \c1 4 \wedge \c1 3}\\
        \Downarrow \\
        \wick{\c1 4 \wedge \c1 5 \wedge 5}\\
        \Downarrow \\
        \wick{\c1 6 \wedge \c1 5}\\
        \Downarrow \\
        7
\end{gathered}
\end{equation}
Where the contractions represent taking the maximum and adding one. At each step, we effectively halve the number of formulas. When we have an odd number of formulas, one of them carries over to the next step and effectively acts as if it had also been increased by 1. We needed 4 steps, because $\lceil 9/2^3 \rceil>1$ but $\lceil 9/2^4 \rceil =1$, and so the format is given by the starting format plus the number of steps taken.

We can generalize this to a formula that consists of $k$ subformulas which all have format $\mathcal{F}_0$, and end up with a format $\mathcal{F} =\mathcal{F}_0+m$,
where $m \in \mathbb{Z}$ such that $\frac{k}{2^{m-1}}>1$ and $\frac{k}{2^m}\leq1$. Solving this for $m$ yields $m=\lceil \log_2 k \rceil$, and so we find that a format for a formula which is a conjunction of $k$ subformulas which all have format $\mathcal{F}_0$:
\begin{equation}
\label{eq:format-equal-formats}
    \mathcal{F}(\underbrace{\mathcal{F}_0 \wedge \dots \wedge \mathcal{F}_0}_{k \text{ times}}) = \mathcal{F}_0 + \lceil \log_2 k \rceil.
\end{equation}
This formula is reminiscent of the formula used to count the number of bits in the binary representation of an integer. Instead of an integer, we now have a conjunction of formulas, and the format thus measures how many `bits' of logical information are contained within the expression.

Let us now generalize this to an arbitrary conjunction of formulas. Suppose the formula $\phi$ now consists of $n_1$ formulas of format 1, $n_2$ formulas of format 2 and so on up until $n_k$ formulas of format $k$. We start with the formulas of format 1, and contract them pairwise. This means we will end up with $\lceil n_1/2\rceil$ formulas of format 2. Then we combine those with the $n_2$ formulas of format 2 that were already present, which leaves us with $\left\lceil \frac{n_2 + \lceil n_1/2 \rceil}{2} \right \rceil$
formulas of format 3. Before continuing, we can simplify this expression, since 
$\left\lceil \frac{ \lceil x \rceil + m}{n} \right\rceil = \left\lceil \frac{x+m}{n} \right\rceil,$
for $n,m$ positive integers \cite{knuth1989concrete}, which leaves us with $\left\lceil \frac{n_2}{2} +  \frac{n_1}{2^2} \right\rceil$ formulas of format 3.
We can continue this process and find that we finally end up with $N$ formulas of format $k$, with $N$ given by:
\begin{equation}
    N=\left\lceil \frac{n_k}{2} + \frac{n_{k-1}}{2^2} + \dots + \frac{n_2}{2^{k-1}} + \frac{n_1}{2^k} \right\rceil.
\end{equation}
Since all formulas now have the same format, we can use equation \eqref{eq:format-equal-formats} to find:\footnote{We have also used the fact that $\lceil f(\lceil x\rceil) \rceil = \lceil f(x) \rceil$ for any monotonic function $f$ \cite{knuth1989concrete}.}
\begin{equation}
\label{eq:format-conjunction}
    \mathcal{F}\left(\bigwedge_i\phi_i\right) = k + \left\lceil \log_2 \left(n_k + \frac{n_{k-1}}{2} + \dots + \frac{n_2}{2^{k-2}} + \frac{n_1}{2^{k-1}} \right) \right\rceil.
\end{equation}

\section{Connection Matrices and Chains around $z=\infty$}
\label{app: connection matrices}
The chain around $z=1$ is summarized for all families of curves as
\begin{equation}
\begin{aligned}
    z \partial_z F_1 &= \frac{n}{2\pi i}F_2 + F_3,\\
    z \partial_z F_2 &= F_4,\\
    z \partial_z F_3 &= \alpha_1 \alpha_2 F_1 F_5+ F_3 F_5 + \frac{n}{2\pi i} F_4,\\
    z \partial_z F_4 &= \alpha_1 \alpha_2 F_2 F_5 + F_4 F_5,\\
    z \partial_z F_5 &= F_5^2 + F_5,
\end{aligned}
\end{equation}
which is solved by $(F_1, \dots, F_5) = (Y_{11}, Y_{12}, Y_{21}, Y_{22}, \frac{z}{1-z})$.

In order to find the corresponding chains around $z=\infty$, we first give 
the connection matrices \eqref{eq:connection-y} as
\begin{align}
    C_\infty^{(\Gamma_1(1))} &= \begin{pmatrix}
        -\frac{i}{6\sqrt{3}} & \frac{i}{3\sqrt{3}} & 1 & 0\\
        -\frac{i}{3\sqrt{3}} & \frac{i}{6\sqrt{3}} & 0 & 1\\
        \frac{5}{36} \frac{1}{1-z} & 0 & -(\frac{i}{6\sqrt{3}} + \frac{1}{1-z}) & \frac{i}{3\sqrt{3}}\\
        0 & \frac{5}{36} \frac{1}{1-z} & -\frac{i}{3\sqrt{3}} & \frac{i}{6\sqrt{3}} - \frac{1}{1-z}
    \end{pmatrix}, \\
    C_\infty^{(\Gamma_1(2))} &= \begin{pmatrix}
        -\frac i 4 & \frac i 2 & 1 & 0\\
        -\frac i 4 & \frac i 4 &  0 & 1\\
        \frac{3}{16} \frac{1}{1-z} & 0 & -(\frac i 4 + \frac{1}{1-z}) & \frac i 2\\
        0 & \frac{3}{16} \frac{1}{1-z} & -\frac i 4 & \frac i 4  - \frac{1}{1-z}
    \end{pmatrix},\\
    C_\infty^{(\Gamma_1(3))} &= \begin{pmatrix}
        -\frac{i}{\sqrt{3}} & \frac{2i}{\sqrt{3}} & 1 & 0\\
        -\frac{2i}{3\sqrt{3}} & \frac{i}{\sqrt{3}} & 0 & 1\\
        \frac{2}{9} \frac{1}{1-z} & 0 & -(\frac{i}{\sqrt{3}} - \frac{1}{1-z}) & \frac{2i}{\sqrt{3}} \\
        0 & \frac{2}{9} \frac{1}{1-z} & -\frac{2i}{3\sqrt{3}} & \frac{i}{\sqrt{3}} - \frac{1}{1-z}
    \end{pmatrix},\\
    C_\infty^{(\Gamma_1(4))} &= \begin{pmatrix}
        \frac 1 2 + \frac{i}{\pi} & \frac{2}{\pi i} & 1 & 0\\
        -\frac{1}{2\pi i} & \frac 1 2 - \frac i \pi &  0 & 1\\
        \frac 1 4 \frac{1}{1-z} & 0 &  \frac 1 2 + \frac i \pi  -\frac{1}{1-z} & \frac{2}{\pi i} \\
        0 & \frac 1 4 \frac{1}{1-z} & -\frac{1}{2\pi i} & \frac 1 2 - \frac i \pi - \frac{1}{1-z}
    \end{pmatrix}.
\end{align}
For $\Gamma_1(1)$ this leads to the chain
\begin{equation}
    \begin{aligned}
        z \partial_z F_1 &= -\frac{i}{6 \sqrt{3}} F_1 + \frac{i}{3\sqrt{3}} F_2 + F_3,\\
        z \partial_z F_2 &= -\frac{i}{3\sqrt{3}} F_1 + \frac{i}{6\sqrt{3}} F_2 + F_4,\\
        z \partial_z F_3 &= \frac{i}{3\sqrt{3}} F_4 - \frac{i}{6 \sqrt{3}} F_3 - F_3 F_5 + \frac{5}{36}  F_1 F_5,\\
        z \partial_z F_4 &= -\frac{i}{3\sqrt{3}} F_3 + \frac{i}{6\sqrt{3}} F_4  - F_4 F_5 + \frac{5}{36} F_2 F_5,\\
        z \partial_z F_5 &= F_5^2 - F_5,
    \end{aligned}
\end{equation}
where $F_1,F_2,F_3,F_4$ are the components of $Y$ as discussed in the text, and $F_5=\frac{1}{1-z}$. This will be the case for the other chains as well. For $\Gamma_1(2)$ we instead obtain
\begin{equation}
    \begin{aligned}
        z \partial_z F_1  &= -\frac i 4 F_1 + \frac i 2 F_2 + F_3,\\
        z \partial_z F_2 &= -\frac i 4 F_1 + \frac i 4 F_2 + F_4,\\
        z \partial_z F_3 &= \frac i 2 F_4 - \frac i 4 F_3 -F_3 F_5 + \frac{3}{16} F_1 F_5,\\
        z \partial_z F_4 &= -\frac i 4 F_3 + \frac i 4 F_4 - F_4 F_5 + \frac{3}{16} F_2 F_5,\\
        z \partial_z F_5 &= F_5^2 - F_5,
    \end{aligned}
\end{equation}
and for $\Gamma_1(3)$ the chain reads
\begin{equation}
    \begin{aligned}
        z \partial_z F_1 &= -\frac{i}{\sqrt{3}} F_1 + \frac{2i}{\sqrt{3}} F_2 + F_3,\\
        z \partial_z F_2 &= -\frac{2i}{3\sqrt{3}} F_1 + \frac{i}{\sqrt{3}} F_2 + F_4,\\
        z \partial_z F_3 &= \frac{2i}{\sqrt{3}} F_4 - \frac{i}{\sqrt{3}} F_3 - F_3 F_5 + \frac 2 9 F_1 F_5,\\
        z \partial_z F_4 &= -\frac{2i}{3\sqrt{3}} F_3 + \frac{i}{\sqrt{3}} F_4 - F_4 F_5 + \frac 2 9 F_2 F_5,\\
        z  \partial_z F_5 &= F_5^2 - F_5.
    \end{aligned}
\end{equation}
Finally for $\Gamma_1(4)$ we find
\begin{equation}
\begin{aligned}
    z \partial_z F_1 &= F_3 + \frac{2}{\pi i} F_2 + \left(\frac 1 2 + \frac i \pi\right) F_1,\\
    z \partial_z F_2 &= F_4 + \left(\frac 1 2 - \frac i \pi \right) F_2  - \frac{1}{2 \pi i} F_1,\\
    z \partial_z F_3 &= \frac 1 4 F_1 F_5 + \frac{2}{\pi i} F_4 + \left(\frac 1 2 + \frac i \pi \right) F_3 - F_3 F_5,\\
    z \partial_z F_4 &= \frac 1 4 F_2 F_5 + \left(\frac 1 2 - \frac i \pi \right) F_4 - F_4 F_5 - \frac{1}{2\pi i} F_3,\\
    z \partial_z F_5 &= F_5^2 - F_5.
\end{aligned}
\end{equation}

\bibliographystyle{utphys}
\bibliography{biblio.bib}

\providecommand{\href}[2]{#2}\begingroup\raggedright\begin{thebibliography}{10}

\bibitem{Grimm:2021vpn}
T.~W. Grimm, ``{Taming the landscape of effective theories},'' \href{http://dx.doi.org/10.1007/JHEP11(2022)003}{{\em JHEP} {\bfseries 11} (2022) 003}, \href{http://arxiv.org/abs/2112.08383}{{\ttfamily arXiv:2112.08383 [hep-th]}}.

\bibitem{Douglas:2022ynw}
M.~R. Douglas, T.~W. Grimm, and L.~Schlechter, ``{The Tameness of Quantum Field Theory, Part I -- Amplitudes},'' \href{http://dx.doi.org/10.4310/ATMP.241119035402}{{\em Adv. Theor. Math. Phys.} {\bfseries 28} no.~8, (2024) 2603--2656}, \href{http://arxiv.org/abs/2210.10057}{{\ttfamily arXiv:2210.10057 [hep-th]}}.

\bibitem{Douglas:2023fcg}
M.~R. Douglas, T.~W. Grimm, and L.~Schlechter, ``{The Tameness of Quantum Field Theory, Part II -- Structures and CFTs},'' \href{http://arxiv.org/abs/2302.04275}{{\ttfamily arXiv:2302.04275 [hep-th]}}.

\bibitem{Grimm:2023xqy}
T.~W. Grimm, L.~Schlechter, and M.~van Vliet, ``{Complexity in tame quantum theories},'' \href{http://dx.doi.org/10.1007/JHEP05(2024)001}{{\em JHEP} {\bfseries 05} (2024) 001}, \href{http://arxiv.org/abs/2310.01484}{{\ttfamily arXiv:2310.01484 [hep-th]}}.

\bibitem{Grimm:2024elq}
T.~W. Grimm and M.~van Vliet, ``{On the complexity of quantum field theory},'' \href{http://dx.doi.org/10.1007/JHEP06(2025)215}{{\em JHEP} {\bfseries 06} (2025) 215}, \href{http://arxiv.org/abs/2410.23338}{{\ttfamily arXiv:2410.23338 [hep-th]}}.

\bibitem{van1998tame}
L.~van~den Dries, {\em Tame topology and o-minimal structures}, vol.~248.
\newblock Cambridge University Press, 1998.

\bibitem{Seiberg:1994rs}
N.~Seiberg and E.~Witten, ``{Electric - magnetic duality, monopole condensation, and confinement in N=2 supersymmetric Yang-Mills theory},'' \href{http://dx.doi.org/10.1016/0550-3213(94)90124-4}{{\em Nucl. Phys. B} {\bfseries 426} (1994) 19--52}, \href{http://arxiv.org/abs/hep-th/9407087}{{\ttfamily arXiv:hep-th/9407087}}. [Erratum: Nucl.Phys.B 430, 485--486 (1994)].

\bibitem{Seiberg:1994aj}
N.~Seiberg and E.~Witten, ``{Monopoles, duality and chiral symmetry breaking in N=2 supersymmetric QCD},'' \href{http://dx.doi.org/10.1016/0550-3213(94)90214-3}{{\em Nucl. Phys. B} {\bfseries 431} (1994) 484--550}, \href{http://arxiv.org/abs/hep-th/9408099}{{\ttfamily arXiv:hep-th/9408099}}.

\bibitem{binyamini2023tameness}
G.~Binyamini and D.~Novikov, ``Tameness in geometry and arithmetic: beyond o-minimality,'' in {\em International congress of mathematicians}, pp.~1440--1461.
\newblock 2023.

\bibitem{binyamini2022sharplyominimalstructuressharp}
G.~Binyamini, D.~Novikov, and B.~Zack, ``Sharply o-minimal structures and sharp cellular decomposition,'' \href{http://arxiv.org/abs/2209.10972}{{\ttfamily arXiv:2209.10972 [math.LO]}}.

\bibitem{binyamini2024log}
G.~Binyamini, ``{Log-Noetherian functions},'' \href{http://arxiv.org/abs/2405.16963}{{\ttfamily arXiv:2405.16963 [math.AG]}}.

\bibitem{binyamini2019complex}
G.~Binyamini and D.~Novikov, ``Complex cellular structures,'' {\em Annals of Mathematics} {\bfseries 190} no.~1, (2019) 145--248, \href{http://arxiv.org/abs/1802.07577}{{\ttfamily arXiv:1802.07577 [math.CV]}}.

\bibitem{Delgado:2024skw}
M.~Delgado, D.~van~de Heisteeg, S.~Raman, E.~Torres, C.~Vafa, and K.~Xu, ``{Finiteness and the emergence of dualities},'' \href{http://dx.doi.org/10.21468/SciPostPhys.19.2.047}{{\em SciPost Phys.} {\bfseries 19} no.~2, (2025) 047}, \href{http://arxiv.org/abs/2412.03640}{{\ttfamily arXiv:2412.03640 [hep-th]}}.

\bibitem{Grimm:2025lip}
T.~W. Grimm, D.~Prieto, and M.~van Vliet, ``{Tame embeddings, volume growth, and complexity of moduli spaces},'' \href{http://dx.doi.org/10.1103/d51c-j1s9}{{\em Phys. Rev. D} {\bfseries 112} no.~10, (2025) 106015}, \href{http://arxiv.org/abs/2503.15601}{{\ttfamily arXiv:2503.15601 [hep-th]}}.

\bibitem{Grimm_2022}
T.~W. Grimm, S.~Lanza, and C.~Li, ``{Tameness, Strings, and the Distance Conjecture},'' \href{http://dx.doi.org/10.1007/JHEP09(2022)149}{{\em JHEP} {\bfseries 09} (2022) 149}, \href{http://arxiv.org/abs/2206.00697}{{\ttfamily arXiv:2206.00697 [hep-th]}}.

\bibitem{Gabrielov-1968}
A.~Gabrielov, ``Projections of semi-analytic sets.'' {\em Funct. Anal. Appl.} {\bfseries 2} no.~4, (1968) 282--291.

\bibitem{yomdin-2004}
Y.~Yomdin and G.~Comte, {\em Tame geometry with application in smooth analysis}, vol.~1834 of {\em Lecture Notes in Mathematics}.
\newblock Springer, 2004.

\bibitem{khovanskiĭfewnomials}
A.~G. Khovanski{\u\i}, {\em Fewnomials}, vol.~88 of {\em Translations of Mathematical Monographs}.
\newblock American Mathematical Soc., 1991.

\bibitem{Wilkie1999ATO}
A.~J. Wilkie, ``A theorem of the complement and some new o-minimal structures,'' {\em Selecta Mathematica} {\bfseries 5} (1999) 397--421. \url{https://api.semanticscholar.org/CorpusID:122959059}.

\bibitem{binyamini2020effectivecylindricalcelldecompositions}
G.~Binyamini and N.~Vorobjov, ``Effective cylindrical cell decompositions for restricted sub-pfaffian sets,'' \href{http://dx.doi.org/10.1093/imrn/rnaa285}{{\em International Mathematics Research Notices} {\bfseries 2022} no.~5, (11, 2020) 3493--3510}, \href{http://arxiv.org/abs/2004.06411}{{\ttfamily arXiv:2004.06411 [math.LO]}}.

\bibitem{Grimm:2024mbw}
T.~W. Grimm, A.~Hoefnagels, and M.~van Vliet, ``{Structure and complexity of cosmological correlators},'' \href{http://dx.doi.org/10.1103/PhysRevD.110.123531}{{\em Phys. Rev. D} {\bfseries 110} no.~12, (2024) 123531}, \href{http://arxiv.org/abs/2404.03716}{{\ttfamily arXiv:2404.03716 [hep-th]}}.

\bibitem{Carlson:period-mappings}
J.~Carlson, S.~M{\"u}ller-Stach, and C.~Peters, {\em Period Mappings and Period Domains}.
\newblock Cambridge Studies in Advanced Mathematics. Cambridge University Press, 2~ed., 2017.

\bibitem{Vafa_1996}
C.~Vafa, ``{Evidence for F theory},'' \href{http://dx.doi.org/10.1016/0550-3213(96)00172-1}{{\em Nucl. Phys. B} {\bfseries 469} (1996) 403--418}, \href{http://arxiv.org/abs/hep-th/9602022}{{\ttfamily arXiv:hep-th/9602022}}.

\bibitem{Gaiotto_2012}
D.~Gaiotto, ``{N=2 dualities},'' \href{http://dx.doi.org/10.1007/JHEP08(2012)034}{{\em JHEP} {\bfseries 08} (2012) 034}, \href{http://arxiv.org/abs/0904.2715}{{\ttfamily arXiv:0904.2715 [hep-th]}}.

\bibitem{vandeheisteeg:2024chartingcomplexstructurelandscape}
D.~van~de Heisteeg, ``{Charting the complex structure landscape of F-theory},'' \href{http://dx.doi.org/10.1007/JHEP05(2025)150}{{\em JHEP} {\bfseries 05} (2025) 150}, \href{http://arxiv.org/abs/2404.03456}{{\ttfamily arXiv:2404.03456 [hep-th]}}.

\bibitem{Schimannek:2022ellipticcurves}
T.~Schimannek, ``{Modular curves, the Tate-Shafarevich group and Gopakumar-Vafa invariants with discrete charges},'' \href{http://dx.doi.org/10.1007/JHEP02(2022)007}{{\em JHEP} {\bfseries 02} (2022) 007}, \href{http://arxiv.org/abs/2108.09311}{{\ttfamily arXiv:2108.09311 [hep-th]}}.

\bibitem{NIST:DLMF}
``{\it NIST Digital Library of Mathematical Functions}.'' \url{https://dlmf.nist.gov/}, release 1.2.4 of 2025-03-15.
\newblock \url{https://dlmf.nist.gov/}. F.~W.~J. Olver, A.~B. {Olde Daalhuis}, D.~W. Lozier, B.~I. Schneider, R.~F. Boisvert, C.~W. Clark, B.~R. Miller, B.~V. Saunders, H.~S. Cohl, and M.~A. McClain, eds.

\bibitem{Bengtsson:2006rfv}
I.~Bengtsson and K.~{\.Z}yczkowski, {\em Geometry of Quantum States: An Introduction to Quantum Entanglement}.
\newblock Cambridge University Press, 2006.

\bibitem{verrill2001algorithm}
H.~Verrill, ``Algorithm for drawing fundamental domains.'' 2001.

\bibitem{bakker2020tametopologyarithmeticquotients}
B.~Bakker, B.~Klingler, and J.~Tsimerman, ``Tame topology of arithmetic quotients and algebraicity of hodge loci,'' {\em Journal of the American Mathematical Society} {\bfseries 33} no.~4, (2020) 917--939, \href{http://arxiv.org/abs/1810.04801}{{\ttfamily arXiv:1810.04801 [math.AG]}}.

\bibitem{Klemm:1995wp}
A.~Klemm, W.~Lerche, and S.~Theisen, ``{Nonperturbative effective actions of N=2 supersymmetric gauge theories},'' \href{http://dx.doi.org/10.1142/S0217751X96001000}{{\em Int. J. Mod. Phys. A} {\bfseries 11} (1996) 1929--1974}, \href{http://arxiv.org/abs/hep-th/9505150}{{\ttfamily arXiv:hep-th/9505150}}.

\bibitem{Lerche:1996xu}
W.~Lerche, ``{Introduction to Seiberg-Witten theory and its stringy origin},'' \href{http://dx.doi.org/10.1016/S0920-5632(97)00073-X}{{\em Nucl. Phys. B Proc. Suppl.} {\bfseries 55} (1997) 83--117}, \href{http://arxiv.org/abs/hep-th/9611190}{{\ttfamily arXiv:hep-th/9611190}}.

\bibitem{Seiberg:1988ur}
N.~Seiberg, ``{Supersymmetry and Nonperturbative beta Functions},'' \href{http://dx.doi.org/10.1016/0370-2693(88)91265-8}{{\em Phys. Lett. B} {\bfseries 206} (1988) 75--80}.

\bibitem{complexityconj}
T.~W. Grimm, D.~Prieto, and M.~van Vliet, ``{Complexity of Effective Field Theories in the Quantum Gravity Landscape},'' \href{http://arxiv.org/abs/2601.xxxx}{{\ttfamily 2601.xxxx}}. (work in progress).

\bibitem{Klemm:1996bj}
A.~Klemm, W.~Lerche, P.~Mayr, C.~Vafa, and N.~P. Warner, ``{Selfdual strings and N=2 supersymmetric field theory},'' \href{http://dx.doi.org/10.1016/0550-3213(96)00353-7}{{\em Nucl. Phys. B} {\bfseries 477} (1996) 746--766}, \href{http://arxiv.org/abs/hep-th/9604034}{{\ttfamily arXiv:hep-th/9604034}}.

\bibitem{Katz:1996fh}
S.~H. Katz, A.~Klemm, and C.~Vafa, ``{Geometric engineering of quantum field theories},'' \href{http://dx.doi.org/10.1016/S0550-3213(97)00282-4}{{\em Nucl. Phys. B} {\bfseries 497} (1997) 173--195}, \href{http://arxiv.org/abs/hep-th/9609239}{{\ttfamily arXiv:hep-th/9609239}}.

\bibitem{schmid}
W.~Schmid, ``{Variation of Hodge structure: the singularities of the period mapping},'' \href{http://dx.doi.org/10.1007/BF01389674}{{\em Invent. Math. , 22:211--319, 1973} (1973) }.

\bibitem{Kerr2017}
M.~{Kerr}, G.~J. {Pearlstein}, and C.~{Robles}, ``{Polarized relations on horizontal \(\operatorname{SL}(2)\)'s.},'' \href{http://dx.doi.org/10.25537/dm.2019v24.1295-1360}{{\em {Doc. Math.}} {\bfseries 24} (2019) 1295--1360}.

\bibitem{Grimm:2018ohb}
T.~W. Grimm, E.~Palti, and I.~Valenzuela, ``{Infinite Distances in Field Space and Massless Towers of States},'' \href{http://dx.doi.org/10.1007/JHEP08(2018)143}{{\em JHEP} {\bfseries 08} (2018) 143}, \href{http://arxiv.org/abs/1802.08264}{{\ttfamily arXiv:1802.08264 [hep-th]}}.

\bibitem{Grimm:2018cpv}
T.~W. Grimm, C.~Li, and E.~Palti, ``{Infinite Distance Networks in Field Space and Charge Orbits},'' \href{http://dx.doi.org/10.1007/JHEP03(2019)016}{{\em JHEP} {\bfseries 03} (2019) 016}, \href{http://arxiv.org/abs/1811.02571}{{\ttfamily arXiv:1811.02571 [hep-th]}}.

\bibitem{Corvilain:2018lgw}
P.~Corvilain, T.~W. Grimm, and I.~Valenzuela, ``{The Swampland Distance Conjecture for K{\"a}hler moduli},'' \href{http://dx.doi.org/10.1007/JHEP08(2019)075}{{\em JHEP} {\bfseries 08} (2019) 075}, \href{http://arxiv.org/abs/1812.07548}{{\ttfamily arXiv:1812.07548 [hep-th]}}.

\bibitem{Grimm:2019ixq}
T.~W. Grimm, C.~Li, and I.~Valenzuela, ``{Asymptotic Flux Compactifications and the Swampland},'' \href{http://dx.doi.org/10.1007/JHEP06(2020)009}{{\em JHEP} {\bfseries 06} (2020) 009}, \href{http://arxiv.org/abs/1910.09549}{{\ttfamily arXiv:1910.09549 [hep-th]}}. [Erratum: JHEP 01, 007 (2021)].

\bibitem{Lee:2021qkx}
S.-J. Lee and T.~Weigand, ``{Elliptic K3 surfaces at infinite complex structure and their refined Kulikov models},'' \href{http://dx.doi.org/10.1007/JHEP09(2022)143}{{\em JHEP} {\bfseries 09} (2022) 143}, \href{http://arxiv.org/abs/2112.07682}{{\ttfamily arXiv:2112.07682 [hep-th]}}.

\bibitem{Hassfeld:2025uoy}
B.~Hassfeld, J.~Monnee, T.~Weigand, and M.~Wiesner, ``{Emergent Strings in Type IIB Calabi--Yau Compactifications},'' \href{http://arxiv.org/abs/2504.01066}{{\ttfamily arXiv:2504.01066 [hep-th]}}.

\bibitem{Klemm:1994qs}
A.~Klemm, W.~Lerche, S.~Yankielowicz, and S.~Theisen, ``{Simple singularities and N=2 supersymmetric Yang-Mills theory},'' \href{http://dx.doi.org/10.1016/0370-2693(94)01516-F}{{\em Phys. Lett. B} {\bfseries 344} (1995) 169--175}, \href{http://arxiv.org/abs/hep-th/9411048}{{\ttfamily arXiv:hep-th/9411048}}.

\bibitem{Argyres:1994xh}
P.~C. Argyres and A.~E. Faraggi, ``{The vacuum structure and spectrum of N=2 supersymmetric SU(n) gauge theory},'' \href{http://dx.doi.org/10.1103/PhysRevLett.74.3931}{{\em Phys. Rev. Lett.} {\bfseries 74} (1995) 3931--3934}, \href{http://arxiv.org/abs/hep-th/9411057}{{\ttfamily arXiv:hep-th/9411057}}.

\bibitem{Vafa:2005ui}
C.~Vafa, ``{The String landscape and the swampland},'' \href{http://arxiv.org/abs/hep-th/0509212}{{\ttfamily arXiv:hep-th/0509212}}.

\bibitem{Kim:2019ths}
H.-C. Kim, H.-C. Tarazi, and C.~Vafa, ``{Four-dimensional $\mathbf{\mathcal{N}=4}$ SYM theory and the swampland},'' \href{http://dx.doi.org/10.1103/PhysRevD.102.026003}{{\em Phys. Rev. D} {\bfseries 102} no.~2, (2020) 026003}, \href{http://arxiv.org/abs/1912.06144}{{\ttfamily arXiv:1912.06144 [hep-th]}}.

\bibitem{Kumar:2010ru}
V.~Kumar, D.~R. Morrison, and W.~Taylor, ``{Global aspects of the space of 6D N = 1 supergravities},'' \href{http://dx.doi.org/10.1007/JHEP11(2010)118}{{\em JHEP} {\bfseries 11} (2010) 118}, \href{http://arxiv.org/abs/1008.1062}{{\ttfamily arXiv:1008.1062 [hep-th]}}.

\bibitem{Morrison:2011mb}
D.~R. Morrison and W.~Taylor, ``{Matter and singularities},'' \href{http://dx.doi.org/10.1007/JHEP01(2012)022}{{\em JHEP} {\bfseries 01} (2012) 022}, \href{http://arxiv.org/abs/1106.3563}{{\ttfamily arXiv:1106.3563 [hep-th]}}.

\bibitem{Grimm:2012yq}
T.~W. Grimm and W.~Taylor, ``{Structure in 6D and 4D N=1 supergravity theories from F-theory},'' \href{http://dx.doi.org/10.1007/JHEP10(2012)105}{{\em JHEP} {\bfseries 10} (2012) 105}, \href{http://arxiv.org/abs/1204.3092}{{\ttfamily arXiv:1204.3092 [hep-th]}}.

\bibitem{Lee:2019skh}
S.-J. Lee and T.~Weigand, ``{Swampland Bounds on the Abelian Gauge Sector},'' \href{http://dx.doi.org/10.1103/PhysRevD.100.026015}{{\em Phys. Rev. D} {\bfseries 100} no.~2, (2019) 026015}, \href{http://arxiv.org/abs/1905.13213}{{\ttfamily arXiv:1905.13213 [hep-th]}}.

\bibitem{Kim:2019vuc}
H.-C. Kim, G.~Shiu, and C.~Vafa, ``{Branes and the Swampland},'' \href{http://dx.doi.org/10.1103/PhysRevD.100.066006}{{\em Phys. Rev. D} {\bfseries 100} no.~6, (2019) 066006}, \href{http://arxiv.org/abs/1905.08261}{{\ttfamily arXiv:1905.08261 [hep-th]}}.

\bibitem{Katz:2020ewz}
S.~Katz, H.-C. Kim, H.-C. Tarazi, and C.~Vafa, ``{Swampland Constraints on 5d $\mathcal{N}=1$ Supergravity},'' \href{http://dx.doi.org/10.1007/JHEP07(2020)080}{{\em JHEP} {\bfseries 07} (2020) 080}, \href{http://arxiv.org/abs/2004.14401}{{\ttfamily arXiv:2004.14401 [hep-th]}}.

\bibitem{Tarazi:2021duw}
H.-C. Tarazi and C.~Vafa, ``{On The Finiteness of 6d Supergravity Landscape},'' \href{http://arxiv.org/abs/2106.10839}{{\ttfamily arXiv:2106.10839 [hep-th]}}.

\bibitem{Martucci:2022krl}
L.~Martucci, N.~Risso, and T.~Weigand, ``{Quantum gravity bounds on $ \mathcal{N} $ = 1 effective theories in four dimensions},'' \href{http://dx.doi.org/10.1007/JHEP03(2023)197}{{\em JHEP} {\bfseries 03} (2023) 197}, \href{http://arxiv.org/abs/2210.10797}{{\ttfamily arXiv:2210.10797 [hep-th]}}.

\bibitem{reid1987moduli}
M.~Reid, ``The moduli space of 3-folds with k= 0 may nevertheless be irreducible,'' {\em Mathematische Annalen} {\bfseries 278} no.~1, (1987) 329--334.

\bibitem{yau2008survey}
S.-T. Yau, ``A survey of calabi-yau manifolds,'' {\em Surveys in differential geometry} {\bfseries 13} no.~1, (2008) 277--318.

\bibitem{Gross1993AFT}
M.~Gross, ``A finiteness theorem for elliptic calabi-yau threefolds,'' {\em Duke Mathematical Journal} {\bfseries 74} (1993) 271--299. \url{https://api.semanticscholar.org/CorpusID:18131031}.

\bibitem{MR4939522}
S.~Filipazzi, C.~D. Hacon, and R.~Svaldi, ``Boundedness of elliptic {C}alabi-{Y}au threefolds,'' \href{http://dx.doi.org/10.4171/jems/1467}{{\em J. Eur. Math. Soc. (JEMS)} {\bfseries 27} no.~9, (2025) 3583--3650}. \url{https://doi.org/10.4171/jems/1467}.

\bibitem{MR4801611}
C.~Birkar, G.~Di~Cerbo, and R.~Svaldi, ``Boundedness of elliptic {C}alabi-{Y}au varieties with a rational section,'' \href{http://dx.doi.org/10.4310/jdg/1727712887}{{\em J. Differential Geom.} {\bfseries 128} no.~2, (2024) 463--519}. \url{https://doi.org/10.4310/jdg/1727712887}.

\bibitem{Kim:2024eoa}
H.-C. Kim, C.~Vafa, and K.~Xu, ``{Finite Landscape of 6d N=(1,0) Supergravity},'' \href{http://arxiv.org/abs/2411.19155}{{\ttfamily arXiv:2411.19155 [hep-th]}}.

\bibitem{Birkar:2025gvs}
C.~Birkar and S.-J. Lee, ``{A Picard rank bound for base surfaces of elliptic Calabi-Yau 3-folds},'' \href{http://arxiv.org/abs/2507.06317}{{\ttfamily arXiv:2507.06317 [hep-th]}}.

\bibitem{knuth97}
D.~E. Knuth, {\em The Art of Computer Programming, Volume 2: Seminumerical Algorithms}.
\newblock Addison-Wesley, Boston, third~ed., 1997.

\bibitem{knuth1989concrete}
R.~L. Graham, D.~E. Knuth, and O.~Patashnik, {\em Concrete Mathematics: A Foundation for Computer Science}.
\newblock Addison-Wesley, Reading, 1989.

\end{thebibliography}\endgroup

\end{document}